\documentclass[a4paper,11pt]{article}

\usepackage{amsmath}    
\usepackage{amssymb}
\usepackage{amsthm} 
\usepackage{bbm}
\usepackage{booktabs}
\usepackage{microtype}
\usepackage{a4wide}
\usepackage{todonotes}

\usepackage[ruled,vlined]{algorithm2e}

\newtheorem{theorem}{Theorem}[section]
\newtheorem{corollary}[theorem]{Corollary}
\newtheorem{lemma}[theorem]{Lemma}
\newtheorem{proposition}[theorem]{Proposition}

\theoremstyle{definition}

\newtheorem{remark}[theorem]{Remark}

\newtheorem{assumption}{Assumption}  
\newtheorem{condition}{Condition}

\usepackage{enumitem}
\usepackage{comment}
\usepackage[toc,page]{appendix}
\usepackage{color}
\usepackage{dsfont}
\usepackage{extarrows}
\usepackage{graphicx}   
\usepackage{mathrsfs}
\usepackage{verbatim}   
\usepackage{subfigure}  
\usepackage{hyperref}   

\newcommand{\Tr}{\mathrm{Tr}}

\newcommand\scalemath[2]{\scalebox{#1}{\mbox{\ensuremath{\displaystyle #2}}}}

\title{\Large \bf
Parameter Estimation for Jump-Diffusion Stochastic Master Equations 
}
\date{\today}

\author{Weichao Liang
\thanks{
{\small School of Automation Science and Engineering, Faculty of Electronic and Information Engineering, Xi’an Jiaotong University, Xianning West Road, Xi'an, 710049, Shaanxi, P.R. China, (weichao.liang@xjtu.edu.cn).}}
\and Shuixin Xiao
\thanks{
{\small Department of Electrical and Electronic Engineering, University of Melbourne, Grattan Street, Parkville, 3010, Victoria, Australia, (shuixin.xiao@anu.edu.au).}}
\and Daoyi Dong
\thanks{
{\small Australian Artificial Intelligence Institute, Faculty of Engineering and Information Technology, University of Technology Sydney, Broadway, Ultimo, 2007, New South Wales, Australia, (daoyidong@gmail.com).}}
\and Ian R. Petersen
\thanks{
{\small School of Engineering, Australian National University, North Road, Canberra, 2601, Australian Capital Territory, Australia, (i.r.petersen@gmail.com).}}
}

\begin{document}

\maketitle

\abstract{This paper investigates parameter estimation for open quantum systems under continuous observation, whose conditional dynamics are governed by jump–diffusion stochastic master equations (SMEs) associated with quantum nondemolition (QND) measurements. Estimation of model parameters such as coupling strengths or measurement efficiencies is essential, yet in practice these parameters are often uncertain. We first establish the existence and well-posedness of a reduced quantum filter: for an $N$-level system, the conditional evolution can be represented in an $\mathcal{O}(N)$-dimensional real state space rather than the full $\mathcal{O}(N^2)$ density-matrix state space. Building on this, we extend the stability theory of quantum filters, showing that exponential convergence persists not only under mismatched initial states but also in the presence of parameter mismatch. Finally, we develop an estimation framework for continuous parameter domains and prove almost sure consistency of the estimator in the long-time limit. These results yield a rigorous treatment of parameter estimation for jump–diffusion SMEs, combining structural reduction with stability and identifiability analysis, and thereby extend the mathematical theory of parameter estimation for open quantum systems.}

\section{Introduction}\label{sec1}

The theory of quantum trajectories~\cite{barchielli1995constructing,wiseman2009quantum,carmichael1993open} provides a stochastic description of open quantum systems under continuous measurements~\cite{breuer2002theory}. In this setting, the quantum filtering approach initiated by Belavkin~\cite{belavkin1992quantum} describes the conditional evolution of the system’s state via stochastic master equations (SMEs)~\cite{barchielli2009quantum}. For quantum nondemolition (QND) measurement schemes~\cite{haroche2006exploring}, where the measured observable commutes with the system Hamiltonian and is preserved by the measurement process, the conditional state evolves according to a jump–diffusion stochastic differential equation~\cite{rouchon2022tutorial} incorporating both continuous diffusive terms (from homodyne or heterodyne detection) and discontinuous jump terms (from photon counting). The asymptotic stability of quantum filters under QND measurements has been rigorously established~\cite{benoist2014large}, providing a foundation for parameter estimation on quantum trajectories.

In practice, however, precise knowledge of system parameters such as coupling strengths or measurement efficiencies is often unavailable. Parameter estimation in continuously observed quantum systems has therefore become an important area of research. Early theoretical work by Mabuchi~\cite{mabuchi1996dynamical} pioneered the real-time estimation of Hamiltonian parameters from measurement records. Subsequent contributions introduced Bayesian and particle filtering techniques for quantum parameter estimation~\cite{gambetta2001state,chase2009single,negretti2013estimation,ralph2017multiparameter}. Related approaches have been proposed for discrete-time quantum experiments and hypothesis testing scenarios~\cite{six2015parameter,bompais2022parameter}, as well as adaptive filtering schemes based on stochastic approximation and control~\cite{enami2021proposal,clausen2024online}. Despite these advances, rigorous performance guarantees for continuous parameter estimation remain limited. Most existing methods assume that the parameter belongs to a finite set and propagate multiple filters in parallel~\cite{chase2009single,six2015parameter,bompais2022parameter}. While such multiple-model approaches may identify the most likely discrete candidate, they do not guarantee that the posterior probability concentrates on the true parameter as time tends to infinity. This limitation is particularly severe when the true parameter lies outside the predefined grid.

In this work we develop a framework for continuous parameter estimation in jump–diffusion quantum filters that overcomes these finite–set limitations. We consider a general SME model for an open quantum system with uncertain initial state, coupling strength, and measurement efficiencies, continuously monitored via QND measurements involving both diffusive and counting observations. Our approach employs a reduced filter representation, which allows one to track parameters without resorting to a full Bayesian update over the $O(N^2)$-dimensional density-matrix space. Instead, a bank of reduced filtering equations is propagated for a strategically chosen set of candidate parameters, each producing a likelihood weight according to the observed measurement record and thereby yielding an online estimator.

We establish two main theoretical results underpinning this framework. We prove that quantum filters remain exponentially stable under model uncertainty, encompassing both mismatched initial states and incorrect parameter values. Even when the filter is initialized with an incorrect parameter, the conditional state converges exponentially fast toward that of the filter driven by the true dynamics, extending previous stability analyses~\cite{handel2009stability,amini2014stability,benoist2021invariant,amini2021asymptotic} to the case of mismatched models. Building on this stability property, we show that the estimation scheme achieves almost sure consistency in the following sense: as $t\to\infty$, the estimator $\hat\lambda(t)$ converges almost surely to a predetermined subinterval of the parameter domain containing the true value. Partitioning the admissible interval $[\underline{\lambda},\bar{\lambda}]$ into finitely many subintervals, each associated with a candidate filter, ensures that one subinterval contains the true parameter. By refining the partition, the estimation accuracy can be made arbitrarily fine, at the expense of increasing the number of candidate filters.

These results provide a rigorous framework for continuous online parameter estimation in SMEs with both Poisson jump and diffusive measurements. Earlier studies have largely focused on purely diffusive measurements (e.g., homodyne detection) or discrete-time settings, whereas our analysis addresses the realistic case of simultaneous photon counting and diffusion, as encountered in quantum optics~\cite{wiseman2009quantum}. The reduced-filter representation avoids the intractable propagation of full probability densities over the parameter space and is rigorously justified within our framework, together with quantitative convergence guarantees. This yields a strong form of consistency that was absent in prior methods.

Finally, the paper is organized as follows. Section~\ref{sec:model} formulates the jump–diffusion SME with uncertain parameters and presents the construction of the reduced filter together with the proof of robust exponential stability. Section~\ref{sec:para_esti} is devoted to the analysis of parameter estimation, addressing convergence of the estimator for quantities such as coupling strengths and measurement efficiencies in both homodyne/heterodyne detection and photon counting.

\textbf{Notation.}
The imaginary unit is denoted by $\mathfrak{i}$. For any finite positive integer $n$, let $[n]:=\{1,\dots,n\}$.
$\mathcal{B}(\mathbb{H})$ denotes the set of all linear operators on a finite-dimensional Hilbert space $\mathbb{H}$. Define $\mathcal{B}_{*}(\mathbb{H}):=\{X\in\mathcal{B}(\mathbb{H})|X=X^*\}$. 
Take $\mathbf{I}_{\mathbb{H}}$ as the identity operator on $\mathbb{H}$. 
The commutator of $A,B\in\mathcal{B}(\mathbb{H})$ is denoted by $[A,B]:=AB-BA.$
The function $\mathrm{Tr}(A)$ corresponds to the trace of $A\in\mathcal{B}(\mathbb{H})$. The Hilbert–Schmidt inner product is denoted by
$\langle A,B\rangle_{\mathrm{HS}}=\Tr(A^\dagger B)$ with $A,B\in \mathcal{B}(\mathbb{H})$. The Hilbert-Schmidt norm of $A\in\mathcal{B}(\mathbb{H})$ is denoted by $\|A\|:=\mathrm{Tr}(AA^*)^{1/2}$.

\section{Asymptotic properties of jump–diffusion quantum filters}\label{sec:model}
\subsection{Stochastic master equations and model setup}
We consider quantum systems described on a finite dimensional Hilbert space $\mathbb{H}$. The system state is represented by a density operator on $\mathbb{H}$,  
$$
\mathcal{S}(\mathbb{H}):=\{\rho\in\mathcal{B}(\mathbb{H})|\,\rho=\rho^*\geq 0,\mathrm{Tr}(\rho)=1\}.
$$
Let $H \in \mathcal{B}_*(\mathbb{H})$ and $L_k,C_k,A_k \in \mathcal{B}(\mathbb{H})$ for $k \in [\mathbf{j}]$. We define the following maps on $\mathcal{B}(\mathbb{H})$:
\begin{align*}
    &\mathcal{L}(\rho)=-\mathfrak{i}[H,\rho]+\textstyle\sum^{N_D}_{k=1}\gamma_k\mathcal{D}_{L_k}(\rho)+\sum^{N_J}_{k=1}\iota_k\mathcal{D}_{C_k}(\rho)+\sum^{N_P}_{k=1}\mathcal{D}_{A_k}(\rho);\\
    &\mathcal{G}_{k}(\rho)=\sqrt{\eta_k \gamma_k}\big(L_k\rho+\rho L_k^*-\mathrm{Tr}(L_k\rho+\rho L_k^*)\rho\big);\\
    &\mathcal{Q}_{k}(\rho) = \mathcal{J}_{k}(\rho)/\mathcal{T}_{k}(\rho) - \rho, 
\end{align*}
where 
\begin{align*}
    &\mathcal{D}_X(\rho)= X\rho X^*-\tfrac12X^*X\rho-\tfrac12 \rho X^*X, \\
    &\mathcal{J}_{k}(\rho)=\theta_k \rho+\textstyle\sum^{N_J}_{\bar{k}=1}\zeta_{k,\bar{k}}\iota_{\bar{k}} C_{\bar{k}}\rho C_{\bar{k}}^*,\\
    &\mathcal{T}_{k}(\rho)=\Tr(\mathcal{J}_{k}(\rho)).
\end{align*}
Here $\gamma_k\geq 0$ and $\iota_k\geq 0$ are coupling strengths, and $\eta_k\in[0,1]$, $\theta_k\geq 0$ and $\zeta_{k,\bar{k}}\geq 0$ with $\zeta_{\bar{k}}=\sum_{k}\zeta_{k,\bar{k}}\leq 1$ describe the measurement efficiency.

The dynamics of this open quantum system can be described by the following Jump-Diffusion Stochastic Master Equation (SME)~\cite{belavkin1992quantum,barchielli1995constructing,bouten2007introduction,rouchon2022tutorial}:
\begin{align}
    d\rho(t) = \mathcal{L}(\rho(t-))dt+\sum^{N_D}_{k=1}\mathcal{G}_{k}(\rho(t-))dW_k(t) + \sum^{N_J}_{k=1}\mathcal{Q}_{k}(\rho(t-))\big(d\mathsf{N}_k(t)-\mathcal{T}_{k}(\rho(t-))dt \big),\label{Eq:J-D SME}   
\end{align}
where $\{W_k\}$ are independent Wiener processes and $\{\mathsf{N}_k\}$ are counting processes with stochastic intensities $\Tr(\mathcal{J}_{k}(\rho(t-)))$. The operator $\mathcal{L}$ is the Lindblad generator, describing the averaged (unconditional) dynamics.  

The associated homodyne/heterodyne measurement record is
\begin{equation}\label{Eq:innovation}
    Y_k(t) = W_k(t) + \int_0^t \sqrt{\eta_k \gamma_k}\,
    \Tr\big((L_k+L_k^*)\rho(s-)\big) \,ds,
\end{equation}
while the photon detection record is given by $\mathsf{N}_k(t)$.

Fix a filtered probability space $(\Omega,\mathcal{F},\mathbb{F}=(\mathcal{F}_t)_{t\ge 0},\mathbb{P})$ satisfying the usual hypotheses of completeness and right continuity, in which
\begin{itemize}
    \item $W_1,\ldots, W_{N_D}$ are independent real valued Wiener processes;
    \item $N_1,\ldots, N_{N_J}$ are independent Poisson processes of intensity $dzdt$ that are independent of $W_1,\ldots, W_{N_J}$.
\end{itemize}
We assume the $(\mathcal{F}_t)$ is the natural filtration of the processes $W$ and $N$. 
On this space, the SME admits the representation
\begin{align}
    d\rho(t) = \mathcal{L}(\rho(t-))dt&+\sum^{N_D}_{k=1}\mathcal{G}_{k}(\rho(t-))dW_k(t) \nonumber\\
    &+ {\sum^{N_J}_{k=1}}\int_{\mathbb{R}}\mathcal{Q}_{k}(\rho(t-))\mathds{1}_{\{0<z<\mathcal{T}_k(\rho(t-))\}}\big(N_k(dz,dt)-dz dt \big).\label{Eq:J-D SDE}   
\end{align}
For all $k\in[N_J]$, define 
\begin{equation*}
    \mathsf{N}_k(t)=\int^t_0\int_{\mathbb{R}}\mathds{1}_{\{0<z<\mathcal{T}_k(\rho(t-))\}}N_k(dz,dt),
\end{equation*}
which is a counting process with the stochastic intensity $\int^t_0\Tr(\mathcal{J}_{k}(\rho(s-)))ds$. Hence, $\mathsf{N}_k(t)-\int^t_0\Tr(\mathcal{J}_{k}(\rho(s-)))ds$ is a $\mathcal{F}_t$ martingale under $\mathbb{P}$. Then, Equation~\eqref{Eq:J-D SDE} can be written as Equation~\eqref{Eq:J-D SME} in terms of $\mathsf{N}_k(t)$.
Existence, uniqueness, and invariance of $\mathcal{S}(\mathbb{H})$ for solutions to~\eqref{Eq:J-D SME} follow from~\cite{barchielli1995constructing,pellegrini2010markov}.

\medskip

We assume throughout the dynamics admit a generalized quantum non‐demolition (QND) structure. We consider the decomposition
$$
 \mathbb{H}=\mathbb{H}_1\oplus \dots \oplus \mathbb{H}_{\mathbf{j}}
$$
and let $\Pi_j:\mathbb{H}\to\mathbb{H}_j$ be the corresponding orthogonal projections, so that $\sum^{\mathbf{j}}_{j=1}\Pi_j=\mathbf{I}_{\mathbb{H}}$.
\begin{assumption}[QND Measurement Setting]\label{asm:qnd}
All system operators are simultaneously block–diagonal with respect to $\{\Pi_j\}$, namely
\begin{align*}
    &H=\mathrm{diag}[H_1,\dots,H_{\mathbf{j}}] \text{ with } H_j\in \mathcal{B}_*(\mathbb{H}_j), \quad \forall j\in[\mathbf{j}];\\
    &L_k=\textstyle\sum^{\mathbf{j}}_{j=1}l_{k,j}\Pi_j \text{ with } l_{k,j}\in\mathbb{C}, \quad \forall k\in[N_D];\\
    &C_k=\textstyle\sum^{\mathbf{j}}_{j=1}c_{k,j}\Pi_j \text{ with } c_{k,j}\in\mathbb{C}, \quad \forall k\in[N_J];\\
    &A_k=\mathrm{diag}[A_{k,1},\dots,A_{k,\mathbf{j}}] \text{ with } A_{k,j}\in \mathcal{B}(\mathbb{H}_j), \quad \forall j\in[\mathbf{j}], \,\forall k\in[N_P].
\end{align*}
\end{assumption}

Define
\begin{equation*}
\mathcal{I}(\mathbb{H}_j):=\{\rho\in\mathcal{S}(\mathbb{H})| \mathrm{Tr}(\Pi_j\rho)=1\}, \quad j\in[\mathbf{j}],
\end{equation*}
whose support is $\mathbb{H}_j$ or a subspace of $\mathbb{H}_j$.

\begin{lemma}\label{Lemma:Invariance}
Suppose that Assumption~\ref{asm:qnd} holds. Then, for each $j \in [\mathbf{j}]$, the set $\mathcal{I}(\mathbb{H}_j)$ is invariant in mean and $\mathbb{P}$-almost surely for the system~\eqref{Eq:J-D SME}. In other words, if $\rho_0 \in \mathcal{I}(\mathbb{H}_j)$, then $\rho(t) \in \mathcal{I}(\mathbb{H}_j)$ for all $t \geq 0$, both in mean and $\mathbb{P}$-almost surely.
\end{lemma}

Moreover, we impose the following \textit{Identifiability Assumption} on the measurement operators $L_k$ and $C_k$:
\begin{assumption}[Identifiability of measurement operators]\label{asm:identifiability}
For the diffusive measurement operators $L_k=\sum^{\mathbf{j}}_{j=1}l_{k,j}\Pi_j$ with  $k\in[N_D]$ and jump measurement operators $ C_k=\sum^{\mathbf{j}}_{j=1}c_{k,j}\Pi_j$ with  $k\in[N_J]$, we require:
\begin{enumerate}[label=\textbf{A\theassumption.\arabic*},ref=\textbf{A\theassumption.\arabic*},wide]
  \item (Diffusive distinguishability)\label{itm:A\theassumption.1}: For every $i\neq j$, there exists some $k\in[N_D]$ such that $\Re\{l_{k,i}\} \neq \Re\{l_{k,j}\}$.
  \item (Jump distinguishability)\label{itm:A\theassumption.2}: For all $i>j$ and all $k\in[N_J]$, it holds $|c_{k,i}| > |c_{k,j}|$.
\end{enumerate}
Assumption~\ref{itm:A\theassumption.2} is more restrictive than Assumption~\ref{itm:A\theassumption.1} due to the presence of cross-talk errors, i.e., $\zeta_{k,\bar{k}}\neq 0$ for some $k\neq\bar{k}$. In contrast, \ref{itm:A\theassumption.1} implicitly assumes the absence of such cross-talk. In the idealized case of no cross-talk ($\zeta_{k,\bar{k}}=0$ for all $k\neq\bar{k}$), \ref{itm:A\theassumption.2} can be relaxed to:
\begin{enumerate}[label=\textbf{A\theassumption.2'},
  ref=\textbf{A\theassumption.2'},wide]
  \item For every $i\neq j$, there exists some $k\in[N_J]$ such that $|c_{k,i}| \neq |c_{k,j}|$.
\end{enumerate}
\end{assumption}

For all $j\in[\mathbf{j}]$, define
\begin{align*}
    &\mathfrak{L}_{k}(q):= \textstyle\sum^{\mathbf{j}}_{j=1}\sqrt{\eta_k\gamma_k} \Re\{l_{k,j}\}q_j,\\
    &\Gamma_{k,j} := \theta_k + \textstyle\sum^{N_J}_{\bar{k}=1}\zeta_{k,\bar{k}}\iota_{\bar{k}} |c_{\bar{k},j}|^2,\\
    &\mathfrak{C}_{k}(q):=\textstyle\sum^{\mathbf{j}}_{j=1}\Gamma_{k,j}q_j.
\end{align*}
Consider the following Doléans–Dade type of stochastic differential equations (SDE)
\begin{align}
    dq_j(t) = q_j(t-)\Bigg[ &2\sum^{N_D}_{k=1} \big(\sqrt{\eta_k\gamma_k} \Re\{l_{k,j}\}-\mathfrak{L}_{k}(q(t-))\big)dW_k(t)\nonumber\\
    &+ \sum^{N_J}_{k=1}\Big(\frac{\Gamma_{k,j}}{\mathfrak{C}_{k}(q(t-))}-1\Big)\big(d\mathsf{N}_k(t)-\mathfrak{C}_{k}(q(t-))dt\big)\Bigg], \quad \forall j\in[\mathbf{j}].\label{Eq:dq_j}
\end{align} 
Also, we define the following simplices:
\begin{equation*}
    \mathcal{O}_{\mathbf{j}}:=\left\{q\in(0,1)^{\mathbf{j}}\,\left|\,\textstyle\sum^{\mathbf{j}}_{j=1}q_j=1\right.\right\},\quad \overline{\mathcal{O}}_{\mathbf{j}}:=\left\{q\in[0,1]^{\mathbf{j}}\,\left|\,\textstyle\sum^{\mathbf{j}}_{j=1}q_j=1\right.\right\}.
\end{equation*}
\begin{proposition}\label{Prop:Dade-SDE}
    The system~\eqref{Eq:dq_j} has a unique global solution $q(t) \in \mathcal{O}_{\mathbf{j}}$ (resp. $q(t) \in \overline{\mathcal{O}}_{\mathbf{j}}$) for all $t \geq 0$ when $q(0) \in \mathcal{O}_{\mathbf{j}}$ (resp. $q(0) \in \overline{\mathcal{O}}_{\mathbf{j}}$) $\mathbb{P}$-almost surely. This solution satisfies the following form:
    \begin{align}
    &q_j(t) = q_j(0) \nonumber \\
    &\times \exp\Bigg[ 2 \sum_{k=1}^{N_D} \int_0^t \left( \sqrt{\eta_k \gamma_k} \, \Re\{l_{k,j}\} - \mathfrak{L}_k(q(s-)) \right) \Big( dW_k(s) - \left( \sqrt{\eta_k \gamma_k} \, \Re\{l_{k,j}\} - \mathfrak{L}_k(q(s-)) \right) ds \Big) \nonumber \\
    &~~~~~~~~~ + \sum_{k=1}^{N_J} \left( \int_0^t \log \frac{\Gamma_{k,j}}{\mathfrak{C}_k(q(s-))} d\mathsf{N}_k(s) - \int_0^t \left( \Gamma_{k,j} - \mathfrak{C}_k(q(s-)) \right) ds \right) \Bigg] \label{Eq:q_j}
    \end{align}
    for all $t \geq 0$.

    Moreover, $q(t)$ is an $\mathcal{F}_t$-martingale whenever $q(0) \in \overline{\mathcal{O}}_{\mathbf{j}}$. Under Assumption~\ref{asm:qnd}, $q_j(t) = \Tr(\Pi_j \rho(t))$ for all $t \geq 0$, $\mathbb{P}$-almost surely, when $q_j(0) = \Tr(\Pi_j \rho_0)$, where $\rho(t)$ is the solution to the SME~\eqref{Eq:J-D SME}.
\end{proposition}

\textit{Proof.}
By the usual representation of the solution to Doléans–Dade SDEs~\cite[Chapter 5.1]{applebaum2009levy}, $q_j(t)$ can be written in the stochastic exponential form~\eqref{Eq:q_j}. Thus, for all $j\in[\mathbf{j}]$, $q_j(t)>0$ if $q_j(0)>0$, and $q_j(t)=0$ if $q_j(0)=0$, for all $t\geq 0$ $\mathbb{P}$-almost surely. Summing the equations~\eqref{Eq:dq_j} over all $j\in[\mathbf{j}]$ and performing a straightforward computation, we obtain the following SDE:
\begin{align*}
    d\sum^{\mathbf{j}}_{j=1}q_j(t) = \Big(1-\sum^{\mathbf{j}}_{j=1}q_j(t-) \Big)\Big[&2\sum^{N_D}_{k=1}\mathfrak{L}_k(q(t-))dW_k(t)+ \sum^{N_J}_{k=1}\big(d\mathsf{N}_k(t)-\mathfrak{C}_{k}(q(t-))dt\big)\Big].
\end{align*} 
Since the coefficients of the above equation are locally Lipschitz continuous, for any given initial state $q(0)\in \overline{\mathcal{O}}_{\mathbf{j}}$, there is a unique local solution on $t\in [0,\tau_e)$, where $\tau_e$ is the explosion time (see~\cite[Theorem 5.38]{protter2004stochastic}). Clearly, $\sum^{\mathbf{j}}_{j=1}q_j(t)=1$ when $q(0)\in \overline{\mathcal{O}}_{\mathbf{j}}$ for all $t\in [0,\tau_e)$ $\mathbb{P}$-almost surely. Thus, $\mathcal{O}_{\mathbf{j}}$ and $\overline{\mathcal{O}}_{\mathbf{j}}$ are $\mathbb{P}$-almost surely invariant for the system~\eqref{Eq:dq_j}, and $\tau_e=\infty$. 

Moreover, since $W_k(t)$ and $\mathsf{N}_k(t)-\int^t_0\mathfrak{C}_{k}(q(s-))ds$ are martingales, together with boundedness of $q(t)$ (from the invariance of $\overline{\mathcal{O}}_{\mathbf{j}}$), it follows that $q(t)$ is martingale. Under assumption \textbf{A0}, we can derive that the dynamics of $\Tr(\Pi_j \rho(t))$ satisfies the SDE~\eqref{Eq:dq_j} when $q_j(t)$ is replaced with $\Tr(\Pi_j \rho(t))$. Hence, due to the uniqueness of the solution,  $q_j(t)=\Tr(\Pi_j\rho(t))$ for all $t\geq 0$ $\mathbb{P}$-almost surely when $q_j(0)=\Tr(\Pi_j\rho_0)$. 
\hfill$\square$

To address the issue of well-definedness when $\Gamma_{k,j}=0$ for some $k,j$, we refer to the proof of \cite[Theorem 2]{benoist2014large}.

\subsection{Exponential state reduction under QND measurements}
\label{sec:EQSR}

The long-time behavior of solutions to~\eqref{Eq:J-D SME} has been studied in several works (see, e.g.,~\cite{benoist2014large,amini2021asymptotic,liang2023model}). This phenomenon, known as \emph{exponential quantum state reduction}, describes the almost-sure convergence of trajectories towards the invariant set
\begin{equation*}
\mathcal{I}(\mathbb{H}):=\cup^{\mathbf{j}}_{j=1}\mathcal{I}(\mathbb{H}_j). 
\end{equation*}

In contrast to earlier proofs based on martingale techniques~\cite{benoist2014large,amini2021asymptotic}, our argument employs a Lyapunov approach. This method not only establishes convergence but also highlights structures that can inform feedback stabilization strategies~\cite{cardona2020exponential,liang2023model}. Define the constant
$$
\mathsf{E}_{l,c}=\min_{i\neq j}\left[ \sum^{N_D}_{k=1}\eta_k \gamma_k (\Re\{l_{k,i}\}-\Re\{l_{k,j}\})^2+\sum^{N_J}_{k=1}\big(\sqrt{\Gamma_{k,i}}-\sqrt{\Gamma_{k,j}}\big)^2\right].
$$

\begin{theorem}[Exponential quantum state reduction]
Suppose Assumptions~\ref{asm:qnd} and \ref{asm:identifiability} hold. For any initial state $\rho_0\in\mathcal{S}(\mathbb{H})$, the system~\eqref{Eq:J-D SME} converges towards the invariant set $\mathcal{I}(\mathbb{H})$ in mean and $\mathbb{P}$-almost surely with average and sample Lyapunov exponent less than or equal to $-\mathfrak{E}_{l,c}/2$. 
Moreover, the probability of convergence to $\mathcal{I}(\mathbb{H}_j)$ is $\mathrm{Tr}(\Pi_j\rho_0)$ for $j\in[\mathbf{j}]$.
\label{Thm:QSR}
\end{theorem}
\textit{Proof.}
Let $I:=\{j| \, \mathrm{Tr}(\Pi_j\rho_0)=0 \}$ and $\mathcal  S_I:=\{\rho\in \mathcal S(\mathbb{H})| \,\mathrm{Tr}(\Pi_j\rho)=0\mbox{ if and only if } j\in I \}.$ Then by Proposition~\ref{Prop:Dade-SDE}, $\mathcal  S_I$ is $\mathbb{P}$-almost surely invariant for the system~\eqref{Eq:J-D SME}.
 Consider the function
\begin{equation*}
V(\rho)=\sum^{\mathbf{j}}_{\substack{i,j=1\\i \neq j}}\sqrt{\mathrm{Tr}(\Pi_i\rho)\mathrm{Tr}(\Pi_j\rho)} \geq 0,
\end{equation*}
as a candidate Lyapunov function. Note that $V(\rho)=0$ if and only if $\rho\in\mathcal{I}(\mathbb{H})$. Due to the invariance of  $\mathcal  S_I$, $V$ is twice continuously differentiable when restricted to $S_I$. Based on Lemma~\ref{Lemma:GeneratorSME}, a straightforward computation shows that
\begin{align*}
\mathscr{L} V(\rho) &= -\frac{1}{2}\textstyle\sum_{i\neq j}  \sqrt{\mathrm{Tr}(\rho \Pi_i)\mathrm{Tr}(\rho \Pi_j)} \\
&~~~~~~~~~~~~~~~~~~\times\Big[\textstyle\sum^{N_D}_{k=1}\eta_k \gamma_k (\Re\{l_{k,i}\}-\Re\{l_{k,j}\})^2+\sum^{N_J}_{k=1}\big(\sqrt{\Gamma_{k,i}}-\sqrt{\Gamma_{k,j}}\big)^2\Big]\\
&\leq -\frac{\mathsf{E}_{l,c}}{2} V(\rho).
\end{align*}
Assumption~\ref{asm:identifiability} guarantees $\mathsf{E}_{l,c}>0$. Standard stochastic Lyapunov arguments (see~\cite[Thm.~5]{liang2019exponential}, \cite[Thm.~2.5]{liang2023model}) then imply exponential convergence to $\mathcal{I}(\mathbb{H})$ with exponent $\le -\mathsf{E}_{l,c}/2$ in mean and almost surely. The distribution of the limit is determined by~\cite{benoist2014large,liang2019exponential}, yielding $\mathbb{P}(\rho(t)\to \mathcal{I}(\mathbb{H}_j))=\Tr(\Pi_j\rho_0)$.
\hfill$\square$

Theorem~\ref{Thm:QSR}, together with Proposition~\ref{Prop:Dade-SDE} and~\cite[Thm.~4]{benoist2014large}, leads to the following asymptotic characterization, which is crucial for applying Girsanov’s theorem~\cite[Ch.~19.2]{liptser2013statistics} on infinite horizons.

\begin{corollary}[Asymptotic convergence and limiting distributions]\label{Cor:Convergence}
Under Assumptions~\ref{asm:qnd} and \ref{asm:identifiability}, for any initial state $\rho_0 \in \mathcal{S}(\mathbb{H})$ and initial condition $q_j(0) = \Tr(\Pi_j \rho_0)$ for all $j \in [\mathbf{j}]$, there exists a random variable $\mathcal{R}$ taking values in $[\mathbf{j}]$ such that
\begin{align*}
    &\lim_{t \rightarrow \infty} \Tr(\Pi_j \rho(t)) = \lim_{t \rightarrow \infty} q_j(t) = \mathds{1}_{\{\mathcal{R} = j\}}, \quad \mathbb{P}\text{-a.s. and in } L^1,\\
    &\lim_{t \rightarrow \infty} \Tr(\Pi_{\mathcal{R}} \rho(t)) = \lim_{t \rightarrow \infty} q_{\mathcal{R}}(t) = 1, \quad \mathbb{P}\text{-a.s.}
\end{align*}
where $\rho(t)$ solves the SME~\eqref{Eq:J-D SME}, and $q(t)$ solves the SDE~\eqref{Eq:dq_j}. Furthermore, for every $j \in [\mathbf{j}]$,
\begin{align*}
    &\mathbb{P}(\mathcal{R} = j \mid \mathcal{F}_t) = \Tr(\Pi_j \rho(t)) = q_j(t), \quad \mathbb{P}\text{-a.s.,}\\
    &\mathbb{P}(\mathcal{R} = j) = \Tr(\Pi_j \rho_0) = q_j(0).
\end{align*}
\end{corollary}

Next, following the approach in \cite{benoist2014large}, we introduce a change of measure, which is essential in establishing the robust stability of the estimated quantum filter in the subsequent subsections. Under Assumption~\ref{asm:qnd} and~\ref{asm:identifiability},  for all $j\in[\mathbf{j}]$ such that $q_j(0)=\Tr(\Pi_j\rho_0)>0$, we define a probability measure $\mathbb{Q}^{j}_t$ on $(\Omega,\mathcal{F}_t)$ by
\begin{align*}
    d\mathbb{Q}^{j}_t=\frac{q_j(t)}{q_j(0)} d\mathbb{P}.
\end{align*}
By Proposition~\ref{Prop:Dade-SDE}, the process $q_j(t)/q_j(0)$ is a nonnegative $\mathbb{P}$-martingale with unit mean, hence $(\mathbb{Q}^j_t)_{t\ge0}$ is consistent. This consistency extends to a unique probability measure $\mathbb{Q}^j$ on $\mathcal{F}_\infty$
since $\mathbb{E}(\mathds{1}_{\{\mathcal{R} = j\}}|\mathcal{F}_t)=q_j(t)$ for all $t\geq 0$, as ensured by Corollary \ref{Cor:Convergence}. In particular, we have 
\begin{equation}
    d\mathbb{Q}^{j}=\frac{\mathds{1}_{\{\mathcal{R} = j\}}}{q_j(0)}d\mathbb{P}.
    \label{Eq:ChangeMeasure}
\end{equation}
For all $j\in [\mathbf{j}]$, define 
\begin{align}
    &\overline{W}^{j}_k(t) = W_k(t)-\int^t_0 2  \big(\sqrt{\eta_k \gamma_k}\Re\{l_{k,j}\}-\mathfrak{L}_{k}(q(s-))\big)ds, \quad \forall k\in[N_D],\label{Eq:W^j_k}\\
    &\overline{\mathsf{N}}^{j}_k(t) = \mathsf{N}_k(t)-\Gamma_{k,j}t, \quad \forall k\in[N_J]\label{Eq:N^j_k}.
\end{align}
By applying Girsanov's theorem~\cite[Chapter 19.2]{liptser2013statistics}, we obtain the following Lemma.
\begin{lemma}\label{Lemma:Girsanov}
    For any $j\in [\mathbf{j}]$ such that $q_j(0)>0$, under probability measure $\mathbb{Q}^{j}$, $\overline{W}^{j}_k(t)$ with $k\in [N_D]$ are Wiener processes and $\mathsf{N}_k(t)$ with $k\in[N_J]$ are Poisson processes with intensities $\Gamma_{k,j}$.
\end{lemma}

\subsection{Robust exponential stability under parameter mismatch}

In practice, neither the initial state $\rho_0$ nor the physical parameters of the system can be known exactly.  Hence the true conditional state $\rho(t)$ solving~\eqref{Eq:J-D SME} is not directly available, although the measurement records $\{Y_k(t)\}_{k\in[N_D]}$ and $\{\mathsf{N}_k(t)\}_{k\in[N_J]}$ are observable. Most existing results address filter stability under perfectly known parameters~\cite{handel2009stability,amini2014stability,amini2021asymptotic,benoist2014large}. By contrast, following the framework of~\cite{liang2023model}, we establish robust exponential stability for an estimated quantum filter $\hat{\rho}(t)$ that operates with possibly mismatched parameters.

We assume the following estimated model parameters:
\begin{itemize}
    \item Estimated couplings: $\hat{\gamma}_k,\hat{\iota}_k\ge0$ for all $k$;
    \item Estimated Hamiltonian: $\hat{H}=\mathrm{diag}[\hat{H}_1,\dots,\hat{H}_{\mathbf{j}}]$ with $\hat{H}_j\in\mathcal{B}_*(\mathbb{H}_j)$;
    \item Estimated noise operators: $\hat{A}_k=\mathrm{diag}[\hat{A}_{k,1},\dots,\hat{A}_{k,\mathbf{j}}]$ with $\hat{A}_{k,j}\in\mathcal{B}(\mathbb{H}_j)$;
    \item Estimated measurement efficiencies: $\hat{\eta}_k\in[0,1]$, $\hat{\theta}_k\ge0$, $\hat{\zeta}_{k,\bar{k}}\ge0$ with $\sum_{k=1}^{N_J}\hat{\zeta}_{k,\bar{k}}\le1$.
\end{itemize}

The estimated conditional state $\hat{\rho}(t)$ evolves according to the jump–diffusion SME:
\begin{align}
    d\hat{\rho}(t) = &\hat{\mathcal{L}}(\hat{\rho}(t-))dt+\sum^{N_D}_{k=1}\hat{\mathcal{G}}_{k}(\hat{\rho}(t-))\big(dY_k(t)-\sqrt{\hat{\eta}_k \hat{\gamma}_k}\mathrm{Tr}(L_k\hat{\rho}(t-)+\hat{\rho}(t-) L_k^*)dt\big) \nonumber\\
    &+ {\sum^{N_J}_{k=1}}\int_{\mathbb{R}}\hat{\mathcal{Q}}_{k}(\hat{\rho}(t-))\big(d\mathsf{N}_k(t)-\Tr(\hat{\mathcal{J}}_{k}(\hat{\rho}(t-)))dt \big), \quad \hat{\rho}(0)=\hat{\rho}_0\in\mathcal{S}(\mathbb{H}),\label{Eq:J-D filter}
\end{align}
with
\begin{align*}
    &\hat{\mathcal{L}}(\rho)=-\mathfrak{i}[\hat{H},\rho]+\textstyle\sum^{N_D}_{k=1}\hat{\gamma}_k\mathcal{D}_{L_k}(\rho)+\sum^{N_J}_{k=1}\hat{\iota}_k\mathcal{D}_{C_k}(\rho)+\sum^{N_P}_{k=1}\mathcal{D}_{\hat{A}_k}(\rho),\\
    &\hat{\mathcal{G}}_{k}(\rho)=\sqrt{\hat{\eta}_k\hat{\gamma}_k}\big(L_k \rho+\rho L_k^*-\Tr(L_k \rho+ \rho L_k^*)\rho \big), \\
    &\hat{\mathcal{Q}}_{k}(\rho) = \hat{\mathcal{J}}_{k}(\rho)/\hat{\mathcal{T}}_{k}(\rho) - \rho, \\
    &\hat{\mathcal{J}}_{k}(\rho)=\hat{\theta}_k \rho+\textstyle\sum^{N_J}_{\bar{k}=1}\hat{\zeta}_{k,\bar{k}}\hat{\iota}_{\bar{k}} C_{\bar{k}}\rho C_{\bar{k}}^*,\\
    &\hat{\mathcal{T}}_{k}(\rho)=\Tr(\hat{\mathcal{J}}_{k}(\rho)).
\end{align*}
Existence, uniqueness, and invariance of $\mathcal{S}(\mathbb{H})$ follow from~\cite{barchielli1995constructing,pellegrini2010markov}.  Using~\eqref{Eq:innovation}, one can rewrite~\eqref{Eq:J-D filter} in terms of the driving Wiener processes:
\begin{align}
    d\hat{\rho}(t) = &\hat{\mathcal{L}}(\hat{\rho}(t-))dt+\sum^{N_D}_{k=1}\hat{\mathcal{G}}_{k}(\hat{\rho}(t-))\big(dW_k(t)+\mathcal{E}_k(\rho(t-),\hat{\rho}(t-))dt\big) \nonumber\\
    &+ {\sum^{N_J}_{k=1}}\int_{\mathbb{R}}\hat{\mathcal{Q}}_{k}(\hat{\rho}(t-))\big(d\mathsf{N}_k(t)-\Tr(\hat{\mathcal{J}}_{k}(\hat{\rho}(t-)))dt \big), \quad \hat{\rho}(0)=\hat{\rho}_0\in\mathcal{S}(\mathbb{H}),\label{Eq:J-D filter-W}
\end{align}    
where $\mathcal{E}_k(\rho,\hat{\rho})=\sqrt{\eta_k \gamma_k}\mathrm{Tr}(L_k\rho+\rho L_k^*)-\sqrt{\hat{\eta}_k \hat{\gamma}_k}\mathrm{Tr}(L_k\hat{\rho}+\hat{\rho}L_k^*)$.

Analogously to Proposition~\ref{Prop:Dade-SDE}, one derives reduced dynamics for the block probabilities $\hat{q}_j(t)=\Tr(\Pi_j\hat{\rho}(t))$. Define
\begin{align*}
    &\hat{\mathfrak{L}}_{k}(\hat{q}):= \sum^{\mathbf{j}}_{j=1}\sqrt{\hat{\eta}_k\hat{\gamma}_k}\Re\{l_{k,j}\}\hat{q}_j,\quad \hat{\Gamma}_{k,j} := \hat{\theta}_k + \sum^{N_J}_{\bar{k}=1}\hat{\zeta}_{k,\bar{k}}\hat{\iota}_{\bar{k}}|c_{\bar{k},j}|^2,\quad \hat{\mathfrak{C}}_{k}(\hat{q}):=\sum^{\mathbf{j}}_{j=1}\hat{\Gamma}_{k,j}\hat{q}_j.
\end{align*}
Then $\hat{q}(t)=(\hat{q}_1(t),\dots,\hat{q}_{\mathbf{j}}(t))$ satisfies the Doléans–Dade type SDE
\begin{align}
    d\hat{q}_j(t) = \hat{q}_j(t-)\Big[ & 2\sum^{N_D}_{k=1} \big(\sqrt{\hat{\eta}_k\hat{\gamma}_k} \Re\{l_{k,j}\}-\hat{\mathfrak{L}}_{k}(\hat{q}(t-))\big)\big(dY_k(t)-2\hat{\mathfrak{L}}_{k}(\hat{q}(t-))dt\big)\nonumber\\
    &+ \sum^{N_J}_{k=1}\Big(\frac{\hat{\Gamma}_{k,j}}{\hat{\mathfrak{C}}_{k}(\hat{q}(t-))}-1\Big)\big(d\mathsf{N}_k(t)-\hat{\mathfrak{C}}_{k}(\hat{q}(t-))dt\big)\Big]\label{Eq:d hat q_j}\\
    =\hat{q}_j(t-)\Big[ &2\sum^{N_D}_{k=1} \big(\sqrt{\hat{\eta}_k\hat{\gamma}_k} \Re\{l_{k,j}\}-\hat{\mathfrak{L}}_{k}(\hat{q}(t-))\big)\big(dW_k(t)+\mathfrak{E}_k(q(t-),\hat{q}(t-))dt\big)\nonumber\\
    &+ \sum^{N_J}_{k=1}\Big(\frac{\hat{\Gamma}_{k,j}}{\hat{\mathfrak{C}}_{k}(\hat{q}(t-))}-1\Big)\big(d\mathsf{N}_k(t)-\hat{\mathfrak{C}}_{k}(\hat{q}(t-))dt\big)\Big],\nonumber
\end{align}
where
$\mathfrak{E}_k(q,\hat{q})=2\mathfrak{L}_{k}(q)-2\hat{\mathfrak{L}}_{k}(\hat{q})$ and $q(t)$ is the solution to the SDE~\eqref{Eq:dq_j} with $q_j(0)=\Tr(\Pi_j \rho_0)$. 

By similar arguments as in the proof of Proposition~\ref{Prop:Dade-SDE}, we can obtain the following results.
\begin{proposition}\label{Prop:Dade-SDE-F}
    The system~\eqref{Eq:d hat q_j} has a unique global solution $\hat{q}(t)\in \mathcal{O}_{\mathbf{j}}$ (resp. $\hat{q}(t)\in \overline{\mathcal{O}}_{\mathbf{j}}$) for all $t\geq 0$ when $\hat{q}(0)\in \mathcal{O}_{\mathbf{j}}$ (resp. $\hat{q}(0)\in \overline{\mathcal{O}}_{\mathbf{j}}$), $\mathbb{P}$-almost surely, which satisfies the following 
    \begin{align}
    \hat{q}_j(t)\nonumber\\
    =&\hat{q}_j(0)\nonumber\\
    &\times\exp\Bigg[ 2\sum^{N_D}_{k=1} \Bigg(\int^t_0 \big(\sqrt{\hat{\eta}_k\hat{\gamma}_k}\Re\{l_{k,j}\}-\hat{\mathfrak{L}}_{k}(\hat{q}(t-))\big)\big(dY_k(s)-2\hat{\mathfrak{L}}_{k}(\hat{q}(s-))ds\big)\nonumber\\ 
    &~~~~~~~~~~~~~~~~~~~~~-\int^t_0\big(\sqrt{\hat{\eta}_k\hat{\gamma}_k}\Re\{l_{k,j}\}-\hat{\mathfrak{L}}_{k}(\hat{q}(s-))\big)^2ds\Bigg)\nonumber\\ 
    &~~~~~~~~~+\sum^{N_J}_{k=1}\Bigg(\int^t_0\log \frac{\hat{\Gamma}_{k,j}}{\hat{\mathfrak{C}}_{k}(\hat{q}(s-))}d\mathsf{N}_k(s)-\int^t_0\big(\hat{\Gamma}_{k,j}-\hat{\mathfrak{C}}_{k}(\hat{q}(s-))\big) ds\Bigg)\Bigg]\label{Eq:Y-qSDE-F}\\
    =&\hat{q}_j(0)\nonumber\\
        &\times\exp\Bigg[ 2\sum^{N_D}_{k=1} \Bigg(\int^t_0 \big(\sqrt{\hat{\eta}_k\hat{\gamma}_k}\Re\{l_{k,j}\}-\hat{\mathfrak{L}}_{k}(\hat{q}(t-))\big)\big(dW_k(s)+\mathfrak{E}_k(q(s-),\hat{q}(s-))ds\big)\nonumber\\ 
    &~~~~~~~~~~~~~~~~~~~~~- \int^t_0\big(\sqrt{\hat{\eta}_k\hat{\gamma}_k}\Re\{l_{k,j}\}-\hat{\mathfrak{L}}_{k}(\hat{q}(s-))^2\big)ds\Bigg)\nonumber\\ 
    &~~~~~~~~~+\sum^{N_J}_{k=1}\Bigg(\int^t_0\log \frac{\hat{\Gamma}_{k,j}}{\hat{\mathfrak{C}}_{k}(\hat{q}(s-))}d\mathsf{N}_k(s)-\int^t_0\big(\hat{\Gamma}_{k,j}-\hat{\mathfrak{C}}_{k}(\hat{q}(s-))\big) ds\Bigg)\Bigg],\label{Eq:W-qSDE-F}
\end{align}
for all $t\geq 0$. Moreover, under Assumption~\ref{asm:qnd}, $\hat{q}_j(t)=\Tr(\Pi_j\hat{\rho}(t))$ for all $t\geq 0$ $\mathbb{P}$-almost surely when $\hat{q}_j(0)=\Tr(\Pi_j\hat{\rho}_0)$ where $\hat{\rho}(t)$ is the solution to SME~\eqref{Eq:J-D filter}.
\end{proposition}

We next give sufficient conditions on the estimated initial state $\hat{\rho}_0\in\mathcal{S}(\mathbb{H})$ and on the estimated model parameters that ensure the estimated filter $\hat{\rho}(t)$ asymptotically tracks the true conditional trajectory $\rho(t)$, despite initial and parametric mismatches.

We introduce the following mismatch functionals
\begin{align*}
    &\Phi^k_{i,j}(\eta,\gamma)=2\hat{\eta}_k\hat{\gamma}_k\Big[ (\Re\{l_{k,i}\}-\Re\{l_{k,j}\})^2+2(\Re\{l_{k,i}\}-\Re\{l_{k,j}\}) \Re\{l_{k,j}\}\Big(1-\sqrt{\frac{\eta_k \gamma_k}{\hat{\eta}_k\hat{\gamma}_k}}\Big)\Big],\\
    &\Psi^k_{i,j}(\Gamma)=\log\frac{\hat{\Gamma}_{k,i}}{\hat{\Gamma}_{k,j}}\Big(\hat{\Gamma}_{k,j}-\Gamma_{k,j}\Big)-\hat{\Gamma}_{k,j}\Big(1-\frac{\hat{\Gamma}_{k,i}}{\hat{\Gamma}_{k,j}}+\log\frac{\hat{\Gamma}_{k,i}}{\hat{\Gamma}_{k,j}}\Big) .
\end{align*}
Heuristically, $\Phi$ quantifies the diffusive (homodyne/heterodyne) mismatch and $\Psi$ the jump (counting) mismatch; both collapse to strictly nonnegative “Fisher–type” gaps at perfect calibration (see Remark~\ref{Rem:known parameter}).

\begin{condition}[Parameter mismatch tolerance]\label{con:robust_SME}
    For all $i\neq j$, $$\sum^{N_D}_{k=1}\Phi^k_{i,j}(\eta,\gamma)+\sum^{N_D}_{k=1}\Psi^k_{i,j}(\Gamma)>0.$$
\end{condition}

\begin{theorem}[Robust exponential stability]\label{Thm:QSR-F}
    Suppose Assumptions~\ref{asm:qnd} and~\ref{asm:identifiability} hold,  along with Condition~\ref{con:robust_SME}. Assume that $\hat{\Gamma}_{k,j}>0$ for all $k\in[N_J]$ and for all $j\in[\mathbf{j}]$. Then, for any initial state $\hat{\rho}_0\in\mathrm{int}\{\mathcal{S}(\mathbb{H})\}$, 
    \begin{align*}
        &\mathbb{P}\big(\lim_{t\rightarrow \infty}\Tr(\Pi_j\hat{\rho}(t))=1\big) = \mathbb{P}(\mathcal{R}=j),\\
        &\lim_{t \rightarrow \infty} \Tr(\Pi_{\mathcal{R}} \hat{\rho}(t)) = \lim_{t \rightarrow \infty} \hat{q}_{\mathcal{R}}(t) = 1, \quad \mathbb{P}\text{-a.s.}\\
        &\lim_{t\rightarrow \infty} \frac{1}{t}\log \frac{\hat{q}_j(t)}{\hat{q}_{\mathcal{R}}(t)}=\textstyle-\sum^{N_D}_{k=1}\Phi^k_{j,\mathcal{R}}(\eta,\gamma)-\sum^{N_J}_{k=1}\Psi^k_{j,\mathcal{R}}(\Gamma),\quad \mathbb{P}\text{-a.s.}
    \end{align*}
where the random variable $\mathcal{R}$ is defined in Corollary~\ref{Cor:Convergence}. 
\end{theorem}

\proof
If $\hat{\rho}_0\in\operatorname{int}\mathcal{S}(\mathbb{H})$, then $\hat{q}_j(0)=\Tr(\Pi_j\hat{\rho}_0)>0$ for all $j$; by the stochastic exponential form~\eqref{Eq:W-qSDE-F}, $\hat{q}_j(t)>0$ for all $t$, $\mathbb{P}$-a.s.

Let $\mathcal{P} := \{j \mid \Tr(\Pi_j \rho_0) > 0\}$ where $\rho_0 \in \mathcal{S}(\mathbb{H})$ is the initial state of the system~\eqref{Eq:J-D SME}, and set $q_j(0) = \Tr(\Pi_j \rho_0)$ for all $j \in [\mathbf{j}]$, where $q(t)$ is the solution to SDE~\eqref{Eq:dq_j}. Then, for any $i \neq j$ with $j \in \mathcal{P}$, we can derive that
\begin{align*}
    \frac{\hat{q}_i(t)}{\hat{q}_j(t)} = \frac{\hat{q}_i(0)}{\hat{q}_j(0)} \exp \Bigg[ &\sum_{k=1}^{N_D} \Big(2 \sqrt{\hat{\eta}_k \hat{\gamma}_k} (\Re\{l_{k,i}\} - \Re\{l_{k,j}\}) \overline{W}^{j}_k(t) - \Phi^k_{i,j}(\eta,\gamma) t \Big) \\
    & + \sum_{k=1}^{N_J} \Big(\log \frac{\hat{\Gamma}_{k,i}}{\hat{\Gamma}_{k,j}} \overline{\mathsf{N}}^{j}_k(t) - \Psi^k_{i,j}(\Gamma) t \Big) \Bigg],
\end{align*}
where, under the probability measure $\mathbb{Q}^{j}$, $\overline{W}^{j}_k(t)$ for $k \in [N_D]$ are Wiener processes and $\overline{\mathsf{N}}^{j}_k(t)$ for $k \in [N_J]$ are centered Poisson processes (Lemma~\ref{Lemma:Girsanov}).  By the strong law of large numbers for martingales, $\overline{W}^j_k(t)/t\to0$ and $\overline{\mathsf{N}}^j_k(t)/t\to0$, $\mathbb{Q}^j$-a.s., hence
\begin{align*}
    \lim_{t \rightarrow \infty} \frac{1}{t} \log \frac{\hat{q}_i(t)}{\hat{q}_j(t)} = & \lim_{t \rightarrow \infty} \frac{1}{t} \Bigg( \log \frac{\hat{q}_i(0)}{\hat{q}_j(0)} + \sum_{k=1}^{N_D} 2 \sqrt{\hat{\eta}_k \hat{\gamma}_k} (\Re\{l_{k,i}\} - \Re\{l_{k,j}\}) \overline{W}^{j}_k(t) \\
    & + \sum_{k=1}^{N_J} \log \frac{\hat{\Gamma}_{k,i}}{\hat{\Gamma}_{k,j}} \overline{\mathsf{N}}^{j}_k(t) \Bigg) - \sum_{k=1}^{N_D} \Phi^k_{i,j} - \sum_{k=1}^{N_J} \Psi^k_{i,j} \\
    = & - \sum_{k=1}^{N_D} \Phi^k_{i,j}(\eta,\gamma) - \sum_{k=1}^{N_J} \Psi^k_{i,j}(\Gamma) < 0, \quad \mathbb{Q}^{j}\text{-a.s.}
\end{align*}
where the negativity is ensured by condition~\ref{con:robust_SME}.
This implies that, for all $i \neq j$, $\hat{q}_i(t)$ converges exponentially to zero $\mathbb{Q}^{j}$-almost surely with the sample Lyapunov exponent equal to $- \sum_{k=1}^{N_D} \Phi^k_{i,j}(\eta,\gamma) - \sum_{k=1}^{N_J} \Psi^k_{i,j}(\Gamma)$. Hence, $\hat{q}_j(t)$ converges exponentially to one $\mathbb{Q}^{j}$-almost surely since $\hat{q}_j(t) = 1 - \sum_{i \neq j} \hat{q}_i(t)$ (Proposition~\ref{Prop:Dade-SDE-F}). Letting $E_{i} = \{\omega \in \Omega \mid \lim_{t \rightarrow \infty} \hat{q}_i(t) = 1\}$, we have $\mathbb{Q}^{j}(E_{i}) = \delta_{i,j}$ for all $j \in \mathcal{P}$.

Using~\eqref{Eq:ChangeMeasure} and Corollary~\ref{Cor:Convergence}, for any event $X \in \mathbb{F}$,
\begin{align*}
    \mathbb{Q}^{j}(X) = \frac{\mathbb{E}\big(\mathds{1}_{\{X\}} \mathds{1}_{\{\mathcal{R} = j\}}\big)}{q_j(0)} = \frac{\mathbb{P}(X \mid \mathcal{R} = j) \mathbb{P}(\mathcal{R} = j)}{q_j(0)} = \mathbb{P}(X \mid \mathcal{R} = j), \quad j \in \mathcal{P},
\end{align*}
and $\sum_{j \in \mathcal{P}} \mathbb{P}(\mathcal{R} = j) = \sum_{j \in \mathcal{P}} \Tr(\Pi_j \rho_0) = 1$. Hence
\begin{align*}
    \mathbb{P}(E_{j}) &= \sum_{i \in \mathcal{P}} \mathbb{P}(\mathcal{R} = i) \mathbb{P}(E_{j} \mid \mathcal{R} = i) = \mathbb{P}(\mathcal{R} = j), \quad \forall j \in \mathcal{P},
\end{align*}
and
\begin{align*}
    \lim_{t \rightarrow \infty} \frac{1}{t} \log \frac{\hat{q}_i(t)}{\hat{q}_{\mathcal{R}}(t)} = -\sum^{N_D}_{k=1}\Phi^k_{j,\mathcal{R}}(\eta,\gamma)-\sum^{N_J}_{k=1}\Psi^k_{j,\mathcal{R}}(\Gamma), \quad \mathbb{P}\text{-a.s.}
\end{align*}
Due to the invariance of $\overline{\mathcal{O}}_{\mathbf{j}}$ for $\hat{q}(t)$ (see Proposition~\ref{Prop:Dade-SDE-F}), we deduce that
\begin{align*}
    \lim_{t \rightarrow \infty} \hat{q}_j(t) = 0, \quad \forall j \notin \mathcal{P}, \quad \mathbb{P}\text{-a.s.}
\end{align*}
Combining this with the fact that $q_j(t) = 0$ for all $t \geq 0$ and for all $j \notin \mathcal{P}$, $\mathbb{P}$-almost surely (see Proposition~\ref{Prop:Dade-SDE}), we can conclude the proof.
\hfill$\square$

\begin{remark}\label{Rem:known parameter}
    In the perfectly calibrated case  $\eta_k \gamma_k=\hat{\eta}_k\hat{\gamma}_k$ and $\Gamma_{k,j}=\hat{\Gamma}_{k,j}$, 
    \begin{align*}
    &\Phi^k_{i,j}(\eta,\gamma)=2\hat{\eta}_k\hat{\gamma}_k(\Re\{l_{k,i}\}-\Re\{l_{k,j}\})^2\geq 0, \quad \Psi^k_{i,j}(\Gamma)=-\hat{\Gamma}_{k,j}\Big(1-\frac{\hat{\Gamma}_{k,i}}{\hat{\Gamma}_{k,j}}+\log\frac{\hat{\Gamma}_{k,i}}{\hat{\Gamma}_{k,j}}\Big)\geq 0,
\end{align*}
with strict positivity whenever Assumption~\ref{asm:identifiability} is satisfied. Thus Condition~\ref{con:robust_SME} is automatically satisfied in the identifiable, perfectly known-parameter regime.
\end{remark}

\section{Parameter estimation via  maximum likelihood}\label{sec:para_esti}
We now develop a parameter estimation framework for jump–diffusion SMEs under QND measurements (Assumption~\ref{asm:qnd}). Let $\lambda\in[\underline{\lambda},\bar{\lambda}]$ denote the true parameter and $\hat{\lambda}$ its estimate. The estimate is required to satisfy a prescribed accuracy criterion, e.g., $\lambda/\hat{\lambda}\in[\bar{a},\bar{b}]$ with $0<\bar{a}<1<\bar{b}$. Inspired by system dilation methods~\cite{chase2009single,negretti2013estimation,six2015parameter,bompais2022parameter}, we propagate $\mathbf{n}$ parallel filters, each associated with a candidate parameter, and assign to each a likelihood $\pi_n(t)$ based on the measurement record up to time $t$. The maximum likelihood estimator then selects $\hat{\lambda}_{n^*}$ corresponding to the dominant likelihood. 


Earlier approaches~\cite{chase2009single,six2015parameter,bompais2022parameter} exploited the stability of quantum filters~\cite{handel2009stability,amini2014stability,benoist2014large} but were restricted to finite parameter sets $\{\lambda_1,\dots,\lambda_{\mathbf{n}}\}$. In practice, physical parameters vary over continuous intervals. Discretization methods~\cite{six2015parameter} or likelihood-based refinements~\cite{bompais2022parameter} provide approximations, yet without rigorous guarantees of consistency when the true parameter lies outside the candidate grid.


Here we establish a mathematical foundation for maximum likelihood estimation on continuous parameter domains. The key ingredient is the robust exponential stability of quantum filters under parameter mismatch (Theorem~\ref{Thm:QSR-F}), quantified by the tolerance Condition~\ref{con:robust_SME}. This allows us to partition the admissible interval $[\underline{\lambda},\bar{\lambda}]$ into subintervals, each ensuring stability of the associated filter. One of such subintervals contains the true parameter, and by refining the partition the estimation accuracy can be made arbitrarily high. The MLE then selects $\hat{\lambda}_{n^*}$ corresponding to that subinterval, yielding an almost sure consistent estimator in the long-time limit.


In the following, we detail this construction for scalar parameters  (e.g., $\gamma_k$, $\eta_k$, $\theta_k$, $\zeta_{k,\bar{k}}$, $\iota_k$).
This can be extended to a vector of parameters by further dilating the system.

\subsection{Augmented stochastic master equations}
According to Theorem~\ref{Thm:QSR-F}, robust stability of the quantum filter depends on tolerances between the true and estimated parameters, notably the measurement efficiencies and coupling strengths. Our objective is to estimate parameters $\lambda$ such that
\[
\lambda/\hat{\lambda} \in [\bar{a},\bar{b}], \qquad 0<\bar{a}<1<\bar{b},
\]
where the class of admissible parameters includes $\gamma_k,\eta_k$ for $k\in[N_D]$ and $\theta_k,\zeta_{k,\bar{k}},\iota_k$ for $k,\bar{k}\in[N_J]$.

Given the coarse prior information that \(\lambda \in [\underline{\lambda}, \bar{\lambda}]\) with \(\underline{\lambda}>0\), we partition this interval into \(\mathbf{n}\) subintervals:
\[[a\hat{\lambda}_1, b\hat{\lambda}_1),\,(a\hat{\lambda}_2, b\hat{\lambda}_2),\,\dots,\,(a\hat{\lambda}_{\mathbf{n}}, b\hat{\lambda}_{\mathbf{n}}],\]
where \(\underline{\lambda} \in [a\hat{\lambda}_{1}, b\hat{\lambda}_{1})\), \(\bar{\lambda} \in (a\hat{\lambda}_{\mathbf{n}}, b\hat{\lambda}_{\mathbf{n}}]\), and the parameters satisfy \(0<\bar{a}\leq a<1<b\leq \bar{b}\). The true parameter \(\lambda\) belongs to one of these subintervals, say \(\lambda \in (a\hat{\lambda}_{n^*}, b\hat{\lambda}_{n^*})\). The problem of estimation therefore reduces to identifying the unknown index $n^*$. Since the partition may not cover the full admissible range, refinements of $a,b$, and $\mathbf{n}$ are addressed in subsequent subsections.

To formalize this construction we embed the candidate parameters into an augmented Hilbert space
\(
\overline{\mathbb{H}} = \mathbb{R}^{\mathbf{n}} \otimes \mathbb{H},
\)
and define an augmented SME on $\overline{\mathbb{H}}$. This allows us to describe simultaneously the evolution of all likelihood weights $\pi_n(t)$ for $n\in[\mathbf{n}]$. Let $\{|n\rangle\}_{n=1}^{\mathbf{n}}$ be the canonical basis of $\mathbb{R}^{\mathbf{n}}$, with projectors $|n\rangle\langle n|$. We then define block–diagonal operators acting on $\overline{\mathbb{H}}$, e.g.,
\begin{align*}
    &\mathbf{H}(\mu):=\textstyle\sum^{\mathbf{n}}_{n=1} |n\rangle\langle n|\otimes H(\mu),\quad \mu\in\mathbb{R}, \\    
    &\mathbf{A}_k(\nu_k):=\textstyle\sum^{\mathbf{n}}_{n=1}|n\rangle\langle n|\otimes A_k(\nu_{k}),\quad \nu_{k}\in\mathbb{R}, \quad  \forall k\in[N_P],\\
    &\mathbf{L}_k(\gamma_k):=\textstyle\sum^{\mathbf{n}}_{n=1}|n\rangle\langle n|\otimes\sqrt{\gamma_{k}} L_k,\quad \gamma_{k}\geq 0, \quad \forall k\in[N_D],\\    
    &\bar{\mathbf{L}}_k(\eta_k,\gamma_k):=\textstyle\sum^{\mathbf{n}}_{n=1}|n\rangle\langle n|\otimes \sqrt{\eta_{k}\gamma_{k}} L_k,\quad \eta_{k}\in(0,1] \text{ and }\gamma_k> 0, \quad \forall k\in[N_D],\\
    &\boldsymbol{\Theta}_k(\theta_k) :=\textstyle\sum^{\mathbf{n}}_{n=1}|n\rangle\langle n|\otimes \theta_{k}\mathbf{I}_{\mathbb{H}},\quad \theta_{k}\geq 0, \quad\forall k\in[N_J],\\
    &\mathbf{C}_k(\iota_k):=\textstyle\sum^{\mathbf{n}}_{n=1}|n\rangle\langle n|\otimes \sqrt{\iota_{k}}C_k,\quad \iota_{k}> 0, \quad\forall k\in[N_J],\\    
    &\bar{\mathbf{C}}_{k,\bar{k}}(\zeta_{k,\bar{k}},\iota_{\bar{k}} ):=\textstyle\sum^{\mathbf{n}}_{n=1}|n\rangle\langle n|\otimes\sqrt{\zeta_{k,\bar{k}}\iota_{\bar{k}}} C_{\bar{k}},\quad  \zeta_{k,\bar{k}}\geq 0 \text{ and } \iota_{k}> 0, \quad\forall k,\bar{k}\in[N_J].
\end{align*}
To highlight the dependency on parameters, we explicitly denote the Hamiltonian as \(H(\mu)\) and noise operators as \(A_k(\nu_{k})\).
In parallel, we introduce analogous operators built from the estimated parameters $\{\hat{\mu}_n,\hat{\nu}_{k,n},\hat{\gamma}_{k,n},\hat{\eta}_{k,n},\hat{\theta}_{k,n},\hat{\iota}_{k,n},\hat{\zeta}_{k,\bar{k},n}\}$, e.g.,
\begin{align*}
    &\mathbf{H}(\hat{\mu}):=\textstyle\sum^{\mathbf{n}}_{n=1} |n\rangle\langle n|\otimes H(\hat{\mu}_n),\quad \hat{\mu}_n\in\mathbb{R},\\    
    &\mathbf{A}_k(\hat{\nu}_k):=\textstyle\sum^{\mathbf{n}}_{n=1}|n\rangle\langle n|\otimes A_k(\hat{\nu}_{k,n}), \quad \hat{\nu}_{k,n}\in \mathbb{R}, \quad \forall k\in[N_P],\\    &\mathbf{L}_k(\hat{\gamma}_k):=\textstyle\sum^{\mathbf{n}}_{n=1}|n\rangle\langle n|\otimes \sqrt{\hat{\gamma}_{k,n}}L_k,\quad \hat{\gamma}_{k,n}\geq 0,  \quad \forall k\in[N_D],\\    &\bar{\mathbf{L}}_k(\hat{\eta}_k,\hat{\gamma}_k):=\textstyle\sum^{\mathbf{n}}_{n=1}|n\rangle\langle n|\otimes \sqrt{\hat{\eta}_{k,n}\hat{\gamma}_{k,n}} L_k,\quad \hat{\eta}_{k,n}\in (0,1] \text{ and }\hat{\gamma}_{k,n}\geq 0, \quad \forall k\in[N_D],\\
    &\boldsymbol{\Theta}_k(\hat{\theta}_k) :=\textstyle\sum^{\mathbf{n}}_{n=1}|n\rangle\langle n|\otimes \hat{\theta}_{k,n}\mathbf{I}_{\mathbb{H}},\quad \hat{\theta}_{k,n}\geq 0, \quad\forall k\in[N_J],\\    
    &\mathbf{C}_k(\hat{\iota}_k):=\textstyle\sum^{\mathbf{n}}_{n=1}|n\rangle\langle n|\otimes \sqrt{\hat{\iota}_{k,n}}C_k,\quad \hat{\iota}_{k,n}>0, \quad\forall k\in[N_J],\\    &\bar{\mathbf{C}}_{k,\bar{k}}(\hat{\zeta}_{k,\bar{k}},\hat{\iota}_{\bar{k}} ):=\textstyle\sum^{\mathbf{n}}_{n=1}|n\rangle\langle n|\otimes \sqrt{\hat{\zeta}_{k,\bar{k},n}\hat{\iota}_{\bar{k},n}}C_{\bar{k}},\quad \hat{\zeta}_{k,\bar{k},n}\geq 0\text{ and }\hat{\iota}_{k,n}>0, \quad\forall k,\bar{k}\in[N_J].
\end{align*}
For simplicity, throughout the rest of this paper, we will interchangeably use $\mathbf{H}$ and $\mathbf{H}(\mu)$, $\mathbf{A}_k$ and $\mathbf{A}_k(\nu_k)$, etc., for the operators related to the augmented system, and $\hat{\mathbf{H}}$ and $\mathbf{H}(\hat{\mu})$, $\hat{\mathbf{A}}_k$ and $\mathbf{A}_k(\hat{\nu}_k)$, etc., for the operators related to the augmented filter.

The dynamics of the augmented system can be described by the following SME, 
\begin{align}
 d\Xi(t)= &\bar{\mathcal{L}}[\mathbf{H},\mathbf{L},\mathbf{C},\mathbf{A}](\Xi(t-))dt\nonumber\\
 &+\textstyle\sum^{N_D}_{k=1}\bar{\mathcal{G}}[\bar{\mathbf{L}}_k](\Xi(t-))\big(dY_k(t) - \Tr(\bar{\mathbf{L}}_k \Xi(t-)+\Xi(t-) \bar{\mathbf{L}}_k^*)dt\big)\nonumber\\
 &+\textstyle\sum^{N_J}_{k=1}\Big(\frac{\bar{\mathcal{J}}[\boldsymbol{\Theta}_k,\bar{\mathbf{C}}_k](\Xi(t-))}{\bar{\mathcal{T}}[\boldsymbol{\Theta}_k,\bar{\mathbf{C}}_k](\Xi(t-))} - \Xi(t-) \Big)\big(d\mathsf{N}_k(t)-\bar{\mathcal{T}}[\boldsymbol{\Theta}_k,\bar{\mathbf{C}}_k](\Xi(t-)) \big),\label{Eq:A-SME-Ori}
\end{align}
with initial condition $\Xi(0)=\Xi_0\in \mathcal{S}(\overline{\mathbb{H}})$, where
\begin{align*}
    &\bar{\mathcal{L}}[\mathbf{H},\mathbf{L},\mathbf{C},\mathbf{A}](\Xi)=-\mathfrak{i}[\mathbf{H},\Xi] + \textstyle\sum^{N_D}_{k=1}\mathcal{D}_{\mathbf{L}_k}(\Xi)+\sum^{N_J}_{k=1}\mathcal{D}_{\mathbf{C}_k}(\Xi)+\sum^{N_P}_{k=1}\mathcal{D}_{\mathbf{A}_k}(\Xi),\\
    &\bar{\mathcal{G}}[\bar{\mathbf{L}}_k](\Xi)=\bar{\mathbf{L}}_k \Xi+\Xi \bar{\mathbf{L}}_k^*-\Tr(\bar{\mathbf{L}}_k \Xi+\Xi \bar{\mathbf{L}}_k^*)\Xi,\\
    &\bar{\mathcal{J}}[\boldsymbol{\Theta}_k,\bar{\mathbf{C}}_k](\Xi) =\boldsymbol{\Theta}_k \Xi+\textstyle\sum^{N_J}_{\bar{k}=1}\bar{\mathbf{C}}_{k,\bar{k}}\Xi\bar{\mathbf{C}}_{k,\bar{k}}^*,\\
    &\bar{\mathcal{T}}[\boldsymbol{\Theta}_k,\bar{\mathbf{C}}_k](\Xi)=\Tr\big(\bar{\mathcal{J}}[\boldsymbol{\Theta}_k,\bar{\mathbf{C}}_k](\Xi)\big).
\end{align*}
Due to the block-diagonal structure of $\mathbf{H}$, $\mathbf{L}_k$, $\mathbf{C}_k$, $\mathbf{A}_k$, $\bar{\mathbf{L}}_k$, $\boldsymbol{\Theta}_k$ and $\bar{\mathbf{C}}_{k,\bar{k}}$,
if $\Xi_0=|n^*\rangle\langle n^*|\otimes \rho_0$ with $\rho_0\in\mathcal{S}(\mathbb{H})$, then
\[
\Xi(t)=|n^*\rangle\langle n^*|\otimes \rho(t), \qquad \mathbb{P}\text{-a.s.},
\]
where $\rho(t)$ solves the SME~\eqref{Eq:J-D SME}. In particular, the Wiener processes can be expressed as
\begin{align}
   W_k(t) &= Y_k(t) -\int^t_0\sqrt{\eta_k\gamma_k}\Tr(L_k \rho(s-)+\rho(s-)L_k^*)ds \nonumber\\
   &= Y_k(t) - \int^t_0\Tr(\bar{\mathbf{L}}_k \Xi(s-)+\Xi(s-) \bar{\mathbf{L}}_k^*)ds. \label{Eq:W-Y-A}
\end{align}
Substituting this expression into~\eqref{Eq:A-SME-Ori} yields the equivalent SME
\begin{align}
 d\Xi(t)= &\bar{\mathcal{L}}[\mathbf{H},\mathbf{L},\mathbf{C},\mathbf{A}](\Xi(t-))dt+\textstyle\sum^{N_D}_{k=1}\bar{\mathcal{G}}[\bar{\mathbf{L}}_k](\Xi(t-))dW_k(t)\nonumber\\
 &+\textstyle\sum^{N_J}_{k=1}\Big(\frac{\bar{\mathcal{J}}[\boldsymbol{\Theta}_k,\bar{\mathbf{C}}_k](\Xi(t-))}{\bar{\mathcal{T}}[\boldsymbol{\Theta}_k,\bar{\mathbf{C}}_k](\Xi(t-))} - \Xi(t-) \Big)\big(d\mathsf{N}_k(t)-\bar{\mathcal{T}}[\boldsymbol{\Theta}_k,\bar{\mathbf{C}}_k](\Xi(t-)) \big),\label{Eq:A-SME}
\end{align}
with initial condition $\Xi(0)=\Xi_0\in \mathcal{S}_{n^*}(\overline{\mathbb{H}})$, where $$\mathcal{S}_{n^*}(\overline{\mathbb{H}}):=\{|n^*\rangle\langle n^*| \otimes \rho \text{ with } \rho\in\mathcal{S}(\mathbb{H})\}.$$

Define
\begin{align*}
    \bar{\mathfrak{L}}_{k}(\mathsf{q})= \sum^{\mathbf{n}}_{n=1}\sum^{\mathbf{j}}_{j=1}\sqrt{\eta_{k}\gamma_{k}} \Re\{l_{k,j}\}\mathsf{q}_{n,j},\quad \bar{\mathfrak{C}}_{k}(\mathsf{q})=\sum^{\mathbf{n}}_{n=1}\sum^{\mathbf{j}}_{j=1}\Gamma_{k,j}\mathsf{q}_{n,j}
\end{align*}
and consider the following Doléans–Dade type SDEs describing the evolution of the diagonal elements of $\Xi(t)$:
\begin{align}
    d\mathsf{q}_{n,j}(t) = \mathsf{q}_{n,j}(t-)\Big[ &2\textstyle\sum^{N_D}_{k=1} \big(\sqrt{\eta_{k}\gamma_{k}}\Re\{l_{k,j}\}-\bar{\mathfrak{L}}_{k}(\mathsf{q}(t-)\big)dW_k(t)\nonumber\\
    &+ \textstyle\sum^{N_J}_{k=1}\Big(\frac{\Gamma_{k,j} }{\bar{\mathfrak{C}}_{k}(\mathsf{q}(t-))}-1\Big)\big(d\mathsf{N}_k(t)-\bar{\mathfrak{C}}_{k}(\mathsf{q}(t-))dt\big)\Big],  \label{Eq:dq_nj} 
\end{align} 
where $\Gamma_{k,j}= \theta_{k} + \sum^{N_J}_{\bar{k}=1}\zeta_{k,\bar{k}}\iota_{\bar{k}} |c_{\bar{k},j}|^2$.

Define the projection $M_{n,j}:=|n\rangle \langle n|\otimes \Pi_j$ and introduce subset $\overline{\mathcal{O}}^{n^*}_{\mathbf{j}\times\mathbf{n}}:=\{\mathsf{q}\in \overline{\mathcal{O}}_{\mathbf{j}\times\mathbf{n}}|\mathsf{q}_{n,j}=0,\,\forall n\neq n^*\}$ of the $\mathbf{j}\times\mathbf{n}$-dimensional simplex $\overline{\mathcal{O}}_{\mathbf{j}\times\mathbf{n}}$.

Using the almost sure invariance of $\mathcal{S}_{n^*}(\overline{\mathbb{H}})$ for~\eqref{Eq:A-SME} and arguments analogous to Proposition~\ref{Prop:Dade-SDE}, we obtain:
\begin{proposition}\label{Prop:Dade-A SDE}
    The systems \eqref{Eq:dq_nj} has a unique global solution $\mathsf{q}(t) \in \mathcal{O}_{\mathbf{j}\times\mathbf{n}}$ (resp. $q(t) \in \overline{\mathcal{O}}_{\mathbf{j}\times\mathbf{n}}$) for all $t \geq 0$ when $\mathsf{q}(0) \in \mathcal{O}_{\mathbf{j}\times\mathbf{n}}$  (resp. $\mathsf{q}(0) \in \overline{\mathcal{O}}_{\mathbf{j}\times\mathbf{n}}$), $\mathbb{P}$-almost surely. 

    Under Assumption~\ref{asm:qnd}, $\mathsf{q}_{n,j}(t) = \Tr(M_{n,j} \Xi(t))\in \overline{\mathcal{O}}^{n^*}_{\mathbf{j}\times\mathbf{n}}$  and $\mathsf{q}_{n^*,j}(t)=q_j(t)$ for all $t \geq 0$, $\mathbb{P}$-almost surely, when $\mathsf{q}_{n,j}(0) = \Tr(M_{n,j} \Xi_0)$ and $q_{j}(0)=\Tr(M_{n^*,j} \Xi_0)$ for all $j\in[\mathbf{j}]$, where $\Xi(t)$ is the solution to the augmented SME \eqref{Eq:A-SME} with $\Xi_0=|n^*\rangle\langle n^*| \otimes \rho_0 $ and $q(t)$ is the solution to SDE~\eqref{Eq:dq_j}.
\end{proposition}

If $q_j(0)=\mathsf{q}_{n^*,j}(0)=\Tr(M_{n^*,j}\Xi_0)$ for all $j$ with $\Xi_0\in\mathcal{S}_{n^*}(\overline{\mathbb{H}})$, then Proposition~\ref{Prop:Dade-A SDE} yields
\begin{align*}
    \overline{W}^{j}_k(t) &= W_k(t)-\int^t_0 2  \big(\sqrt{\eta_{k,n^*} \gamma_{k,n^*}}\Re\{l_{k,j}\}-\mathfrak{L}_{k}(q(s-))\big)ds\\
    &=W_k(t)-\int^t_0 2  \big(\sqrt{\eta_{k,n^*} \gamma_{k,n^*}}\Re\{l_{k,j}\}-\bar{\mathfrak{L}}_{k}(\mathsf{q}(s-))\big)ds, \quad \forall k\in[N_D],
\end{align*}
which are Wiener processes under $\mathbb{Q}^{j}$ (Lemma~\ref{Lemma:Girsanov}).

The dynamics of the augmented filter associated with~\eqref{Eq:A-SME} are given by 
\begin{align}
 d\hat{\Xi}(t)= &\textstyle\bar{\mathcal{L}}[\hat{\mathbf{H}},\hat{\mathbf{L}},\hat{\mathbf{C}},\hat{\mathbf{A}}](\hat{\Xi}(t-))dt+\sum^{N_D}_{k=1}\bar{\mathcal{G}}[\hat{\bar{\mathbf{L}}}_k](\hat{\Xi}(t-))\big(dY_k(t) - \Tr(\hat{\bar{\mathbf{L}}}_k \hat{\Xi}(t-)+\hat{\Xi}(t-) \hat{\bar{\mathbf{L}}}_k^*)dt\big)\nonumber\\
 &\textstyle+\sum^{N_J}_{k=1}\Big(\frac{\bar{\mathcal{J}}[\hat{\boldsymbol{\Theta}}_k,\hat{\bar{\mathbf{C}}}_k](\hat{\Xi}(t-))}{\bar{\mathcal{T}}[\hat{\boldsymbol{\Theta}}_k,\hat{\bar{\mathbf{C}}}_k](\hat{\Xi}(t-))} - \hat{\Xi}(t-) \Big)\big(d\mathsf{N}_k(t)-\bar{\mathcal{T}}[\hat{\boldsymbol{\Theta}}_k,\hat{\bar{\mathbf{C}}}_k](\hat{\Xi}(t-))dt \big)\label{Eq:A-SME-F}\\
 =&\textstyle\bar{\mathcal{L}}[\hat{\mathbf{H}},\hat{\mathbf{L}},\hat{\mathbf{C}},\hat{\mathbf{A}}](\hat{\Xi}(t-))dt+\sum^{N_D}_{k=1}\bar{\mathcal{G}}[\hat{\bar{\mathbf{L}}}_k](\hat{\Xi}(t-))\big(dW_k(t)+\bar{\mathcal{E}}_k(\Xi(t-),\hat{\Xi}(t-))dt\big)\nonumber\\
 &\textstyle+\sum^{N_J}_{k=1}\Big(\frac{\bar{\mathcal{J}}[\hat{\boldsymbol{\Theta}}_k,\hat{\bar{\mathbf{C}}}_k](\hat{\Xi}(t-))}{\bar{\mathcal{T}}[\hat{\boldsymbol{\Theta}}_k,\hat{\bar{\mathbf{C}}}_k](\hat{\Xi}(t-))} - \hat{\Xi}(t-) \Big)\big(d\mathsf{N}_k(t)-\bar{\mathcal{T}}[\hat{\boldsymbol{\Theta}}_k,\hat{\bar{\mathbf{C}}}_k](\hat{\Xi}(t-))dt \big)\nonumber,
\end{align}
where $\hat{\Xi}(0)=\hat{\Xi}_0\in \mathcal{S}(\overline{\mathbb{H}})$ and $\bar{\mathcal{E}}_k(\Xi,\hat{\Xi})=\Tr(\bar{\mathbf{L}}_k\Xi+\Xi\bar{\mathbf{L}}_k^*)-\Tr(\hat{\bar{\mathbf{L}}}_k\hat{\Xi}+\hat{\Xi} \hat{\bar{\mathbf{L}}}_k^*)$.
Existence, uniqueness, and invariance of solutions to~\eqref{Eq:A-SME}–\eqref{Eq:A-SME-F} follow from~\cite{barchielli1995constructing,pellegrini2010markov}.

Define 
\begin{align*}
    &\hat{\Gamma}_{k,n,j} := \hat{\theta}_{k,n} + \textstyle\sum^{N_J}_{\bar{k}=1}\hat{\zeta}_{k,\bar{k},n}\hat{\iota}_{\bar{k},n} |c_{\bar{k},j}|^2,\\
    &\hat{\bar{\mathfrak{L}}}_{k}(\hat{\mathsf{q}}):= \textstyle\sum^{\mathbf{n}}_{n=1}\sum^{\mathbf{j}}_{j=1}\sqrt{\hat{\eta}_{k,n}\hat{\gamma}_{k,n}} \Re\{l_{k,j}\}\hat{\mathsf{q}}_{n,j},\\
    &\hat{\bar{\mathfrak{C}}}_{k}(\hat{\mathsf{q}}):=\textstyle\sum^{\mathbf{n}}_{n=1}\sum^{\mathbf{j}}_{j=1}\hat{\Gamma}_{k,n,j}\hat{\mathsf{q}}_{n,j}.
\end{align*}
Using~\eqref{Eq:W-Y-A} and Proposition~\ref{Prop:Dade-A SDE}, the diagonal entries of $\hat{\Xi}(t)$ solve the Doléans–Dade type SDEs
\begin{align}
    d\hat{\mathsf{q}}_{n,j}(t) = \hat{\mathsf{q}}_{n,j}(t-)\Big[ &\textstyle 2\sum^{N_D}_{k=1} \Big(\sqrt{\hat{\eta}_{k,n}\hat{\gamma}_{k,n}}\Re\{l_{k,j}\}-\hat{\bar{\mathfrak{L}}}_{k}(\hat{\mathsf{q}}(t-))\Big)\big(dY_k(t)-2\hat{\bar{\mathfrak{L}}}_{k}(\hat{\mathsf{q}}(t-))dt\big)\nonumber\\
    &+ \textstyle\sum^{N_J}_{k=1}\Big(\frac{\hat{\Gamma}_{k,n,j} }{\hat{\bar{\mathfrak{C}}}_{k}(\hat{\mathsf{q}}(t-))}-1\Big)\big(d\mathsf{N}_k(t)-\hat{\bar{\mathfrak{C}}}_{k}(\hat{\mathsf{q}}(t-))dt\big)\Big]\label{Eq:d hat q_nj}\\
    = \hat{\mathsf{q}}_{n,j}(t-)\Big[ &2\textstyle \sum^{N_D}_{k=1} \Big(\sqrt{\hat{\eta}_{k,n}\hat{\gamma}_{k,n}}\Re\{l_{k,j}\}-\hat{\bar{\mathfrak{L}}}_{k}(\hat{\mathsf{q}}(t-)\Big)\big(dW_k(t)+\bar{\mathfrak{E}}_k(\mathsf{q}(t-),\hat{\mathsf{q}}(t-))dt\big)\nonumber\\
    &+ \textstyle\sum^{N_J}_{k=1}\Big(\frac{\hat{\Gamma}_{k,n,j} }{\hat{\bar{\mathfrak{C}}}_{k}(\hat{\mathsf{q}}(t-))}-1\Big)\big(d\mathsf{N}_k(t)-\hat{\bar{\mathfrak{C}}}_{k}(\hat{\mathsf{q}}(t-))dt\big)\Big],\nonumber
\end{align} 
where $\bar{\mathfrak{E}}_k(\mathsf{q},\hat{\mathsf{q}})=2\bar{\mathfrak{L}}_{k}(\mathsf{q})-2\hat{\bar{\mathfrak{L}}}_{k}(\hat{\mathsf{q}})$ and $\mathsf{q}(t)$ solves~\eqref{Eq:dq_nj} with $\mathsf{q}_{n,j}(0)=\Tr(M_{n,j}\Xi_0)$. By applying the similar arguments as in Proposition~\ref{Prop:Dade-SDE-F}, we have the following proposition.
\begin{proposition}\label{Prop:Dade-A-F SDE}
    The system \eqref{Eq:d hat q_nj} has a unique global solution $\hat{\mathsf{q}}(t) \in \mathcal{O}_{\mathbf{j}\times\mathbf{n}}$ (resp. $\hat{\mathsf{q}}(t) \in \overline{\mathcal{O}}_{\mathbf{j}\times\mathbf{n}}$) for all $t \geq 0$ when $\hat{\mathsf{q}}(0) \in \mathcal{O}_{\mathbf{j}\times\mathbf{n}}$ (resp. $\hat{\mathsf{q}}(0) \in \overline{\mathcal{O}}_{\mathbf{j}\times\mathbf{n}}$), $\mathbb{P}$-almost surely, which satisfies the following 
    \begin{align}
    &\hat{\mathsf{q}}_{n,j}(t)\nonumber\\
    =&\hat{\mathsf{q}}_{n,j}(0)\nonumber\\
    &\times\exp\Bigg[ 2\sum^{N_D}_{k=1} \Bigg(\int^t_0 
     \Big(\sqrt{\hat{\eta}_{k,n}\hat{\gamma}_{k,n}}\Re\{l_{k,j}\}-\hat{\bar{\mathfrak{L}}}_{k}(\hat{\mathsf{q}}(s-))\Big)\big(dY_k(s)-2\hat{\bar{\mathfrak{L}}}_{k}(\hat{\mathsf{q}}(s-))ds\big)\nonumber\\ 
    &~~~~~~~~~~~~~~~~~~~~~-\int^t_0\big(\sqrt{\hat{\eta}_{k,n}\hat{\gamma}_{k,n}}\Re\{l_{k,j}\}-\hat{\bar{\mathfrak{L}}}_{k}(\hat{\mathsf{q}}(s-))\big)^2ds\Bigg)\nonumber\\ 
    &~~~~~~~~~+\sum^{N_J}_{k=1}\Bigg(\int^t_0\log \frac{\hat{\Gamma}_{k,n,j} }{\hat{\bar{\mathfrak{C}}}_{k}(\hat{\mathsf{q}}(s-))}d\mathsf{N}_k(s)-\int^t_0\big(\hat{\Gamma}_{k,n,j}-\hat{\bar{\mathfrak{C}}}_{k}(\hat{\mathsf{q}}(s-))\big) ds\Bigg)\Bigg]\label{Eq:Y-qSDE-AF}\\
    =&\hat{\mathsf{q}}_{n,j}(0)\nonumber\\
        &\times\exp\Bigg[2\sum^{N_D}_{k=1} \Bigg(\int^t_0 
     \Big(\sqrt{\hat{\eta}_{k,n}\hat{\gamma}_{k,n}}\Re\{l_{k,j}\}-\hat{\bar{\mathfrak{L}}}_{k}(\hat{\mathsf{q}}(s-))\Big)\big(dW_k(s)+\bar{\mathfrak{E}}_k(\mathsf{q}(s-),\hat{\mathsf{q}}(s-))ds\big)\nonumber\\ 
   &~~~~~~~~~~~~~~~~~~~~~-\int^t_0\big(\sqrt{\hat{\eta}_{k,n}\hat{\gamma}_{k,n}}\Re\{l_{k,j}\}-\hat{\bar{\mathfrak{L}}}_{k}(\hat{\mathsf{q}}(s-))\big)^2ds\Bigg)\nonumber\\ 
    &~~~~~~~~~+\sum^{N_J}_{k=1}\Bigg(\int^t_0\log \frac{\hat{\Gamma}_{k,n,j} }{\hat{\bar{\mathfrak{C}}}_{k}(\hat{\mathsf{q}}(s-))}d\mathsf{N}_k(s)-\int^t_0\big(\hat{\Gamma}_{k,n,j}-\hat{\bar{\mathfrak{C}}}_{k}(\hat{\mathsf{q}}(s-))\big) ds\Bigg)\Bigg]\label{Eq:W-qSDE-AF}
    \end{align}
    for all $t\geq 0$.
    
    Under Assumption~\ref{asm:qnd}, $\hat{\mathsf{q}}_{n,j}(t) = \Tr(M_{n,j} \hat{\Xi}(t))$ for all $t \geq 0$, $\mathbb{P}$-almost surely, when $\hat{\mathsf{q}}_{n,j}(0) = \Tr(M_{n,j} \hat{\Xi}_0)$, where $\hat{\Xi}(t)$ is the solution to the augmented SME \eqref{Eq:A-SME-F}.
\end{proposition}
Closed-form stochastic exponential representations of $\mathsf{q}(t)$ and $\hat{\mathsf{q}}(t)$ analogous to~\eqref{Eq:q_j} and~\eqref{Eq:W-qSDE-F} follow from standard Doléans–Dade theory and are omitted for brevity.

We next quantify how parameter mismatches affect the contraction rates of the augmented filter. For $k\in[N_D]$, $m,n\in[\mathbf{n}]$, and $i,j\in[\mathbf{j}]$, define
\begin{align*}
    &\bar{\Phi}^k_{m,i,n,j}(\eta,\gamma):=2\Big[ (\sqrt{\hat{\eta}_{k,m}\hat{\gamma}_{k,m}}\Re\{l_{k,i}\}-\sqrt{\hat{\eta}_{k,n}\hat{\gamma}_{k,n}}\Re\{l_{k,j}\})^2\\
    &~~~~~~~~~~~~+2(\sqrt{\hat{\eta}_{k,m}\hat{\gamma}_{k,m}}\Re\{l_{k,i}\}-\sqrt{\hat{\eta}_{k,n}\hat{\gamma}_{k,n}}\Re\{l_{k,j}\}) \Re\{l_{k,j}\}\big(\sqrt{\hat{\eta}_{k,n}\hat{\gamma}_{k,n}}-\sqrt{\eta_{k,n}\gamma_{k,n}}\big)\Big],\\
    &\bar{\Psi}^k_{m,i,n,j}(\Gamma):=\log\frac{\hat{\Gamma}_{k,m,i}}{\hat{\Gamma}_{k,n,j}}\Big(\hat{\Gamma}_{k,n,j}-\Gamma_{k,j}\Big)-\hat{\Gamma}_{k,n,j}\Big(1-\frac{\hat{\Gamma}_{k,m,i}}{\hat{\Gamma}_{k,n,j}}+\log\frac{\hat{\Gamma}_{k,m,i}}{\hat{\Gamma}_{k,n,j}}\Big) .
\end{align*}
\begin{condition}[Augmented parameter mismatch tolerance]\label{con:robust_ASME}
    There exists an $n^*\in[\mathbf{n}]$ such that, for all $(n^*,j)\neq (m,i)$ where $m\in [\mathbf{n}]$ and $i,j\in[\mathbf{j}]$, the following inequality holds: $$\sum^{N_D}_{k=1}\bar{\Phi}^k_{m,i,n^*,j}(\eta,\gamma)+\sum^{N_D}_{k=1}\bar{\Psi}^k_{m,i,n^*,j}(\Gamma)>0.$$
\end{condition}


\begin{theorem}[Robust exponential stability of the augmented quantum filter]\label{Thm:QSR-A-F}
   Suppose Assumptions~\ref{asm:qnd} and~\ref{asm:identifiability} hold,  along with Condition~\ref{con:robust_ASME}. Assume that $\hat{\Gamma}_{k,n,j}>0$ for all $k\in[N_J]$, $n\in[\mathbf{n}]$ and $j\in[\mathbf{j}]$. Then, for any initial state $\hat{\Xi}_0\in\mathrm{int}\{\mathcal{S}(\overline{\mathbb{H}})\}$ and $\Xi_0=|n^*\rangle\langle n^*| \otimes \rho_0 $, and for all $n\in[\mathbf{n}]$ and $j\in[\mathbf{j}]$,
     \begin{align*}
        &\mathbb{P}\big(\lim_{t\rightarrow \infty}\Tr(M_{n,j}\hat{\Xi}(t))=1\big) = 0,\quad \forall n\neq n^*,\\
        &\mathbb{P}\big(\lim_{t\rightarrow \infty}\Tr(M_{n^*,j}\hat{\Xi}(t))=1\big) = \mathbb{P}\big(\mathcal{R}=j\big),\\
    &\lim_{t \rightarrow \infty} \Tr(M_{n^*,\mathcal{R}} \hat{\Xi}(t)) = \lim_{t \rightarrow \infty} \hat{\mathsf{q}}_{n^*,\mathcal{R}}(t) = 1, \quad \mathbb{P}\text{-a.s.}\\
        &\lim_{t\rightarrow \infty} \frac{1}{t}\log \frac{\hat{\mathsf{q}}_{n,j}(t)}{\hat{\mathsf{q}}_{n^*,\mathcal{R}}(t)}=\textstyle-\sum^{N_D}_{k=1}\bar{\Phi}^k_{n,j,n^*,\mathcal{R}}(\eta,\gamma)-\sum^{N_J}_{k=1}\bar{\Psi}^k_{n,j,n^*,\mathcal{R}}(\Gamma),\quad \mathbb{P}\text{-a.s.}
    \end{align*}
where $\hat{\Xi}(t)$ is the solution to SME~\eqref{Eq:A-SME-F} and the random variable $\mathcal{R}$ is defined in Corollary~\ref{Cor:Convergence}.
\end{theorem}
\proof
For all $\hat{\Xi}_0 \in \mathrm{int}\{\mathcal{S}(\overline{\mathbb{H}})\}$, we have $\hat{\mathsf{q}}_{n,j}(0) = \Tr(\Pi_{n,j}\hat{\Xi}_0) > 0$ for all $n \in [\mathbf{n}]$ and $j \in [\mathbf{j}]$. Due to the stochastic exponential form, $\hat{\mathsf{q}}_{n,j}(t) > 0$ almost surely.

For any $(m,i) \neq (n^*,j)$, we derive  
\begin{align*}
    \frac{\hat{\mathsf{q}}_{m,i}(t)}{\hat{\mathsf{q}}_{n^*,j}(t)} = \frac{\hat{\mathsf{q}}_{m,i}(0)}{\hat{\mathsf{q}}_{n^*,j}(0)} \exp \Bigg[ 
    & \sum_{k=1}^{N_D} \Big( 2 \big( \sqrt{\hat{\eta}_{k,m} \hat{\gamma}_{k,m}} \Re\{l_{k,i}\} - \sqrt{\hat{\eta}_{k,n^*} \hat{\gamma}_{k,n^*}} \Re\{l_{k,j}\} \big) \overline{W}^{j}_k(t)  \\
    &- \bar{\Phi}^k_{m,i,n^*,j}(\eta,\gamma) t \Big) + \sum_{k=1}^{N_J} \Big( \log \frac{\hat{\Gamma}_{k,m,i}}{\hat{\Gamma}_{k,n^*,j}} \overline{\mathsf{N}}^{j}_k(t) - \bar{\Psi}^k_{m,i,n^*,j}(\Gamma) t \Big) \Bigg].
\end{align*}
Here, under $\mathbb{Q}^j$, $\overline{W}^j_k(t)$ are Wiener processes and $\overline{\mathsf{N}}^j_k(t)$ are centered Poisson processes (Lemma~\ref{Lemma:Girsanov}).

By the law of large numbers for Wiener and Poisson processes,
\begin{align*}
    &\lim_{t \rightarrow \infty} \frac{1}{t} \log \frac{\hat{\mathsf{q}}_{m,i}(t)}{\hat{\mathsf{q}}_{n^*,j}(t)} \\
    &= \lim_{t \rightarrow \infty} \frac{1}{t} \log\Bigg( \frac{\hat{\mathsf{q}}_{m,i}(0)}{\hat{\mathsf{q}}_{n^*,j}(0)} + \sum_{k=1}^{N_D} 2 \big( \sqrt{\hat{\eta}_{k,m} \hat{\gamma}_{k,m}} \Re\{l_{k,i}\} - \sqrt{\hat{\eta}_{k,n^*} \hat{\gamma}_{k,n^*}} \Re\{l_{k,j}\} \big) \overline{W}^{j}_k(t) \\
    &~~~~~~~~~~~~~~~~~~~~ + \sum_{k=1}^{N_J} \log \frac{\hat{\Gamma}_{k,m,i}}{\hat{\Gamma}_{k,n^*,j}} \overline{\mathsf{N}}^{j}_k(t) \Bigg) - \sum_{k=1}^{N_D} \bar{\Phi}^k_{m,i,n^*,j}(\eta,\gamma) - \sum_{k=1}^{N_J} \bar{\Psi}^k_{m,i,n^*,j}(\Gamma) \\
    &=  - \sum_{k=1}^{N_D} \bar{\Phi}^k_{m,i,n^*,j}(\eta,\gamma) - \sum_{k=1}^{N_J} \bar{\Psi}^k_{m,i,n^*,j}(\Gamma) < 0, \quad \mathbb{Q}^{j}\text{-a.s.},
\end{align*}
where negativity is guaranteed by Condition~\ref{con:robust_ASME}.  

The result follows by adapting the arguments in the proof of Theorem~\ref{Thm:QSR-F}.
\hfill$\square$

\subsection{Stochastic dynamics of the selection probabilities $\pi_n(t)$}
We initialize the augmented filter as 
\begin{align*}
    \hat{\Xi}_0 = \sum_{n=1}^{\mathbf{n}} |n\rangle \langle n| \otimes\pi_n(0)  \hat{\rho}_0,\quad \hat{\rho}_0 \in \mathcal{S}(\mathbb{H}), \quad \pi_n(0) \in (0,1),  \quad \sum_{n=1}^{\mathbf{n}}\pi_n(0)=1,
\end{align*}
where $\pi_n(0)$ represents the a priori probability of the target parameter $\lambda$ satisfying $\lambda \in (a \hat{\lambda}_n, b \hat{\lambda}_n)$. By the block–diagonal structure of $\hat{\mathbf{H}}$, $\hat{\mathbf{L}}_k$, $\hat{\mathbf{C}}_k$, $\hat{\mathbf{A}}_k$,  $\hat{\bar{\mathbf{L}}}_k$, $\hat{\boldsymbol{\Theta}}_k$, $\hat{\bar{\mathbf{C}}}_{k,\bar{k}}$ and the chosen $\hat{\Xi}_0$, the solution to the augmented filter~\eqref{Eq:A-SME-F} decomposes as
\begin{align*}
\hat{\Xi}(t) = \sum^{\mathbf{n}}_{n=1}|n\rangle\langle n|\otimes \hat{\sigma}_n(t), \quad \mathbb{P}\text{-a.s.}
\end{align*}
Taking the partial trace over the $\mathbb{R}^{\mathbf{n}}$ yields, for each $n\in[\mathbf{n}]$,
\begin{align}
    &d\hat{\sigma}_n(t) \nonumber\\
    &= \Big( -\mathfrak{i}[H(\hat{\mu}_n), \hat{\sigma}_n(t-)] + \textstyle\sum^{N_D}_{k=1} \hat{\gamma}_{k,n} \mathcal{D}_{L_k}(\hat{\sigma}_n(t-)) + \sum^{N_J}_{k=1} \hat{\iota}_{k,n} \mathcal{D}_{C_k}(\hat{\sigma}_n(t-))   \nonumber \\
    &~~ + \textstyle\sum^{N_P}_{k=1} \mathcal{D}_{A_k(\hat{\nu}_{k,n})}(\hat{\sigma}_n(t-))\Big)dt+ \sum^{N_D}_{k=1} \Big( \sqrt{\hat{\eta}_{k,n}\hat{\gamma}_{k,n}} (L_k \hat{\sigma}_n(t-) + \hat{\sigma}_n(t-) L_k^*) \nonumber \\
    &~~ \quad - \Tr(\hat{\bar{\mathbf{L}}}_k \hat{\Xi}(t-) + \hat{\Xi}(t-) \hat{\bar{\mathbf{L}}}^*_k) \hat{\sigma}_n(t-) \Big) \big(dW_k(t) + \bar{\mathcal{E}}_k(\Xi(t-), \hat{\Xi}(t-))dt \big) \nonumber \\
    &~~ + \textstyle\sum^{N_J}_{k=1} \Big( \frac{\tilde{\mathcal{J}}_{k,n}(\hat{\sigma}_n(t-))}{\bar{\mathcal{T}}[\hat{\boldsymbol{\Theta}}_k,\hat{\bar{\mathbf{C}}}_k](\hat{\Xi}(t-))} 
    - \hat{\sigma}_n(t-) \Big) \big(d\mathsf{N}_k(t) - \bar{\mathcal{T}}[\hat{\boldsymbol{\Theta}}_k,\hat{\bar{\mathbf{C}}}_k](\hat{\Xi}(t-))dt \big), \quad 
    \label{Eq:sigma_t}
\end{align}
where $\hat{\sigma}_n(0)=\pi_n(0)\hat{\rho}_0$ and $\tilde{\mathcal{J}}_{k,n}(\hat{\sigma})=\hat{\theta}_{k,n} \hat{\sigma} + \sum^{N_J}_{\bar{k}=1} \hat{\zeta}_{k,\bar{k},n} \hat{\iota}_{\bar{k},n} C_{\bar{k}} \hat{\sigma} C_{\bar{k}}^*$. 

It is natural to look for $\pi_n(t)$ such that 
$
    \hat{\sigma}_n(t) = \pi_n(t)\,\hat{\rho}_n(t),
$
where $\hat{\rho}_n(t)$ is the normalized quantum filter driven by the $n$-th candidate parameters (same form as~\eqref{Eq:J-D filter} with 
$\hat{\mu}_n$, $\hat{\nu}_{k,n}$, $\hat{\gamma}_{k,n}$, $\hat{\iota}_{k,n}$, $\hat{\eta}_{k,n}$, $\hat{\theta}_{k,n}$, and $\hat{\zeta}_{k,\bar{k},n}$), and with initial condition $\hat{\rho}_n(0)=\hat{\rho}_0$. A Doléans-Dade SDE that achieves this factorization is:
\begin{align}
    d\pi_n(t) = \pi_n(t) \Big[ 
    & \textstyle\sum_{k=1}^{N_D} \Big( \sqrt{\hat{\eta}_{k,n}\hat{\gamma}_{k,n}} \Tr(L_k \hat{\rho}_n(t-) 
    + \hat{\rho}_n(t-) L_k^*) \nonumber \\
    & - \Tr(\hat{\bar{\mathbf{L}}}_k \hat{\Xi}(t-) + \hat{\Xi}(t-) \hat{\bar{\mathbf{L}}}^*_k) \Big) 
    \big(dW_k(t) + \bar{\mathcal{E}}_k(\Xi(t-), \hat{\Xi}(t-)) dt \big) \nonumber \\
    & \textstyle+ \sum_{k=1}^{N_J} \Big(\frac{\Tr\big(\tilde{\mathcal{J}}_{k,n}(\hat{\rho}_n(t-))\big)}{\bar{\mathcal{T}}[\hat{\boldsymbol{\Theta}}_k,\hat{\bar{\mathbf{C}}}_k](\hat{\Xi}(t-))} - 1\Big) 
    \big(d\mathsf{N}_k(t) - \bar{\mathcal{T}}[\hat{\boldsymbol{\Theta}}_k,\hat{\bar{\mathbf{C}}}_k](\hat{\Xi}(t-)) dt \big) \Big].
    \label{Eq:proba}
\end{align}
\begin{proposition}\label{Prop:solution}
      The system~\eqref{Eq:proba} has a unique global solution $\pi_n(t)\in\mathcal{O}_{\mathbf{n}}$ (resp. $\pi_n(t)\in\overline{\mathcal{O}}_{\mathbf{n}}$) for all $t> 0$ when $\pi_n(0)\in\mathcal{O}_{\mathbf{n}}$ (resp. $\pi_n(0)\in\overline{\mathcal{O}}_{\mathbf{n}}$ ), $\mathbb{P}$-almost surely. Moreover, for all $t\geq 0$, 
      \begin{align*}
          \hat{\Xi}(t)=\sum^{\mathbf{n}}_{n=1}|n\rangle\langle n|\otimes \pi_n(t)\hat{\rho}_n(t),\quad \pi_n(t)=\Tr\big((|n\rangle \langle n|\otimes \mathbf{I}_{\mathbb{H}})\hat{\Xi}(t)  \big)=\sum^{\mathbf{j}}_{j=1}\hat{\mathsf{q}}_{n,j}(t), \quad \mathbb{P}\text{-a.s.}
      \end{align*}
      where $\hat{\mathsf{q}}_{n,j}(t)$ is given by~\eqref{Eq:Y-qSDE-AF}.
\end{proposition}
\proof
By the standard Doléans–Dade representation~\cite[Chapter 5.1]{applebaum2009levy}, \eqref{Eq:proba} admits a unique strong solution in stochastic exponential form. In particular, if $\pi_n(0)> 0$ implies $\pi_n(t)> 0$, and  $\pi_n(0)=0$ implies $\pi_n(t)=0$ for all $t\geq 0$, $\mathbb{P}$-almost surely. Then, applying It\^o's formula, we have 
\begin{align*}
    \hat{\sigma}_n(t)=\pi_n(t)\hat{\rho}_n(t), \quad \forall t\geq 0, \quad \mathbb{P}\text{-a.s.}
\end{align*}
Comparing with \eqref{Eq:sigma_t} yields the claimed factorization and the identities
\begin{align*}
    &\Tr(\hat{\bar{\mathbf{L}}}_k \hat{\Xi}(t) + \hat{\Xi}(t) \hat{\bar{\mathbf{L}}}^*_k) = \textstyle\sum^{\mathbf{n}}_{n=1}\pi_n(t) \sqrt{\hat{\eta}_{k,n}\hat{\gamma}_{k,n}}\Tr(L_k \hat{\rho}_n(t) 
    + \hat{\rho}_n(t) L_k^*), \quad \forall k\in[N_D], \quad \mathbb{P}\text{-a.s.}\\
    & \bar{\mathcal{T}}[\hat{\boldsymbol{\Theta}}_k,\hat{\bar{\mathbf{C}}}_k](\hat{\Xi}(t)) = \textstyle\sum^{\mathbf{n}}_{n=1}\pi_n(t) \Tr\big(\tilde{\mathcal{J}}_{k,n}(\hat{\rho}_n(t))\big),\quad \forall k\in[N_J], \quad \mathbb{P}\text{-a.s.}
\end{align*}
Summing \eqref{Eq:proba} over $n$ gives
\begin{align*}
    d\sum^{\mathbf{n}}_{n=1} \pi_n(t) = \Big(1-\sum^{\mathbf{n}}_{n=1} \pi_n(t)\Big)\Big(& \sum^{N_D}_{k=1}\Tr(\hat{\bar{\mathbf{L}}}_k \hat{\Xi}(t-) + \hat{\Xi}(t-) \hat{\bar{\mathbf{L}}}^*_k)\big(dW_k(t) + \bar{\mathcal{E}}_k(\Xi(t-), \hat{\Xi}(t-)) \big)\\
    &+\sum^{N_D}_{k=1} \big(d\mathsf{N}_k(t) - \bar{\mathcal{T}}[\hat{\boldsymbol{\Theta}}_k,\hat{\bar{\mathbf{C}}}_k](\hat{\Xi}(t-)) dt \big)\Big),
\end{align*}
whose coefficients are a.s. bounded since $\hat{\Xi}(t)\in\mathcal{S}(\overline{\mathbb{H}})$. The functional Lipschitz conditions~\cite[pp 256]{protter2004stochastic} hold, as a concequence, the existence and uniqueness of the solution can be ensured~\cite[Theorem 5.7]{protter2004stochastic}. Hence, $\sum^{\mathbf{n}}_{n=1}\pi_n(t)=1$ for all $t\geq 0$ almost surely whenever $\sum^{\mathbf{n}}_{n=1}\pi_n(0)=1$. Therefore, $\overline{\mathcal{O}}_{\mathbf{n}}$ and $\mathcal{O}_{\mathbf{n}}$ are almost surely invariant for $\pi(t)$.

Finally, using $\hat{\Xi}(t)=\sum^{\mathbf{n}}_{n=1}|n\rangle\langle n|\otimes \pi_n(t)\hat{\rho}_n(t)$ and Proposition~\ref{Prop:Dade-A SDE} gives
\begin{align*}
    \pi_n(t)=\Tr\big((|n\rangle \langle n|\otimes \mathbf{I}_{\mathbb{H}})\hat{\Xi}(t)  \big)=\sum^{\mathbf{j}}_{j=1}\hat{\mathsf{q}}_{n,j}(t), \quad \mathbb{P}\text{-a.s.}
\end{align*}
The proof is complete.
\hfill$\square$

Next, we state our main theorem on the asymptotic behavior of the selection probability $\pi_{n^*}(t)$.
\begin{theorem}[Asymptotic consistency of parameter selection]\label{Thm:ParameterConverge}
    Let $\pi(0)\in\mathcal{O}_{\mathbf{n}}$ and $\Xi_0=|n^*\rangle\langle n^*|\otimes\rho_0$. Suppose that the conditions of Theorem~\ref{Thm:QSR-A-F} hold.
    Then, for all $n\in[\mathbf{n}]$,
    $\lim_{t\rightarrow \infty}\pi_{n}(t)=\delta_{n,n^*}$, $\mathbb{P}$-almost surely. Moreover, for all $n\neq n^*$, 
    \begin{equation}\label{Eq:Rate_Pi}
    \lim_{t\rightarrow \infty}\frac{1}{t}\log \frac{\pi_n(t)}{\pi_{n^*}(t)}\leq \max_{i,j}\Big( - \sum_{k=1}^{N_D} \bar{\Phi}^k_{n,i,n^*,j} (\eta,\gamma)- \sum_{k=1}^{N_J} \bar{\Psi}^k_{n,i,n^*,j}(\Gamma) \Big)<0, \quad \mathbb{P}\text{-a.s.} 
    \end{equation}
\end{theorem}
\proof
By Theorem~\ref{Thm:QSR-A-F} and Proposition~\ref{Prop:Dade-A SDE}, 
$$
\mathbb{P}(\lim_{t \to \infty} \pi_{n^*}(t) = 1) = \mathbb{P}\left(\lim_{t \to \infty} \textstyle\sum^{\mathbf{j}}_{j=1}\hat{\mathsf{q}}_{n^*,j}(t) = 1\right) = \textstyle\sum_{j=1}^{\mathbf{j}} \hat{\mathsf{q}}_{n^*,j}(0) = 1.
$$
Additionally, by Proposition~\ref{Prop:solution}, $\pi(t) \in \mathcal{O}_{\mathbf{j}}$ for all $t \geq 0$ almost surely. Therefore, for all $n \neq n^*$:
$
\mathbb{P}(\lim_{t \to \infty} \pi_{n}(t) = 0) = 1.
$

Furthermore, for all $n\neq n^*$, we have
\begin{align*}
    \lim_{t\rightarrow \infty}\frac{1}{t}\log \frac{\pi_n(t)}{\pi_{n^*}(t)}&=\lim_{t\rightarrow \infty}\frac{1}{t}\log \frac{\sum^{\mathbf{j}}_{i=1}\hat{\mathsf{q}}_{n,i}(t)}{\sum^{\mathbf{j}}_{j=1}\hat{\mathsf{q}}_{n^*,j}(t)}\\
    &\leq \max_{i\in[\mathbf{j}]}\left( \lim_{t\rightarrow \infty}\frac{1}{t}\log \frac{\mathbf{j}\hat{\mathsf{q}}_{n,i}(t)}{\hat{\mathsf{q}}_{n^*,\mathcal{R}}(t)}\right)\\
    &\leq \max_{i\in[\mathbf{j}]}\Big( \textstyle- \sum_{k=1}^{N_D} \bar{\Phi}^k_{n,i,n^*,\mathcal{R}}(\eta,\gamma) - \sum_{k=1}^{N_J} \bar{\Psi}^k_{n,i,n^*,\mathcal{R}}(\Gamma) \Big)\\
    &\leq \max_{i,j}\Big( \textstyle- \sum_{k=1}^{N_D} \bar{\Phi}^k_{n,i,n^*,j}(\eta,\gamma) - \sum_{k=1}^{N_J} \bar{\Psi}^k_{n,i,n^*,j} (\Gamma)\Big)<0, \quad \mathbb{P}\text{-a.s.}
\end{align*}
where the second inequality follows from the same arguments as in Theorem~\ref{Thm:QSR-A-F}. 
\hfill$\square$

\begin{remark}
Estimate~\eqref{Eq:Rate_Pi} further implies that for each $n\neq n^*$ there exists a random variable $\mathbf{R}_n(\omega)\in[0,1]$ such that 
    \begin{equation*}
        \pi_n(t)\leq \mathbf{R}_n(\omega)\mathrm{exp}\left[\max_{i,j}\left( - \sum_{k=1}^{N_D} \bar{\Phi}^k_{n,i,n^*,j}(\eta,\gamma) - \sum_{k=1}^{N_J} \bar{\Psi}^k_{n,i,n^*,j}(\Gamma) \right)t\right], \quad \forall t\geq 0, \quad \mathbb{P}\text{-}a.s.
    \end{equation*}
\end{remark}

In summary, Theorem~\ref{Thm:ParameterConverge} provides sufficient conditions for parameter estimation that guarantee the robust stability of the augmented filter. Building on Theorem~\ref{Thm:QSR-F}, we next analyze specific parameter-estimation scenarios and identify sufficient conditions under which Condition~\ref{con:robust_ASME} holds.

\subsection{Estimating the diffusion parameters $\eta_k$ and $\gamma_k$}\label{Sec:diff}
We now focus on estimating the parameters $\{\eta_k\}$, $\{\gamma_k\}$, or their product $\{\eta_k\gamma_k\}$ for $k \in [N_D]$. 
We assume that, aside from the parameter of interest, all other quantities appearing in Condition~\ref{con:robust_ASME} are known. 
For clarity, we present the estimation procedure for $\eta_k\gamma_k$, noting that the same reasoning applies to estimating $\eta_k$ or $\gamma_k$ individually.
To simplify the exposition, we restrict attention to the purely diffusive case ($N_J=0$) with a single measurement channel ($N_D=1$). 
The multi-channel setting can be handled by straightforward generalization.

Let $$\lambda = \sqrt{\eta \gamma} \in [\underline{\lambda}, \bar{\lambda}]$$ 
with $\underline{\lambda}>0$, and the requirement $\lambda/\hat{\lambda}\in[\bar{a},\bar{b}]$ with $0<\bar{a}<1<\bar{b}$.
We define a sequence of estimators $\{\hat{\lambda}_n\}_{n\in[\mathbf{n}]}$ via
\begin{align*}
    \hat{\lambda}_{n+1}=\frac{b}{a}\hat{\lambda}_n,\quad \underline{\lambda}=a\hat{\lambda}_1, \quad \bar{\lambda}\in(a\hat{\lambda}_{\mathbf{n}},b\hat{\lambda}_{\mathbf{n}}],
\end{align*}
where $0<\bar{a}\leq a<1<b\leq \bar{b}$, and the total number of estimators is given by: $\mathbf{n}=\lceil \log(\bar{\lambda}/\underline{\lambda})/\log(b/a) \rceil$. 

Here $a$ and $b$ remain unspecified. We next derive admissible bounds for $(a,b)$ that guarantee Condition~\ref{con:robust_ASME} is satisfied.
Define $\Lambda_{a,b}:=\bigcup^{\mathbf{n}}_{n=1}(a\hat{\lambda}_n,b\hat{\lambda}_n)$.
The first objective is to determine ranges of $a$ and $b$ such that for every $\lambda \in \Lambda_{a,b}$ the selection probability converges asymptotically: $\lim_{t\rightarrow \infty}\pi_{n}(t)=\delta_{n,n^*}$, $\mathbb{P}$-almost surely, where $\pi_{n}(t)$ represents the likelihood that $\lambda/\hat{\lambda}_n\in(a,b)$ given the measurement record up to time $t$, and $n^*$ is the index of the subinterval containing the true parameter.

Although it is possible to work under Assumption~\ref{asm:identifiability}, doing so requires substantially more technical analysis. 
For the augmented filter, we instead impose the following stronger but simpler condition.
\begin{assumption}[Identifiability of Diffusive measurement operators]\label{asm:augmented-identifiability}
    For the diffusive measurement operators $L_k=\sum^{\mathbf{j}}_{j=1}l_{k,j}\Pi_j$ with  $k\in[N_D]$, we require:
    \begin{itemize}
     \item For all $i\neq j$ and for all $k\in[N_D]$, $\Re\{l_{k,i}\}\neq \Re\{l_{k,j}\}$. Moreover, $\Re\{l_j\}\neq 0$ for all $j\in[\mathbf{j}]$.
     \end{itemize}
\end{assumption}
The second condition guarantees that each measurement operator contributes nontrivially to the observed output. 
If $\Re\{l_{k,j}\}=0$, then $\bar{\Phi}_{m,i,n,j}(\eta,\gamma)=(\hat{\lambda}_m\Re\{l_{k,i}\})^2$, and in particular $\bar{\Phi}_{m,j,n,j}(\eta,\gamma)=0$ for $m\neq n$. 
In this case, it becomes impossible to distinguish between $\Tr(M_{n,j}\hat{\Xi}(t))$ and $\Tr(M_{m,j}\hat{\Xi}(t))$ from measurement data, violating Theorem~\ref{Thm:QSR-A-F}.

The second condition ensures that each measurement operator contributes nontrivially to the observed output. If $\Re\{l_j\}=0$, then $\bar{\Phi}_{m,i,n,j}(\eta,\gamma)=(\hat{\lambda}_m\Re\{l_i\})^2$, and in particular, $\bar{\Phi}_{m,j,n,j}(\eta,\gamma)=0$ for $m\neq n$. In this case, it becomes impossible to distinguish between $\Tr(M_{n,j}\hat{\Xi}(t))$ and $\Tr(M_{m,j}\hat{\Xi}(t))$ from measurement data, violating Theorem~\ref{Thm:QSR-A-F}.

Under Assumption~\ref{asm:augmented-identifiability}, define 
\begin{align*}
    \Phi_{m,i,n,j}(\kappa):=\left(\frac{\Re\{l_i\}}{\Re\{l_j\}}\Big(\frac{b}{a}\Big)^{m-n}-1\right)\left(\frac{\Re\{l_i\}}{\Re\{l_j\}}\Big(\frac{b}{a}\Big)^{m-n}+1-2\kappa\right),
\end{align*}
where $\kappa=\lambda/\hat{\lambda}_{n^*}\in(a,b)$ with $n^*\in[\mathbf{n}]$.
Then,
\[
\bar{\Phi}_{m,i,n^*,j}(\eta,\gamma)=(\hat{\lambda}_{n^*}\Re\{l_j\})^2\Phi_{m,i,n^*,j}(\kappa).
\]
Thus, under Assumption~\ref{asm:augmented-identifiability}, $\bar{\Phi}_{m,i,n^*,j}(\eta,\gamma)>0$ if and only if $\Phi_{m,i,n^*,j}(\kappa)>0$. 

We now derive sufficient conditions on $a$ and $b$ to guarantee Condition~\ref{con:robust_ASME} by establishing positivity of $\Phi_{m,i,n^*,j}(\kappa)$ in two distinct cases:
\paragraph{Case $i=j$:} Here $m\neq n^*$ and 
    \begin{align*}
    \Phi_{m,j,n^*,j}(\kappa)=\big((b/a)^{m-n^*}-1\big)\big((b/a)^{m-n^*}+1-2\kappa\big), \quad \kappa\in(a,b).
\end{align*}
Then, we have the following two scenarios:
\begin{itemize}
    \item If $m>n^*$, then $(b/a)^{m-n^*}>1$. Thus, $\Phi_{m,j,n^*,j}(\kappa)>0$ whenever $b/a+1-2b\geq 0$. 
    \item If $m<n^*$, then $(b/a)^{m-n^*}<1$. Thus, $\Phi_{m,j,n^*,j}(\kappa)>0$ whenever $a/b+1-2a\geq 0$.
\end{itemize}
Together, these give the sufficient condition: $$b=\frac{a}{2a-1}.$$

\paragraph{Case $i\neq j$:} Define $f^{m,i}_{n,j}(a)=\frac{\Re\{l_i\}}{\Re\{l_j\}}(2a-1)^{n-m}$. Taking into account $b=a/(2a-1)$, we have
\begin{equation*}
    \Phi_{m,j,n,j}(\kappa)=\big(f^{m,i}_{n,j}(a)-1 \big)\big(f^{m,i}_{n,j}(a)+1-2\kappa \big).
\end{equation*}
To avoid degeneracy, i.e., $\Phi_{m,j,n^*,j}(\kappa)=0$, we impose 
$$
f^{m,i}_{n,j}(a)\neq 1, \quad \forall (m,i)\neq(n,j).
$$ 
Next, consider the following two scenarios:
\begin{itemize}
    \item If $\Re\{l_i\}/\Re\{l_j\}<0$, then $f^{m,i}_{n,j}(a)<0$. Since $\kappa\in(a,b)$, we have $f^{m,i}_{n,j}(a)+1-2\kappa<1-2a$. Thus, $\Phi_{m,i,n^*,j}(\kappa)>0$ whenever $a> 1/2$.
    \item If $\Re\{l_i\}/\Re\{l_j\}>0$, we consider the two situations where $f^{m,i}_{n^*,j}(a)>0$ and $f^{m,i}_{n^*,j}(a)\neq 1$:
    \begin{itemize}
        \item If $f^{m,i}_{n^*,j}(a)>1$, $\Phi_{m,i,n^*,j}(\kappa)>0$ whenever $f^{m,i}_{n^*,j}(a)+1-2\kappa>0$, which can be guaranteed if:
        \begin{equation}\label{Eq:Diff_b}
             \bar{f}(a):=\frac{a}{2a-1}-\frac{1}{2}\Big(1+\min_{\substack{n, m \\ i \neq j}}\{f^{m,i}_{n,j}(a) \text{ s.t. } f^{m,i}_{n,j}(a)>1\}\Big)< 0,
        \end{equation}
        with  $\min\{\emptyset\}=\infty$. 
        \item If $f^{m,i}_{n^*,j}(a)\in(0,1)$, $\Phi_{m,i,n^*,j}(\kappa)>0$ whenever $f^{m,i}_{n^*,j}(a)+1-2\kappa<0$, which can be guaranteed if:
        \begin{equation}\label{Eq:Diff_a}
            \underline{f}(a):=a- \frac{1}{2}\Big(1+\max_{\substack{n, m \\ i \neq j}}\{f^{m,i}_{n,j}(a) \text{ s.t. } f^{m,i}_{n,j}(a)\in(0,1)\}\Big)> 0,
        \end{equation}
        with $\max\{\emptyset\}=-\infty$. 
    \end{itemize}
\end{itemize}

In \eqref{Eq:Diff_b}-\eqref{Eq:Diff_a}, equality also suffices. However, for the boundary case $\kappa=b$ (see Proposition~\ref{Prop:Convergence_D_bound}), stricter inequalities are required. Existence of suitable $a\in(1/2,1)$ is ensured since $\bar{f}(1)<0$ and $\underline{f}(1)>0$ (by Assumption~\ref{asm:augmented-identifiability}) and both functions $\bar{f}(a)$ and $\underline{f}(a)$ are continuous in $a$.


Based on the above discussion, we conclude the following condition on $a$ and $b$:
\begin{condition}[Diffusive grid design]\label{con:para_est_diff}
    $0<\bar{a}<a<1<b<\bar{b}$, $b=a/(2a-1)$, $a\in(1/2,1)$, $f^{m,i}_{n,j}(a)\neq 1$ for all $(m,i)\neq(n,j)$, $\bar{f}(a)> 0$ and $\underline{f}(a)< 0$.
\end{condition}

Under Assumptions~\ref{asm:qnd} and~\ref{asm:augmented-identifiability}, Condition~\ref{con:para_est_diff} guarantees Condition~\ref{con:robust_ASME}. Hence Theorem~\ref{Thm:ParameterConverge} yields the following.
\begin{proposition}\label{Prop:Convergence_D}
    Suppose that Assumptions~\ref{asm:qnd} and~\ref{asm:augmented-identifiability} hold. Given prior information $\lambda=\sqrt{\eta \gamma}\in[\underline{\lambda},\bar{\lambda}]$ and the requirement $\lambda/\hat{\lambda}\in[\bar{a},\bar{b}]$, assume $a$ and $b$ satisfy Condition~\ref{con:para_est_diff}. Set $\mathbf{n}=\lceil \log(\bar{\lambda}/\underline{\lambda})/\log(b/a) \rceil$. Then, for all $\lambda\in \Lambda_{a,b}=\bigcup^{\mathbf{n}}_{n=1}(a\hat{\lambda}_n,b\hat{\lambda}_n)$, the following statements hold:
    \begin{align*}
        &\lim_{t\rightarrow \infty}\pi_{n}(t)=\delta_{n,n^*}, \quad \forall n\in[\mathbf{n}],\\
        &\lim_{t\rightarrow \infty}\frac{1}{t}\log \frac{\pi_n(t)}{\pi_{n^*}(t)}\leq -\min_{i,j}\bar{\Phi}_{n,i,n^*,j}(\eta,\gamma)<0, \quad \forall n\neq n^*, \quad \mathbb{P}\text{-a.s.} 
    \end{align*}
\end{proposition}

\smallskip

Thus, $\lambda$ is identifiable within $\Lambda_{a,b}$. For boundary values $\lambda\in [\underline{\lambda},\bar{\lambda}]\setminus \Lambda_{a,b}=\{a\hat{\lambda}_1, b\hat{\lambda}_1, \dots, b\hat{\lambda}_{\mathbf{n}-1}, b\hat{\lambda}_{\mathbf{n}}\}$ we have:
\begin{proposition}\label{Prop:Convergence_D_bound}
    Suppose the conditions of Proposition~\ref{Prop:Convergence_D} are satisfied. Then,
    \begin{itemize}
        \item For all $\lambda\in\{a\hat{\lambda}_1, b\hat{\lambda}_{\mathbf{n}}\}$, $\mathbb{P}\big(\lim_{t\rightarrow \infty}\pi_{n}(t)=\delta_{n,n^*}\big)=1$ for all $n\in[\mathbf{n}]$;
        \item For all $\lambda\in\{ b\hat{\lambda}_1, \dots, b\hat{\lambda}_{\mathbf{n}-1}\}$, 
        \begin{align*}
            &\mathbb{P}\big(\lim_{t\rightarrow \infty}\pi_{n}(t)=0\big)=1,\quad \forall n\in [\mathbf{n}]\setminus\{n^*,n^*+1\},\\
            &\mathbb{P}\big(\lim_{t\rightarrow \infty}\pi_{n^*}(t)+\pi_{n^*+1}(t)=1\big)=1,\\
            &\mathbb{P}\big(\lim_{t\rightarrow \infty}\pi_{n^*}(t)=1\big)\in(0,1), \quad \mathbb{P}\big(\lim_{t\rightarrow \infty}\pi_{n^*+1}(t)=1\big)\in(0,1),
            \end{align*}
            and the asymptotic moments satisfy
            \begin{align*}
            &\lim_{t\rightarrow \infty}\mathbb{E}\big(\pi_{n^*}(t)\big)=\lim_{t\rightarrow \infty}\mathbb{E}\big(\pi_{n^*+1}(t)\big)=1/2,\\
            &\lim_{t\rightarrow \infty}\mathrm{Var}\big(\pi_{n^*}(t)\big)=\lim_{t\rightarrow \infty}\mathrm{Var}\big(\pi_{n^*+1}(t)\big)=1/4.
        \end{align*}
    \end{itemize}
\end{proposition}
\proof
For the first item, direct calculation shows $\Phi_{m,i,1,j}(a)>0$ for all $(m,i)\neq(1,j)$ and $\Phi_{m,i,\mathbf n,j}(b)>0$ for all $(m,i)\neq(\mathbf n,j)$, from which the claim follows.

For the second item, fix $n^*\in\{1,\dots,\mathbf{b}-1\}$ and define $\Delta^{n^*}_{j}:=\{m\in[\mathbf{n}], j\in[\mathbf{j}]|(m,i)\neq (n^*,j) \text{ nor } (n^*+1,j)\}$. Through direct computation, we have $\Phi_{m,i,n^*,j}(b)>0$ for all $(m,i)\in \Delta^{n^*}_{j}$. 
Using similar arguments as in the proof of Theorem~\ref{Thm:QSR-A-F}, $\Tr(M_{m,i}\hat{\Xi}(t))$ converges exponentially to zero $\mathbb{Q}^{j}$-almost surely for all $(m,i)\in \Delta^{n^*}_{j}$. Consequently, $\Tr(M_{n^*,j}\hat{\Xi}(t))+\Tr(M_{n^*+1,j}\hat{\Xi}(t))$ converges exponentially to one $\mathbb{Q}^{j}$-almost surely. This leads to
\begin{align*}
    &\mathbb{Q}^j\big(\lim_{t\rightarrow \infty}\pi_{n}(t)=0\big)=1,\quad \forall n\in [\mathbf{n}]\setminus\{n^*,n^*+1\},\\
    &\mathbb{Q}^j\big(\lim_{t\rightarrow \infty}\pi_{n^*}(t)+\pi_{n^*+1}(t)=1\big)=1.
\end{align*}
Applying the same reasoning as in the proof of Theorem~\ref{Thm:QSR-F}, we obtain 
\begin{align}
    &\mathbb{P}\big(\lim_{t\rightarrow \infty}\pi_{n}(t)=0\big)=1,\quad \forall n\in [\mathbf{n}]\setminus\{n^*,n^*+1\},\nonumber\\
    &\mathbb{P}\big(\lim_{t\rightarrow \infty}\pi_{n^*}(t)+\pi_{n^*+1}(t)=1\big)=1. \label{Eq:Proba_nstar}
\end{align}

Next, consider the case where $\Phi_{n^*+1,j,n^*,j}(b)=0$. We have the ration
\begin{align*}
    \mathfrak{Q}_{n^*,j}(t) :=\frac{\hat{\mathsf{q}}_{n^*+1,j}(t)}{\hat{\mathsf{q}}_{n^*,j}(t)}= \mathfrak{Q}_{n^*,j}(0) \exp\big(  \mathfrak{C}_{n^*,j} \overline{W}^{j}(t)\big),
\end{align*}
where $\mathfrak{C}_{n^*,j}:=2 \Re\{l_{j}\}(\hat{\lambda}_{n^*+1} - \hat{\lambda}_{n^*} )\neq 0$ and $\overline{W}^{j}(t)$ is $\mathbb{Q}^{j}$-Wiener process. 
Define 
\begin{align*}
    \mathsf{R}_{n^*,j}(t)=\sum_{(m,i)\in\Delta^{n^*}_{j}} \hat{\mathsf{q}}_{m,i}(t)\in(0,1),\quad \mathsf{r}_{n^*,j}(t)=\sum_{i\neq j}\hat{\mathsf{q}}_{n^*,i}(t)\in(0,1),
\end{align*}
both of which converge to zero $\mathbb{Q}^{j}$-almost surely. Since $\hat{\mathsf{q}}_{n^*+1,j}(t)+\hat{\mathsf{q}}_{n^*,j}(t)=1-\mathsf{R}_{n^*,j}(t)$ for all $t\geq 0$, we have $\hat{\mathsf{q}}_{n^*,j}(t)=\frac{1-\mathsf{R}_{n^*,j}(t)}{1+\mathfrak{Q}_{n^*,j}(t)}$, $\mathbb{Q}^{j}$-almost surely, and consequently
$$
\pi_{n^*}(t)=\mathsf{r}_{n^*,j}(t)+\frac{1-\mathsf{R}_{n^*,j}(t)}{1+\mathfrak{Q}_{n^*,j}(t)},\quad \mathbb{Q}^{j}\text{-a.s.}
$$

By dominated convergence and the symmetry of the Gaussian,
\begin{align*}
    \lim_{t\rightarrow \infty}\mathbb{E}_{\mathbb{Q}^j}\left(\frac{1}{1+\mathfrak{Q}_{n^*,j}(t)}\right) &=\int_{\mathbb{R}}\lim_{t\rightarrow \infty} \frac{1}{1+\exp(\log \mathfrak{Q}_{n^*,j}(0)+ x\mathfrak{C}_{n^*,j}\sqrt{t})}\frac{\exp(-x^2/2)}{\sqrt{2\pi}}dx=\frac{1}{2},
\end{align*}
and $\mathsf r_{n^*,j}(t),\mathsf R_{n^*,j}(t)\to 0$ in $L^2(\mathbb Q^{\,j})$, we deduce $\lim_{t\rightarrow \infty} \mathbb{E}_{\mathbb{Q}^j}(\pi_{n^*}(t)) = 1/2$, which implies
\begin{equation*}
    \lim_{t\rightarrow \infty} \mathbb{E}(\pi_{n^*}(t))=\lim_{t\rightarrow \infty} \sum_{j}\mathbb{P}(\mathcal{R}=j)\mathbb{E}_{\mathbb{Q}^j}(\pi_{n^*}(t))=\frac{1}{2}.
\end{equation*}
From Equation~\eqref{Eq:Proba_nstar}, we obtain
$
    \lim_{t\rightarrow \infty} \mathbb{E}(\pi_{n^*+1}(t))=1/2.
$
Hence, it follows
\begin{align*}
    \mathbb{P}\big(\lim_{t\rightarrow \infty}\pi_{n^*}(t)=1\big)\in(0,1), \quad \mathbb{P}\big(\lim_{t\rightarrow \infty}\pi_{n^*+1}(t)=1\big)\in(0,1).
\end{align*}

For the variances, using the decomposition above and again dominated convergence,
\begin{align*}
    \scalemath{0.93}{\lim_{t\rightarrow \infty}\mathbb{E}_{\mathbb{Q}^j}\left(\frac{1}{(1+\mathfrak{Q}_{n^*,j}(t))^2}\right)=\int_{\mathbb{R}}\lim_{t\rightarrow \infty} \frac{1}{(1+\exp(\log \mathfrak{Q}_{n^*,j}(0)+ x\mathfrak{C}_{n^*,j}\sqrt{t}))^2}\frac{\exp(-x^2/2)}{\sqrt{2\pi}}dx=\frac{1}{2}.}
\end{align*}
and cross/error terms vanish in the limit, yielding $\lim_{t\rightarrow \infty}\text{Var}_{\mathbb{Q}^j}(\pi_{n^*}(t))=1/4$ which implies
\begin{equation*}
    \lim_{t\rightarrow \infty} \text{Var}(\pi_{n^*}(t))=\lim_{t\rightarrow \infty} \sum_{j}\mathbb{P}(\mathcal{R}=j)\text{Var}_{\mathbb{Q}^j}(\pi_{n^*}(t))=\frac{1}{4}.
\end{equation*}
The same conclusion holds for $\pi_{n^*+1}(t)$.
\hfill$\square$

\medskip

Based on Proposition~\ref{Prop:Convergence_D} and Proposition~\ref{Prop:Convergence_D_bound}, the asymptotic behavior of the trajectories $\{\pi_n(t,\omega)\}$, for each sample path $\omega\in\Omega$,  falls into one of two categories:
\begin{enumerate}
    \item \textit{Single Convergence}: There exists exactly one trajectory that converges to one, while all others converge to zero.
    \item \textit{Oscillatory Behavior}: Two trajectories oscillate between zero and one, while all others converge to zero.
\end{enumerate}
In the case of single convergence, three possibilities arise:
 $\lambda/\hat{\lambda}_{n} \in (a,b)$, $\lambda/\hat{\lambda}_{n} = a$, or $\lambda/\hat{\lambda}_{n} = b$. Thus, we have $\lambda/\hat{\lambda}_{n} \in [a,b]$.
 From a practical perspective, when only finite measurement data are available, the oscillatory behavior case requires further investigation:
\begin{itemize}
    \item It may correspond to the single convergence scenario, where one trajectory should converge to one and the other to zero. However, the available measurement data may be insufficient to observe such long-term behavior.
    \item Alternatively, the two trajectories may never converge and continue oscillating indefinitely.
\end{itemize}
To address these possibilities, we propose the following procedure for estimating $\lambda\in[\underline{\lambda}, \bar{\lambda}]$.

\begin{algorithm}[H]
\caption{Parameter Estimation for $\eta$ or $\gamma$}\label{alg:parameter_estimation_D}




\textit{Step 1:} Collect the measurement data $\{Y(t)\}_{t \leq T}$ for sufficiently large $T>0$\;

\textit{Step 2:} Given prior information $\lambda \in [\underline{\lambda},\bar{\lambda}]$ with $\underline{\lambda}>0$ and the requirement $\lambda/\hat{\lambda}\in[\bar{a},\bar{b}]$, determine $a,b$ satisfying Condition~\ref{con:para_est_diff}. Define the estimator sequence $\hat{\lambda}_n=(b/a)^{n-1}\underline{\lambda}/a$ for $n \in [\mathbf{n}]$, where $\mathbf{n}=\lceil \log(\bar{\lambda}/\underline{\lambda})/\log(b/a) \rceil$\;

\textit{Step 3:} Using $\{\hat{\lambda}_n\}_{n \in [\mathbf{n}]}$ and the measurement data, compute $\hat{\mathsf{q}}_{n,j}(T)$ and then calculate $\pi_n(T) = \sum^{\mathbf{j}}_{j=1}\hat{\mathsf{q}}_{n,j}(T)$ for $n \in [\mathbf{n}]$\;




\textit{Step 4:}\vspace{-1em}
\begin{itemize}
    \item \textit{Single Convergence:} 
    If a unique $n^*$ exists such that $\pi_{n^*}(T)\approx 1$ and $\pi_n(T)\approx 0$ for $n \neq n^*$, then conclude that $\lambda/\hat{\lambda}_{n^*} \in [a,b]$. 
    \item \textit{Two Oscillatory Trajectories:} If both $\pi_{n^*}(T)$ and $\pi_{n^*+1}(T)$ remain non-negligible while others vanish, two explanations are possible:
  \begin{enumerate}
    \item[1.] $\lambda/\hat{\lambda}_{n^*} \approx b$ (i.e., near subinterval boundary);
    \item[2.] Insufficient data. 
  \end{enumerate}
  In this case, refine the admissible interval from $[\underline{\lambda}, \bar{\lambda}]$ to $[a\hat{\lambda}_{n^*}, b\hat{\lambda}_{n^*+1}]$, update $(a,b)$ to $(\mathsf{a},\mathsf{b})$ satisfying Condition~\ref{con:para_est_diff}, and recompute with the new estimator sequence then return to \textit{Step~3}. Persistent oscillation indicates \textit{insufficient data}. 
    \item \textit{Insufficient Data:} If more than two $\pi_n(T)$ remain significant, the data are insufficient. Collect new measurements and return to \textit{Step~1}.
\end{itemize}
\end{algorithm}

\begin{remark}
The measurement horizon $T$ required for reliable estimation can be shortened by increasing the number of measurement channels, which accelerates the convergence of $\pi_n(t)$ (see Theorem~\ref{Thm:ParameterConverge}). Moreover, even in cases where $\Re\{l_j\}=0$, consistent parameter estimation remains feasible by introducing additional measurement channels so that
$$ 
\min_{i,j}\left(\textstyle\sum_{k=1}^{N_D} \bar{\Phi}^k_{n,i,n^*,j} (\eta,\gamma)+ \sum_{k=1}^{N_J} \bar{\Psi}^k_{n,i,n^*,j}(\Gamma) \right)>0.
$$
\end{remark}

\subsection{Estimating the jump parameters $\theta_k$, $\zeta_{k,\bar{k}}$ and $\iota_k$}
We next address the estimation of $\theta_k$, $\zeta_{k,\bar{k}}$, $\iota_k$ and their product $\zeta_{k,\bar{k}}\iota_{\bar{k}}$ for $k \in [N_J]$. All other parameters appearing in Condition~\ref{con:robust_ASME} are assumed to be known. For clarity, we restrict to the purely jump case ($N_D=0$) and distinguish the following settings:
\begin{enumerate}
    \item A single measurement channel ($N_J=1$) with shot noise set to zero ($\theta=0$).
    \item Multiple measurement channels with non-zero shot noise ($\theta_k\neq 0$).
\end{enumerate}

We impose the following \textit{Identifiability Assumption} for jump operators:
\begin{assumption}[Identifiability of Jump measurement operators]\label{asm:augmented-identifiability-jump}
    For the jump measurement operators $C_k=\sum^{\mathbf{j}}_{j=1}c_{k,j}\Pi_j$ with  $k\in[N_J]$, we require:
    \begin{itemize}
     \item For all $i>j$ and for all $k\in[N_J]$, t $|c_{k,i}|>|c_{k,j}|$. Moreover, $|c_{k,j}|>0$ for all $j\in[\mathbf{j}]$ and $k\in[N_J]$.
     \end{itemize}
\end{assumption}
The strict positivity of the coefficients $\hat{\Gamma}_{k,n,j}$ ensured by this assumption is required in Theorem~\ref{Thm:QSR-A-F} to obtain exponential convergence.

\subsubsection{Single-channel case without shot noise}
We consider the estimation of $\zeta \in (0,1]$, $\iota >0$, or their product $\lambda=\zeta\iota$. The estimation procedures are analogous; we focus here on 
\begin{align*}
    \lambda=\zeta\iota\in[\underline{\lambda},\bar{\lambda}]
\end{align*}
with $\underline{\lambda}>0$, and the requirement $\lambda/\hat{\lambda}\in[\bar{a},\bar{b}]$ with $0<\bar{a}<1<\bar{b}$. We construct an estimator sequence $\{\hat{\lambda}_n\}_{n=1}^{\mathbf{n}}$ via
\begin{align*}
    \hat{\lambda}_{n+1}=\frac{b}{a}\hat{\lambda}_n,\quad \underline{\lambda}=a\hat{\lambda}_1, \quad \bar{\lambda}\in(a\hat{\lambda}_{\mathbf{n}},b\hat{\lambda}_{\mathbf{n}}],
\end{align*}
where $0<\bar{a}\leq a<1<b\leq \bar{b}$, and the number of estimators is $\mathbf{n}=\lceil \log(\bar{\lambda}/\underline{\lambda})/\log(b/a) \rceil$.
Next, we determine the ranges of $a$ and $b$ to ensure Condition~\ref{con:robust_ASME}.

Under Assumption~\ref{asm:augmented-identifiability-jump}, define 
\begin{align*}
    \Psi_{m,i,n,j}(\kappa):=1-\left|\frac{c_i}{c_j}\right|^2\left(\frac{b}{a}\right)^{m-n}+\kappa\log\left[ \left|\frac{c_i}{c_j}\right|^2\left(\frac{b}{a}\right)^{m-n} \right],
\end{align*}
where $\kappa=\lambda/\hat{\lambda}_{n^*}\in(a,b)$ with $n^*\in[\mathbf{n}]$. Then,
$$
\bar{\Psi}_{m,i,n^*,j}(\Gamma)=-\hat{\lambda}_{n^*}|c_j|^2\Psi_{m,i,n^*,j}(\kappa).
$$ Thus, $\bar{\Psi}_{m,i,n^*,j}(\Gamma)>0$ if and only if $\Psi_{m,i,n^*,j}(\kappa)<0$.

Next, we provide the sufficient conditions on $a$ and $b$ to guarantee Condition~\ref{con:robust_ASME} by establishing negativity of $\Psi_{m,i,n^*,j}(\kappa)$ in two distinct cases:
\paragraph{Case $i=j$:} Here $m\neq n^*$, and define 
\begin{align*}
    \psi_{d}(\kappa):=1-(b/a)^d+\kappa d\log(b/a), \quad d\in\{-\mathbf{n}+1,\dots,-1,1\dots,\mathbf{n}-1\}.
\end{align*}
We have 
\begin{equation*}
    \Psi_{m,j,n^*,j}(\kappa)=\psi_{d}(\kappa)\quad \text{with}\quad d=m-n^*.
\end{equation*}
Consider the following two scenarios:

$\psi_d(\kappa)<0$ for all $\kappa\in(a,b)$ is ensured if $\psi_1(b)\leq 0$;  
for $d\leq -1$, the corresponding condition is $\psi_{-1}(a)\leq 0$. 

\begin{itemize}
    \item If $d\geq 1$ (i.e., $m>n^*$), $\psi_{d}(\kappa)<0$ for all $\kappa\in(a,b)$ whenever $\psi_{d}(b)<0$. Moreover, we compute:
\begin{align*}
    \psi_{d+1}(b)-\psi_{d}(b)=\Big(\frac{b}{a}\Big)^d\Big(1-\frac{b}{a}\Big)+b\log \Big(\frac{b}{a}\Big)\leq b\log\Big(\frac{b}{a}\Big)\Big(1-\frac{b^{d-1}}{a^d}\Big)\leq 0.
\end{align*}
Thus, $\psi_{d}(\kappa)<0$ with $d\geq 1$ for all $\kappa\in(a,b)$ is ensured if $\psi_{1}(b)\leq 0$.
\item If $d\leq -1$ (i.e., $m<n^*$), $\psi_{d}(\kappa)<0$ for all $\kappa\in(a,b)$ whenever $\psi_{d}(a)<0$. Similarly, we compute
\begin{align*}
    \psi_{d-1}(a)-\psi_{d}(a)=\Big(\frac{b}{a}\Big)^d\Big(1-\frac{a}{b}\Big)+a\log \Big(\frac{a}{b}\Big)\leq a\log\Big(\frac{a}{b}\Big)\Big(1-\frac{b^{d}}{a^{d+1}}\Big)\leq 0.
\end{align*}
Thus, $\psi_{d}(\kappa)<0$ with $d\leq -1$ for all $\kappa\in(a,b)$ is ensured if $\psi_{-1}(a)\leq 0$.
\end{itemize}
Together these imply: $\Psi_{m,j,n^*,j}(\kappa)<0$ with $m\neq n^*$ for all $\kappa\in(a,b)$ if
\begin{align*}
    a-b+ab\log(b/a)=0.
\end{align*}
\begin{remark}
    The relation $a-b+ab\log(b/a)=0$ is transcendental. Writing
    \begin{align*}
        -\frac{1}{b}e^{-\frac{1}{b}} = -\frac{1}{a}e^{-\frac{1}{a}} \in \big(-1/e, 0\big), \quad \forall a \in (0, 1).
    \end{align*}
    we can solve for $b$ in terms of $a$ via the Lambert $W$ function~\cite{corless1996lambert}. Selecting the principal branch $W_0$ ensures $b>1$, yielding
    $$
    b = -\frac{1}{W_{0}\left(-\frac{1}{a}e^{-\frac{1}{a}}\right)} > 1.
    $$
\end{remark}

\paragraph{Case $i\neq j$:} Define $g^{m,i}_{n,j}(a,b)=|c_i/c_j|^2(b/a)^{m-n}>0$. We have
\begin{align*}
    \Psi_{m,i,n,j}(\kappa):=1-g^{m,i}_{n,j}(a,b)+\kappa\log g^{m,i}_{n,j}(a,b).
\end{align*}
To ensure $\Psi_{m,i,n^*,j}(\kappa)\neq 0$, we impose the condition:
$$
g^{m,i}_{n,j}(a,b)\neq 1, \quad \forall (m,i)\neq(n,j).
$$ 
We distinguish two situations:
\begin{itemize}
    \item If $g^{m,i}_{n^*,j}(a,b)>1$, then $\Psi_{m,i,n^*,j}(\kappa)<0$ provided 
\begin{equation}\label{Eq:Jump_b}
    \bar{g}(a,b):=\min_{\substack{n, m \\ i \neq j}}\left\{\frac{g^{m,i}_{n,j}(a,b)-1}{\log g^{m,i}_{n,j}(a,b)} \text{ s.t. } g^{m,i}_{n,j}(a,b)>1\right\}-b > 0,
\end{equation}
with $\min\{\emptyset\}=\infty$.
\item If $g^{m,i}_{n^*,j}(a,b)\in(0,1)$, then $\Psi_{m,i,n^*,j}(\kappa)<0$ provided 
\begin{equation}\label{Eq:Jump_a}
\underline{g}(a,b):=a- \min_{\substack{n, m \\ i \neq j}}\left\{\frac{g^{m,i}_{n,j}(a,b)-1}{\log g^{m,i}_{n,j}(a,b)} \text{ s.t. } g^{m,i}_{n,j}(a,b)\in(0,1)\right\} >0,
\end{equation}
with $\max\{\emptyset\}=-\infty$. 
\end{itemize}

In both~\eqref{Eq:Jump_b} and~\eqref{Eq:Jump_a}, allowing equality still suffices, except when $\kappa=\lambda/\hat{\lambda}_{n^*}=b$ (see Proposition~\ref{Prop:Convergence_J_bound}), where stricter inequalities are required.

The existence of parameters $a,b$ with $\bar{g}(a,b)>0$ and $\underline{g}(a,b)<0$ follows from:
\begin{enumerate}
    \item $\bar{g}(1,1)<0$ and $\underline{g}(1,1)>0$ ensured by Assumption~\ref{asm:augmented-identifiability-jump};
    \item Continuity of $\bar{g}(a,b)$,  $\underline{g}(a,b)$ and $a-b+ab\log(b/a)$ in $(a,b)$. 
\end{enumerate}

 Based on the above discussion, we conclude the following conditions on $a$ and $b$:
\begin{condition}[Jump grid design]\label{con:para_est_jump1}
    $0<\bar{a}<a<1<b<\bar{b}$, $a-b+ab\log(b/a)=0$, $g^{m,i}_{n,j}(a,b)\neq 1$ for all $ (m,i)\neq(n,j)$, $\bar{g}(a,b)> 0$ and $\underline{g}(a,b)> 0$.
\end{condition}

Under Assumptions~\ref{asm:qnd} and~\ref{asm:augmented-identifiability-jump}, Condition~\ref{con:para_est_jump1} guarantees Condition~\ref{con:robust_ASME}. Hence Theorem~\ref{Thm:ParameterConverge} yields the following.
\begin{proposition}\label{Prop:Convergence_J}
    Suppose that Assumptions~\ref{asm:qnd} and~\ref{asm:augmented-identifiability-jump} hold. Given prior information $\lambda=\zeta\iota\in[\underline{\lambda},\bar{\lambda}]$ and the requirement $\lambda/\hat{\lambda}\in[\bar{a},\bar{b}]$, assume $a$ and $b$ satisfy Condition~\ref{con:para_est_jump1}. Set $\mathbf{n}=\lceil \log(\bar{\lambda}/\underline{\lambda})/\log(b/a) \rceil$. Then, for all $\lambda\in \Lambda_{a,b}=\bigcup^{\mathbf{n}}_{n=1}(a\hat{\lambda}_n,b\hat{\lambda}_n)$, the following statements hold:
    \begin{align*}
        &\lim_{t\rightarrow \infty}\pi_{n}(t)=\delta_{n,n^*}, \quad \forall n\in[\mathbf{n}],\\
        &\lim_{t\rightarrow \infty}\frac{1}{t}\log \frac{\pi_n(t)}{\pi_{n^*}(t)}\leq -\min_{i,j}\bar{\Psi}_{n,i,n^*,j}(\Gamma)<0, \quad \forall n\neq n^*, \quad \mathbb{P}\text{-a.s.} 
    \end{align*}
\end{proposition}

Thus, $\lambda$ is identifiable within $\Lambda_{a,b}$. For boundary values $\lambda\in[\underline{\lambda},\bar{\lambda}]\setminus\Lambda_{a,b}=\{a\hat{\lambda}_1, b\hat{\lambda}_1,\dots,b\hat{\lambda}_{\mathbf{n}-1}, b\hat{\lambda}_{\mathbf{n}}\}$, the dynamics of $\pi_n(t)$ require further analysis:

\begin{proposition}\label{Prop:Convergence_J_bound}
    Suppose that the conditions of Proposition~\ref{Prop:Convergence_D} are satisfied. Then,
    \begin{itemize}
        \item For all $\lambda\in\{a\hat{\lambda}_1, b\hat{\lambda}_{\mathbf{n}}\}$, $\mathbb{P}\big(\lim_{t\rightarrow \infty}\pi_{n}(t)=\delta_{n,n^*}\big)=1$ for all $n\in[\mathbf{n}]$;
        \item For all $\lambda\in\{ b\hat{\lambda}_1, \dots, b\hat{\lambda}_{\mathbf{n}-1}\}$, 
        \begin{align*}
            &\mathbb{P}\big(\lim_{t\rightarrow \infty}\pi_{n}(t)=0\big)=1,\quad \forall n\in [\mathbf{n}]\setminus\{n^*,n^*+1\},\\
            &\mathbb{P}\big(\lim_{t\rightarrow \infty}\pi_{n^*}(t)+\pi_{n^*+1}(t)=1\big)=1,\\
            &\mathbb{P}\big(\lim_{t\rightarrow \infty}\pi_{n^*}(t)=1\big)\in(0,1), \quad \mathbb{P}\big(\lim_{t\rightarrow \infty}\pi_{n^*+1}(t)=1\big)\in(0,1),
            \end{align*}
            and the asymptotic moments satisfy
            \begin{align*}
            &\lim_{t\rightarrow \infty}\mathbb{E}\big(\pi_{n^*}(t)\big)=\lim_{t\rightarrow \infty}\mathbb{E}\big(\pi_{n^*+1}(t)\big)=1/2,\\
             &\lim_{t\rightarrow \infty}\mathrm{Var}\big(\pi_{n^*}(t)\big)=\lim_{t\rightarrow \infty}\mathrm{Var}\big(\pi_{n^*+1}(t)\big)=1/4.
        \end{align*}
    \end{itemize}
\end{proposition}
\proof
For the first item, direct calculation shows $\Psi_{m,i,1,j}(a)<0$ for all $(m,i)\neq (1,j)$ and $\Psi_{m,i,\mathbf{n},j}(b)<0$ for all $(m,i)\neq (\mathbf{n},j)$, which concludes the first claim. 


For the second item,  applying similar arguments as in Proposition~\ref{Prop:Convergence_D_bound}, the first two equalities are established.

We now analyze the asymptotic behavior of the expectation. Since $\Psi_{n^*+1,j,n^*,j}(b)=0$, we have the ratio
\begin{align*}
    \mathfrak{Q}_{n^*,j}(t):=\frac{\hat{\mathsf{q}}_{n^*+1,j}(t)}{\hat{\mathsf{q}}_{n^*,j}(t)} = \mathfrak{Q}_{n^*,j}(0) \exp\big[  \log(b/a) (\mathsf{N}(t)-\Gamma_j t)\big],
\end{align*}
where $\mathfrak{Q}_{n^*,j}(0)>0$, $\mathsf{N}(t)$ is Poisson process with intensity $\Gamma_j>0$ under probability measure $\mathbb{Q}^{j}$. Following the arguments in Proposition~\ref{Prop:Convergence_D_bound}, the asymptotic behavior of $\pi_{n^*}(t)$ is determined by analyzing $1/(1+\mathfrak{Q}_{n^*,j}(t))$. For an arbitrary $\delta>0$ and $T>0$, define
\begin{align*}
    &f_t(x) := \frac{1}{1+\mathfrak{Q}_{n^*,j}(0) \exp\big(  \log(b/a) \sqrt{\Gamma_j t} x\big)}, \\
    &\underline{f}_{T,\delta}(x)=
    \begin{cases}
    f_T(-\delta), & x\leq -\delta,\\
    0, & x>-\delta,
    \end{cases}  \quad 
    \bar{f}_{T,\delta}(x)=
    \begin{cases}
    1, & x\leq \delta,\\
    f_T(\delta), & x>\delta.
    \end{cases}
\end{align*}
For $t\geq T$, we have $\underline{f}_{T,\delta}(x)\leq f_t(x)\leq \bar{f}_{T,\delta}(x)$. Thus,
\begin{align*}
    \mathbb{E}_{\mathbb{Q}^j}(\underline{f}_{T,\delta}(X_t))\leq \mathbb{E}_{\mathbb{Q}^j}(f_t(X_t))\leq \mathbb{E}_{\mathbb{Q}^j}(\bar{f}_{T,\delta}(X_t)), \quad \forall t\geq T,
\end{align*}
where $X_t:=(\mathsf{N}(t)-\Gamma_j t)/\sqrt{\Gamma_j t}$. By straightforward computation, we obtain
\begin{align*}
    \mathbb{E}_{\mathbb{Q}^j}(\underline{f}_{T,\delta}(X_t))=f_T(-\delta)\mathbb{Q}^j(X_t\leq -\delta),\quad \mathbb{E}_{\mathbb{Q}^j}(\bar{f}_{T,\delta}(X_t))=\mathbb{Q}^j(X_t\leq \delta)+f_T(\delta)\mathbb{Q}^j(X_t> \delta).
\end{align*}
By the central limit theorem, $X_t$ converges in distribution to  standard normal variable $Z$. Hence, 
\begin{align*}
    &\lim_{t\rightarrow\infty}\mathbb{E}_{\mathbb{Q}^j}(\underline{f}_{T,\delta}(X_t))=f_T(-\delta)\mathbb{Q}^j(Z\leq -\delta), \\
    &\lim_{t\rightarrow\infty}\mathbb{E}_{\mathbb{Q}^j}(\bar{f}_{T,\delta}(X_t))=\mathbb{Q}^j(Z\leq \delta)+f_T(\delta)\mathbb{Q}^j(Z> \delta). 
\end{align*}
For any fixed $\delta>0$, $\lim_{T\rightarrow\infty}f_T(-\delta)=1$ and $\lim_{T\rightarrow\infty}f_T(\delta)=0$, then
\begin{align*}
    \mathbb{Q}^j(Z\leq -\delta)\leq \liminf_{t\rightarrow \infty} \mathbb{E}_{\mathbb{Q}^j}(f_t(X_t))\leq \limsup_{t\rightarrow \infty} \mathbb{E}_{\mathbb{Q}^j}(f_t(X_t))\leq \mathbb{Q}^j(Z\leq \delta).
\end{align*}
Letting $\delta\rightarrow 0$, we have
\begin{align*}
    \lim_{t\rightarrow \infty} \mathbb{E}_{\mathbb{Q}^j}(f_t(X_t))=\mathbb{Q}^j(Z\leq 0)=\frac{1}{2}.
\end{align*}
Following the similar arguments as in Proposition~\ref{Prop:Convergence_D_bound}, it follows:
$
    \lim_{t\rightarrow \infty} \mathbb{E}(\pi_{n^*}(t))= \lim_{t\rightarrow \infty} \mathbb{E}(\pi_{n^*+1}(t))=1/2.
$
Consequently:
\begin{align*}
    \mathbb{P}\big(\lim_{t\rightarrow \infty}\pi_{n^*}(t)=1\big)\in(0,1), \quad \mathbb{P}\big(\lim_{t\rightarrow \infty}\pi_{n^*+1}(t)=1\big)\in(0,1).
\end{align*}

For the variance of $\pi_{n^*}(t)$,  using arguments analogous to those in Proposition~\ref{Prop:Convergence_D_bound},  it suffices to analyze the asymptotic behavior of $\mathrm{Var}_{\mathbb{Q}^j}\big(1/(1+\mathfrak{Q}_{n^*,j}(t))^2 \big)$.
This can be achieved by applying the earlier results to the following inequality: 
\begin{align*}
    \underline{f}^2_{T,\delta}(x)\leq f^2_t(x)\leq \bar{f}^2_{T,\delta}(x),\quad \forall t\geq T.
\end{align*}
From this, we deduce $\lim_{t\rightarrow \infty}\text{Var}_{\mathbb{Q}^j}(\pi_{n^*}(t))=1/4$. Consequently,
\begin{equation*}
    \lim_{t\rightarrow \infty} \text{Var}(\pi_{n^*}(t))=\lim_{t\rightarrow \infty} \sum_{j}\mathbb{P}(\mathcal{R}=j)\text{Var}_{\mathbb{Q}^j}(\pi_{n^*}(t))=\frac{1}{4}.
\end{equation*}
The same conclusion holds for $\pi_{n^*+1}(t)$.
\hfill$\square$

We can estimate $\lambda\in[\underline{\lambda}, \bar{\lambda}]$ by following \textbf{Algorithm~\ref{alg:parameter_estimation_D}} substituting Condition~\ref{con:para_est_diff} with Condition~\ref{con:para_est_jump1}.

\begin{algorithm}[H]
\caption{Parameter Estimation for $\zeta$ or $\iota$}\label{alg:parameter_estimation_J}

\textit{Step 1:} Collect the measurement data $\{\mathsf{N}(t)\}_{t \leq T}$  for sufficiently large $T>0$\;

\textit{Step 2:} Given prior information $\lambda \in [\underline{\lambda}, \bar{\lambda}]$ with $\underline{\lambda}>0$ and the requirement $\lambda/\hat{\lambda}\in[\bar{a},\bar{b}]$ with $0<\bar{a}<1<\bar{b}$,
determine $a$ and $b$ satisfying Condition~\ref{con:para_est_jump1}. Define estimator sequence $\hat{\lambda}_n=(b/a)^{n-1}\underline{\lambda}/a$ for $n \in [\mathbf{n}]$, where $\mathbf{n}=\lceil \log(\bar{\lambda}/\underline{\lambda})/\log(b/a) \rceil$\;

\textit{Step 3:} Using $\{\hat{\lambda}_n\}_{n \in [\mathbf{n}]}$ and the measurement data, compute $\hat{\mathsf{q}}_{n,j}(T)$ and then calculate $\pi_n(T) = \sum^{\mathbf{j}}_{j=1}\hat{\mathsf{q}}_{n,j}(T)$ for $n \in [\mathbf{n}]$\;

\textit{Step 4:}\vspace{-1em}
\begin{itemize}
    \item \textit{Single Convergence:}  If a unique $n^*$ exists such that $\pi_{n^*}(T)\approx 1$ and $\pi_n(T)\approx 0$ for $n \neq n^*$, then conclude that $\lambda/\hat{\lambda}_{n^*} \in [a,b]$. 
    \item \textit{Two Oscillatory Trajectories:}  If both $\pi_{n^*}(T)$ and $\pi_{n^*+1}(T)$ remain non-negligible while others vanish, two explanations are possible:
  \begin{enumerate}
    \item[1.] $\lambda/\hat{\lambda}_{n^*} \approx b$ (i.e., near subinterval boundary);
    \item[2.] Insufficient data.
  \end{enumerate}
  In this case, refine the admissible interval from $[\underline{\lambda}, \bar{\lambda}]$ to $[a\hat{\lambda}_{n^*}, b\hat{\lambda}_{n^*+1}]$, update $(a,b)$ to $(\mathsf{a},\mathsf{b})$ satisfying Condition~\ref{con:para_est_jump1}, and recompute with the new estimator sequence then return to \textit{Step~3}. Persistent oscillation indicates \textit{insufficient data}. 
    \item \textit{Insufficient Data:} If more than two $\pi_n(T)$ remain significant, the data are insufficient. Collect new measurements and return to \textit{Step~1}.
\end{itemize}
\end{algorithm}

\subsubsection{Multiple-channel case with non-zero shot noise}
We now address parameter estimation in the presence of several photon-counting channels with nonzero shot noise.  
In this setting, the additive term in $\Gamma_{k,j}$ complicates the analysis, so the single–channel strategy from the previous subsection no longer applies directly.  
We therefore extend the construction to estimate $\theta_k$, $\zeta_{k,\bar{k}}$, $\iota_k$, or their product $\zeta_{k,\bar{k}}\iota_{\bar{k}}$, assuming that all other quantities entering Condition~\ref{con:robust_ASME} are known.  
For clarity, we detail the procedure for the composite parameter
\begin{align*}
    \lambda=\zeta_{\mathsf{k},\bar{\mathsf{k}}}\iota_{\bar{\mathsf{k}}}\in [\underline{\lambda}, \bar{\lambda}],
\end{align*}
the cases of $\theta_k$, $\zeta_{k,\bar{k}}$, or $\iota_k$ being analogous.  
As before, the admissible ratio constraint is $\lambda/\hat\lambda\in[\bar a,\bar b]$ with $0<\bar a<1<\bar b$.
We define a sequence of estimators $\{\hat{\lambda}_n\}_{n\in[\mathbf{n}]}$ via
\begin{align*}
    \hat{\lambda}_{n+1}=\epsilon\frac{b}{a}\hat{\lambda}_n, \quad \underline{\lambda}=a\hat{\lambda}_1, \quad \bar{\lambda}\in(a\hat{\lambda}_{\mathbf{n}},b\hat{\lambda}_{\mathbf{n}}].
\end{align*}
where 
$0<\bar{a}\leq a/\epsilon<1<\epsilon b\leq \bar{b}$ with $\epsilon\geq 1$
and $\mathbf{n}=\lceil \log(\bar{\lambda}/\underline{\lambda})/\log(\epsilon b/a) \rceil$. 
Here $\epsilon$ is an auxiliary parameter: when $\epsilon=1$ the scheme reduces to the single–channel construction, while $\epsilon>1$ introduces a gap between successive intervals $[a\hat\lambda_n,b\hat\lambda_n]$ and $[a\hat\lambda_{n+1},b\hat\lambda_{n+1}]$.  
In the following, we determine admissible ranges of $(a,b,\epsilon)$ that guarantee Condition~\ref{con:robust_ASME}.

Under Assumption~\ref{asm:augmented-identifiability-jump}, define 
\begin{align*}
    \Psi^{\mathsf{k},\bar{\mathsf{k}}}_{m,i,n,j}(\kappa):=1-\frac{\hat{\lambda}_{m}|c_{\bar{\mathsf{k}},i}|^2+\Upsilon_{\mathsf{k},\bar{\mathsf{k}},i}}{\hat{\lambda}_{n}|c_{\bar{\mathsf{k}},j}|^2+\Upsilon_{\mathsf{k},\bar{\mathsf{k}},j}}+\frac{\kappa\hat{\lambda}_{n^*}|c_{\bar{\mathsf{k}},i}|^2+\Upsilon_{\mathsf{k},\bar{\mathsf{k}},i}}{\hat{\lambda}_{n}|c_{\bar{\mathsf{k}},j}|^2+\Upsilon_{\mathsf{k},\bar{\mathsf{k}},j}}\log\left[\frac{\hat{\lambda}_{m}|c_{\bar{\mathsf{k}},i}|^2+\Upsilon_{\mathsf{k},\bar{\mathsf{k}},i}}{\hat{\lambda}_{n}|c_{\bar{\mathsf{k}},j}|^2+\Upsilon_{\mathsf{k},\bar{\mathsf{k}},j}}\right],
\end{align*}
where $\Upsilon_{\mathsf{k},\bar{\mathsf{k}},i}:= \theta_{\mathsf{k}} + \sum_{\bar{k}\neq \bar{\mathsf{k}}}\zeta_{k,\bar{k}}\iota_{\bar{k}} |c_{\bar{k},i}|^2$ is known, and $\kappa=\lambda/\hat{\lambda}_{n^*}\in[a,b]$ with $n^*\in[\mathbf{n}]$. 
It follows that 
$$\bar{\Psi}^{\mathsf{k}}_{m,i,n^*,j}(\Gamma)=-\hat{\Gamma}_{\mathsf{k},n^*,j} \Psi^{\mathsf{k},\bar{\mathsf{k}}}_{m,i,n^*,j}(\kappa).$$
Thus, $\bar{\Psi}^{\mathsf{k}}_{m,i,n^*,j}(\Gamma)>0$ if and only if $ \Psi^{\mathsf{k},\bar{\mathsf{k}}}_{m,i,n^*,j}(\kappa)<0$.

Next, we we establish sufficient conditions on $a$, $b$ and $\epsilon$ to guarantee Condition~\ref{con:robust_ASME} in the following two cases.
\paragraph{Case $i=j$:} Here $m\neq n^*$, and define 
\begin{align*}
   \scalemath{0.93}{ \psi^{n^*,j}_{d}(\kappa)=1-\frac{(\epsilon b/a)^{d}\hat{\lambda}_{n^*}|c_{\bar{\mathsf{k}},j}|^2+\Upsilon_{\mathsf{k},\bar{\mathsf{k}},j}}{\hat{\lambda}_{n^*}|c_{\bar{\mathsf{k}},j}|^2+\Upsilon_{\mathsf{k},\bar{\mathsf{k}},j}}+\frac{\kappa\hat{\lambda}_{n^*}|c_{\bar{\mathsf{k}},j}|^2+\Upsilon_{\mathsf{k},\bar{\mathsf{k}},j}}{\hat{\lambda}_{n^*}|c_{\bar{\mathsf{k}},j}|^2+\Upsilon_{\mathsf{k},\bar{\mathsf{k}},j}}\log\left[\frac{(\epsilon b/a)^{d}\hat{\lambda}_{n^*}|c_{\bar{\mathsf{k}},j}|^2+\Upsilon_{\mathsf{k},\bar{\mathsf{k}},j}}{\hat{\lambda}_{n^*}|c_{\bar{\mathsf{k}},j}|^2+\Upsilon_{\mathsf{k},\bar{\mathsf{k}},j}}\right],}
\end{align*}
where $d=m-n^*\in\{-\mathbf{n}+1,\dots,-1,1\dots,\mathbf{n}-1\}$ and $\kappa\in[a,b]$.
Then, $\Psi^{\mathsf{k},\bar{\mathsf{k}}}_{m,i,n^*,j}(\kappa)=\psi^{n^*,j}_{d}(\kappa)$.

Consider the following two scenarios:
\begin{itemize}
    \item If $d\geq 1$ (i.e., $m>n^*$), $\psi^{n^*,j}_{d}(\kappa)<0$ for all $\kappa\in[a,b]$ whenever $\psi^{n^*,j}_{d}(b)<0$.
    Moreover, we have
\begin{align*}
    &\psi^{n^*,j}_{d+1}(b)-\psi^{n^*,j}_{d}(b)\\
    &=\frac{b\hat{\lambda}_{n^*}|c_{\bar{\mathsf{k}},j}|^2+\Upsilon_{\mathsf{k},\bar{\mathsf{k}},j}}{\hat{\lambda}_{n^*}|c_{\bar{\mathsf{k}},j}|^2+\Upsilon_{\mathsf{k},\bar{\mathsf{k}},j}}\log\left[\frac{(\epsilon b/a)^{d+1}\hat{\lambda}_{n^*}|c_{\bar{\mathsf{k}},j}|^2+\Upsilon_{\mathsf{k},\bar{\mathsf{k}},j}}{(\epsilon b/a)^{d}\hat{\lambda}_{n^*}|c_{\bar{\mathsf{k}},j}|^2+\Upsilon_{\mathsf{k},\bar{\mathsf{k}},j}}\right]+\frac{(\epsilon b/a)^{d}\hat{\lambda}_{n^*}|c_{\bar{\mathsf{k}},j}|^2(1-\epsilon b/a)}{\hat{\lambda}_{n^*}|c_{\bar{\mathsf{k}},j}|^2+\Upsilon_{\mathsf{k},\bar{\mathsf{k}},j}}\\
    &\leq \frac{b\hat{\lambda}_{n^*}|c_{\bar{\mathsf{k}},j}|^2+\Upsilon_{\mathsf{k},\bar{\mathsf{k}},j}}{\hat{\lambda}_{n^*}|c_{\bar{\mathsf{k}},j}|^2+\Upsilon_{\mathsf{k},\bar{\mathsf{k}},j}} \frac{(\epsilon b/a)^{d}\hat{\lambda}_{n^*}|c_{\bar{\mathsf{k}},j}|^2(\epsilon b/a-1)}{(\epsilon b/a)^{d}\hat{\lambda}_{n^*}|c_{\bar{\mathsf{k}},j}|^2+\Upsilon_{\mathsf{k},\bar{\mathsf{k}},j}}+\frac{(\epsilon b/a)^{d}\hat{\lambda}_{n^*}|c_{\bar{\mathsf{k}},j}|^2(1-\epsilon b/a)}{\hat{\lambda}_{n^*}|c_{\bar{\mathsf{k}},j}|^2+\Upsilon_{\mathsf{k},\bar{\mathsf{k}},j}}\\
    &=\frac{(\epsilon b/a)^{d}\hat{\lambda}_{n^*}|c_{\bar{\mathsf{k}},j}|^2(\epsilon b/a-1)}{\hat{\lambda}_{n^*}|c_{\bar{\mathsf{k}},j}|^2+\Upsilon_{\mathsf{k},\bar{\mathsf{k}},j}}\left(\frac{b\hat{\lambda}_{n^*}|c_{\bar{\mathsf{k}},j}|^2+\Upsilon_{\mathsf{k},\bar{\mathsf{k}},j}}{(\epsilon b/a)^{d}\hat{\lambda}_{n^*}|c_{\bar{\mathsf{k}},j}|^2+\Upsilon_{\mathsf{k},\bar{\mathsf{k}},j}}-1 \right)\\
    &<0.
\end{align*}
where we used the fact $\log(1+x)\leq x$ for all $x\geq 0$. Thus, $\psi^{n^*,j}_{d}(\kappa)<0$ with $d\geq 1$ for all $\kappa\in[a,b]$ is ensured if: $\psi^{n^*,j}_{1}(b)< 0$.
    \item If $d\leq -1$ (i.e., $m<n^*$), $\psi^{n^*,j}_{d}(\kappa)<0$ for all $\kappa\in[a,b]$ whenever $\psi^{n^*,j}_{d}(a)<0$.
    Similarly, we have
\begin{align*}
    &\psi^{n^*,j}_{d-1}(a)-\psi^{n^*,j}_{d}(a)\\
    &=\frac{a\hat{\lambda}_{n^*}|c_{\bar{\mathsf{k}},j}|^2+\Upsilon_{\mathsf{k},\bar{\mathsf{k}},j}}{\hat{\lambda}_{n^*}|c_{\bar{\mathsf{k}},j}|^2+\Upsilon_{\mathsf{k},\bar{\mathsf{k}},j}}\log\left[\frac{(\epsilon b/a)^{d-1}\hat{\lambda}_{n^*}|c_{\bar{\mathsf{k}},j}|^2+\Upsilon_{\mathsf{k},\bar{\mathsf{k}},j}}{(\epsilon b/a)^{d}\hat{\lambda}_{n^*}|c_{\bar{\mathsf{k}},j}|^2+\Upsilon_{\mathsf{k},\bar{\mathsf{k}},j}}\right]+\frac{(\epsilon b/a)^{d}\hat{\lambda}_{n^*}|c_{\bar{\mathsf{k}},j}|^2(1-a/\epsilon b)}{\hat{\lambda}_{n^*}|c_{\bar{\mathsf{k}},j}|^2+\Upsilon_{\mathsf{k},\bar{\mathsf{k}},j}}\\
    &\leq \frac{a\hat{\lambda}_{n^*}|c_{\bar{\mathsf{k}},j}|^2+\Upsilon_{\mathsf{k},\bar{\mathsf{k}},j}}{\hat{\lambda}_{n^*}|c_{\bar{\mathsf{k}},j}|^2+\Upsilon_{\mathsf{k},\bar{\mathsf{k}},j}} \frac{(\epsilon b/a)^{d}\hat{\lambda}_{n^*}|c_{\bar{\mathsf{k}},j}|^2(a/\epsilon b-1)}{(\epsilon b/a)^{d}\hat{\lambda}_{n^*}|c_{\bar{\mathsf{k}},j}|^2+\Upsilon_{\mathsf{k},\bar{\mathsf{k}},j}}+\frac{(\epsilon b/a)^{d}\hat{\lambda}_{n^*}|c_{\bar{\mathsf{k}},j}|^2(1- a/\epsilon b)}{\hat{\lambda}_{n^*}|c_{\bar{\mathsf{k}},j}|^2+\Upsilon_{\mathsf{k},\bar{\mathsf{k}},j}}\\
    &=\frac{(\epsilon b/a)^{d}\hat{\lambda}_{n^*}|c_{\bar{\mathsf{k}},j}|^2(1-a/\epsilon b)}{\hat{\lambda}_{n^*}|c_{\bar{\mathsf{k}},j}|^2+\Upsilon_{\mathsf{k},\bar{\mathsf{k}},j}}\left(1-\frac{a\hat{\lambda}_{n^*}|c_{\bar{\mathsf{k}},j}|^2+\Upsilon_{\mathsf{k},\bar{\mathsf{k}},j}}{(\epsilon b/a)^{d}\hat{\lambda}_{n^*}|c_{\bar{\mathsf{k}},j}|^2+\Upsilon_{\mathsf{k},\bar{\mathsf{k}},j}} \right)\\
    &<0.
\end{align*}
Thus, $\psi^{n^*,j}_{d}(\kappa)<0$ with $d\leq -1$ for all $\kappa\in[a,b]$ is ensured if: $\psi^{n^*,j}_{-1}(a)< 0$.
\end{itemize}

Together these imply: $\Psi^{\mathsf{k},\bar{\mathsf{k}}}_{m,i,n^*,j}<0$ with $m\neq n^*$ if 
\begin{align*}
    \psi^{n,j}_{-1}(a)< 0 \quad \text{and} \quad \psi^{n,j}_{1}(b)< 0, \quad \forall n\in[\mathbf{n}],\, j\in[\mathbf{j}].
\end{align*}

To study existence of admissible parameters $(a,b,\epsilon)$, we introduce the auxiliary functions
\begin{align*}
    \mathfrak{M}^1_{n,j}(a,b,\epsilon)&:=1-\frac{\epsilon b}{a}+\Big(b+\frac{a\varrho_{j}}{(\epsilon b/a)^{n-1}}\Big)\log\left(\frac{\frac{\epsilon b}{a}+\frac{a\varrho_{j}}{(\epsilon b/a)^{n-1}}}{1+\frac{a\varrho_{j}}{(\epsilon b/a)^{n-1}}}\right),\quad n\in[\mathbf{n}], \, j\in[\mathbf{j}],\\
    \mathfrak{M}^2_{n,j}(a,b,\epsilon)&:=1-\frac{a}{\epsilon b}+\Big(a+\frac{a\varrho_{j}}{(\epsilon b/a)^{n-1}}\Big)\log\left(\frac{\frac{a}{\epsilon b}+\frac{a\varrho_{j}}{(\epsilon b/a)^{n-1}}}{1+\frac{a\varrho_{j}}{(\epsilon b/a)^{n-1}}}\right),\quad n\in[\mathbf{n}], \, j\in[\mathbf{j}],
\end{align*}
where $\varrho_{j}=\frac{\Upsilon_{\mathsf{k},\bar{\mathsf{k}},j}}{\underline{\lambda}|c_{\bar{\mathsf{k}},j}|^2}>0$. By direct computation, 
\begin{align*}
    \mathfrak{M}^1_{n,j}(a,b,\epsilon)=\psi^{n,j}_{1}(b)\left(1+\frac{\Upsilon_{\mathsf{k},\bar{\mathsf{k}},j}}{\hat{\lambda}_n|c_{\bar{\mathsf{k}},j}|^2} \right),\quad 
    \mathfrak{M}^2_{n,j}(a,b,\epsilon)=\psi^{n,j}_{-1}(a)\left(1+\frac{\Upsilon_{\mathsf{k},\bar{\mathsf{k}},j}}{\hat{\lambda}_n|c_{\bar{\mathsf{k}},j}|^2} \right).
\end{align*}
Hence, 
\begin{align*}
    \psi^{n,j}_{1}(b)< 0 \Leftrightarrow \mathfrak{M}^1_{n,j}(a,b,\epsilon)< 0, \quad \text{and} \quad \psi^{n,j}_{-1}(a)< 0 \Leftrightarrow \mathfrak{M}^2_{n,j}(a,b,\epsilon)< 0.
\end{align*}
\begin{lemma}\label{Lemma:Neighborhood1}
There exist \( \alpha\in(0,1) \), \( \beta>1 \) such that, setting
\( I_a:=(\alpha,1) \) and \( I_b:=(1,\beta) \), we have
\[
\mathfrak{M}^1_{n,j}(a,b,\epsilon)<0 \ \text{and}\ 
\mathfrak{M}^2_{n,j}(a,b,\epsilon)<0,\qquad
\forall\, a\in I_a,\ b\in I_b,\ \epsilon\in (1,5/2],\ n\in[\mathbf n],\, j\in[\mathbf j].
\]
\end{lemma}
\proof
At $(a,b)=(1,1)$, using $\log(1+x)<x-x^2/2+x^3/3$ for all $x>0$, we have
\begin{align*}
    \mathfrak{M}^1_{n,j}(1,1,\epsilon)&=1-\epsilon+\big(1+\varrho_{j}/\epsilon^{n-1}\big)\log\left(\frac{\epsilon+\varrho_{j}/\epsilon^{n-1}}{1+\varrho_{j}/\epsilon^{n-1}}\right)\\
    &\leq -\frac{1}{2}\frac{(\epsilon-1)^2}{1+\varrho_{j}/\epsilon^{n-1}}\left( 1- \frac{2}{3}\frac{\epsilon-1}{1+\varrho_{j}/\epsilon^{n-1}}\right), \quad \forall \epsilon>1.
\end{align*}
Since $\epsilon-1>\frac{\epsilon-1}{1+\varrho_{j}/\epsilon^{n-1}}$, it follows that for all $\epsilon\in(1,5/2]$, $\mathfrak{M}^1_{n,j}(1,1,\epsilon)<0$ for all $n\in[\mathbf{n}]$ and $j\in[\mathbf{j}]$.
Similarly, using $\log(1-x)<-x-x^2/2$ for all $x\in(0,1)$ yields
\begin{align*}
    \mathfrak{M}^2_{n,j}(1,1,\epsilon)&=1-\frac{1}{\epsilon}+\big(1+\varrho_{j}/\epsilon^{n-1}\big)\log\left(\frac{1/\epsilon+\varrho_{j}/\epsilon^{n-1}}{1+\varrho_{j}/\epsilon^{n-1}}\right)\\
    &\leq -(1-1/\epsilon)\left(1+\frac{1-1/\epsilon}{2(1+\varrho_{j}/\epsilon^{n-1})} \right)<0, \quad \forall \epsilon>1.
\end{align*}

Define
\(
m(\epsilon):=\max_{n,j}\max\big\{\mathfrak{M}^1_{n,j}(1,1,\epsilon),
                                 \mathfrak{M}^2_{n,j}(1,1,\epsilon)\big\}<0.
\)
By continuity of $\mathfrak{M}^{1,2}_{n,j}$ in $(a,b)$ and the finiteness of the index set,
there exist $\delta_a,\delta_b>0$ uniform in $n,j$ such that
$\mathfrak{M}^{1,2}_{n,j}(a,b,\epsilon)<\tfrac12 m(\epsilon)<0$ whenever
$|a-1|<\delta_a$ and $|b-1|<\delta_b$.  
The claim follows by setting $\alpha:=1-\delta_a$ and $\beta:=1+\delta_b$.
\hfill$\square$
\begin{remark}
The factor $\tfrac{a\varrho_{j}}{(\epsilon b/a)^{n-1}}$ varies with $(n,j)$, so the simplifications used in the single-channel case do not apply directly. Introducing the adjustable parameter $\epsilon\geq 1$ provides the necessary flexibility to ensure that the above conditions remain satisfied uniformly across all channels.
\end{remark}

\paragraph{Case $i\neq j$:} Denote
\begin{align*}
    h^{m,i}_{n,j}(a,b,\epsilon):=\frac{\underline{\lambda}\frac{(\epsilon b)^{m-1}}{a^m}|c_{\bar{\mathsf{k}},i}|^2+\Upsilon_{\mathsf{k},\bar{\mathsf{k}},i}}{\underline{\lambda}\frac{(\epsilon b)^{n-1}}{a^n}|c_{\bar{\mathsf{k}},j}|^2+\Upsilon_{\mathsf{k},\bar{\mathsf{k}},j}}=\frac{\hat{\lambda}_{m}|c_{\bar{\mathsf{k}},i}|^2+\Upsilon_{\mathsf{k},\bar{\mathsf{k}},i}}{\hat{\lambda}_{n}|c_{\bar{\mathsf{k}},j}|^2+\Upsilon_{\mathsf{k},\bar{\mathsf{k}},j}}.
\end{align*}
Then,
\begin{align*}
    \Psi^{\mathsf{k},\bar{\mathsf{k}}}_{m,i,n,j}(\kappa):=1-h^{m,i}_{n,j}(a,b,\epsilon)+\frac{\kappa\hat{\lambda}_{n^*}|c_{\bar{\mathsf{k}},i}|^2+\Upsilon_{\mathsf{k},\bar{\mathsf{k}},i}}{\hat{\lambda}_{n}|c_{\bar{\mathsf{k}},j}|^2+\Upsilon_{\mathsf{k},\bar{\mathsf{k}},j}}\log h^{m,i}_{n,j}(a,b,\epsilon).
\end{align*}
To avoid degeneracy $\Psi^{\mathsf{k},\bar{\mathsf{k}}}_{m,i,n^*,j}(\kappa)=0$, we impose:
$$
h^{m,i}_{n,j}(a,b,\epsilon)\neq 1, \quad \forall (m,i)\neq(n,j).
$$ 
We distinguish two cases:
\begin{itemize}
    \item If $h^{m,i}_{n^*,j}(a,b)>1$, $\Psi^{\mathsf{k},\bar{\mathsf{k}}}_{m,i,n^*,j}(\kappa)<0$ is satisfied provided
    \begin{align*}
        \bar{h}(a,b,\epsilon):=\min_{\substack{n, m \\ i \neq j}}\Bigg\{&\frac{\frac{h^{m,i}_{n,j}(a,b,\epsilon)-1}{\log h^{m,i}_{n,j}(a,b,\epsilon)}\left( \underline{\lambda}\frac{(\epsilon b)^{n-1}}{a^n}|c_{\bar{\mathsf{k}},j}|^2+\Upsilon_{\mathsf{k},\bar{\mathsf{k}},j} \right)-\Upsilon_{\mathsf{k},\bar{\mathsf{k}},j}}{\underline{\lambda}\frac{(\epsilon b)^{n-1}}{a^n}|c_{\bar{\mathsf{k}},j}|^2}\\
        &~~~~~~~~~~~~~~~~~~~~~~~~~~~~~~~~~~~~~~~~~\text{ s.t. } h^{m,i}_{n,j}(a,b,\epsilon)>1\Bigg\}-b>0.
    \end{align*}
    \item If $g^{m,i}_{n^*,j}(a,b)\in(0,1)$, $\Psi_{m,i,n^*,j}(\kappa)<0$ is satisfied provided
    \begin{align*}
        \underline{h}(a,b,\epsilon):=a-\max_{\substack{n, m \\ i \neq j}}\Bigg\{&\frac{\frac{1-h^{m,i}_{n,j}(a,b,\epsilon)}{\log h^{m,i}_{n,j}(a,b,\epsilon)}\left( \underline{\lambda}\frac{(\epsilon b)^{n-1}}{a^n}|c_{\bar{\mathsf{k}},j}|^2+\Upsilon_{\mathsf{k},\bar{\mathsf{k}},j} \right)-\Upsilon_{\mathsf{k},\bar{\mathsf{k}},j}}{\underline{\lambda}\frac{(\epsilon b)^{n-1}}{a^n}|c_{\bar{\mathsf{k}},j}|^2} \\
        &~~~~~~~~~~~~~~~~~~~~~~~~~~~~~~~~~~~~~~~~ \text{ s.t. } h^{m,i}_{n,j}(a,b,\epsilon)\in(0,1)\Bigg\}>0.
    \end{align*}
\end{itemize}

The existence of parameters $\bar{a}<a<1<b<\bar{b}$ and $\epsilon\geq 1$ such that
\[
\psi^{n,j}_{-1}(a)< 0 \quad \text{and} \quad \psi^{n,j}_{1}(b)< 0, \quad \forall\, n\in[\mathbf{n}],\; j\in[\mathbf{j}],
\]
as well as $\bar{h}(a,b,\epsilon)> 0$ and $\underline{h}(a,b,\epsilon)< 0$, is guaranteed by Lemma~\ref{Lemma:Neighborhood1} together with the following observations:
 \begin{enumerate}
     \item $\bar{h}(1,1,a)<0$ and $\underline{h}(1,1,1)>0$ ensured by Assumption~\ref{asm:augmented-identifiability-jump}.
     \item Continuity of $\bar{h}(a,b,\epsilon)$ and $\underline{h}(a,b,\epsilon)$ in $(a,b,\epsilon)$. 
 \end{enumerate}

Based on the above discussion, we conclude the following condition on $a$, $b$ and $\epsilon$:
\begin{condition}[Jump grid design]\label{con:para_est_jump2}
    $0<\bar{a}<a<1<b<\bar{b}$, $\max_{n\in[\mathbf{n}],j\in[\mathbf{j}]}\{\mathfrak{M}^1_{n,j}(a,b,\epsilon)\}<0$, $\max_{n\in[\mathbf{n}],j\in[\mathbf{j}]}\{\mathfrak{M}^2_{n,j}(a,b,\epsilon)\}<0$, $h^{m,i}_{n,j}(a,b,\epsilon)\neq 1$ for all $ (m,i)\neq(n,j)$, $\bar{h}(a,b,\epsilon)> 0$ and $\underline{h}(a,b,\epsilon)>0$.
\end{condition}

Under Assumptions~\ref{asm:qnd} and~\ref{asm:augmented-identifiability-jump}, Condition~\ref{con:para_est_jump2} guarantees Condition~\ref{con:robust_ASME}. Hence, Theorem~\ref{Thm:ParameterConverge} yields the following.
\begin{proposition}\label{Prop:Convergence_J_mul}
    Suppose that Assumptions~\ref{asm:qnd} and~\ref{asm:augmented-identifiability-jump} hold. Given prior information $\lambda\in[\underline{\lambda},\bar{\lambda}]$ and the requirement $\lambda/\hat{\lambda}\in[\bar{a},\bar{b}]$, assume $a$, $b$ and $\epsilon$ satisfy Condition~\ref{con:para_est_jump2}. Set $\mathbf{n}=\lceil \log(\bar{\lambda}/\underline{\lambda})/\log(\epsilon b/a) \rceil$. Then, for all $\lambda\in \bar{\Lambda}_{a,b}=\bigcup^{\mathbf{n}}_{n=1}[a\hat{\lambda}_n,b\hat{\lambda}_n]$, the following statements hold:
    \begin{align*}
        &\lim_{t\rightarrow \infty}\pi_{n}(t)=\delta_{n,n^*}, \quad \forall n\in[\mathbf{n}],\\
        &\lim_{t\rightarrow \infty}\frac{1}{t}\log \frac{\pi_n(t)}{\pi_{n^*}(t)}\leq -\min_{i,j}\sum^{N_J}_{k=1}\bar{\Psi}^k_{n,i,n^*,j}(\Gamma)<-\min_{i,j}\bar{\Psi}^{\mathsf{k}}_{n,i,n^*,j}(\Gamma)<0, \quad \forall n\neq n^*, \quad \mathbb{P}\text{-a.s.} 
    \end{align*}
\end{proposition}

Thus, Proposition~\ref{Prop:Convergence_J_mul} guarantees consistent estimation of $\lambda$ whenever $\lambda\in \bar{\Lambda}_{a,b}$. However, for $\lambda\in [\underline{\lambda},\bar{\lambda}]\setminus \bar{\Lambda}_{a,b}=\bigcup^{\mathbf{n}-1}_{n=1}(b\hat{\lambda}_n,\epsilon b\hat{\lambda}_n)$ the trajectories $\{\pi_n(t)\}$ require further investigation. In contrast to the earlier single-channel case, where only isolated boundary singularities arose, here nonzero shot noise and channel cross-talk introduce entire intervals of ambiguity. To control these effects, we introduce a parameter $\epsilon>1$, which confines the problematic region to a finite union of disjoint intervals. A refined analysis (see Proposition~\ref{Prop:Convergence_J_Multi_Gap}) shows that ensuring $\psi^{n^*,j}_2(\epsilon b)<0$ is essential. To this end, we define
\begin{align*}
    \mathfrak{M}^3_{n,j}(a,b,\epsilon)&:=1-\Big(\frac{\epsilon b}{a}\Big)^2+\Big(\epsilon b+\frac{a\varrho_{j}}{(\epsilon b/a)^{n-1}}\Big)\log\left(\frac{\big(\frac{\epsilon b}{a}\big)^2+\frac{a\varrho_{j}}{(\epsilon b/a)^{n-1}}}{1+\frac{a\varrho_{j}}{(\epsilon b/a)^{n-1}}}\right),\quad n\in[\mathbf{n}], \, j\in[\mathbf{j}],
\end{align*}
so that
$$
\mathfrak{M}^3_{n,j}(a,b,\epsilon)=\psi^{n,j}_{2}(\epsilon b)\left(1+\frac{\Upsilon_{\mathsf{k},\bar{\mathsf{k}},j}}{\hat{\lambda}_n|c_{\bar{\mathsf{k}},j}|^2} \right).
$$
We impose the following additional condition ensuring $\psi^{n^*,j}_2(\epsilon b)<0$: 
\begin{condition}\label{con:para_est_jump3}
    $\max_{n\in[\mathbf{n}],j\in[\mathbf{j}]}\{\mathfrak{M}^3_{n,j}(a,b,\epsilon)\}<0$,
\end{condition}

We now establish the existence of parameters $a$, $b$ and $\epsilon$ that jointly satisfy Conditions~\ref{con:para_est_jump2} and~\ref{con:para_est_jump3}.  
\begin{lemma}\label{Lemma:Neighborhood2}
There exist \( \alpha\in(0,1) \), \( \beta>1 \) such that, setting
\( I_a:=(\alpha,1) \) and \( I_b:=(1,\beta) \), for all $a\in I_a,\ b\in I_b,\ \epsilon\in (1,(\sqrt{13}-1)/2),\ n\in[\mathbf n],$ and $ j\in[\mathbf j]$, we have
\[
\mathfrak{M}^1_{n,j}(a,b,\epsilon)<0, \  
\mathfrak{M}^2_{n,j}(a,b,\epsilon)<0 \text{ and }\ 
\mathfrak{M}^3_{n,j}(a,b,\epsilon)<0.
\]
\end{lemma}
\proof
Fix $(a,b)=(1,1)$ and set $r_j:=\varrho_j/\epsilon^{\,n-1}>0$. Then
\[
\mathfrak{M}^3_{n,j}(1,1,\epsilon)
=1-\epsilon^2+(\epsilon+r_j)\log\!\left(\frac{\epsilon^2+r_j}{1+r_j}\right).
\]
Using $\log(1+y)<y-\tfrac{y^2}{2}+\tfrac{y^3}{3}$, we have
\begin{align*}
    \mathfrak{M}^3_{n,j}(1,1,\epsilon)&\leq -\frac{(\epsilon-1)^2(\epsilon+1)}{1+r_j}\left[\frac{(\epsilon+1)(\epsilon+r_j)}{2(1+r_j)}-1+\frac{(\epsilon-1)(\epsilon+1)^2(\epsilon+r_j)}{3(1+r_j)^2} \right]\\
    &\leq -\frac{(\epsilon-1)^3(\epsilon+1)}{(1+r_j)^2}\left[1-\frac{\epsilon(\epsilon+1)}{3}\right], 
\end{align*}
where we used $(\epsilon+r_j)/(1+r_j)\leq \epsilon$ for the second inequality. Thus, for all $\epsilon\in(1,(\sqrt{13}-1)/2)$, $\mathfrak{M}^3_{n,j}(1,1,\epsilon)<0$ uniformly in $n,j$.

From Lemma~\ref{Lemma:Neighborhood1} we already have $\mathfrak{M}^{1,2}_{n,j}(1,1,\epsilon)<0$ for $\epsilon\in(1,5/2]$. Define
\[
M(\epsilon):=\max_{n,j}\max\{\mathfrak{M}^1_{n,j}(1,1,\epsilon),\ \mathfrak{M}^2_{n,j}(1,1,\epsilon),\ \mathfrak{M}^3_{n,j}(1,1,\epsilon)\}<0.
\]
By continuity of $\mathfrak{M}^{1,2,3}_{n,j}$ in $(a,b)$ and finiteness of the index set, there exist $\delta_a,\delta_b>0$ (uniform in $n,j$) such that
\[
\mathfrak{M}^{\ell}_{n,j}(a,b,\epsilon)<\tfrac12 M(\epsilon)<0,\quad \ell=1,2,3,
\]
whenever $|a-1|<\delta_a$ and $|b-1|<\delta_b$. Then, we can conclude the proof by setting $\alpha:=1-\delta_a$ and $\beta:=1+\delta_b$.
\hfill$\square$

For $\lambda\in \bigcup^{\mathbf{n}-1}_{n=1}(b\hat{\lambda}_n,\epsilon b\hat{\lambda}_n)$,  the behavior of $\pi_n(t)$ is characterized below.
\begin{proposition}\label{Prop:Convergence_J_Multi_Gap}
    Suppose that the conditions of Proposition~\ref{Prop:Convergence_J_mul} are satisfied. Moreover, assume that Condition~\ref{con:para_est_jump3} holds. Then, for all $\lambda\in\bigcup^{\mathbf{n}-1}_{n=1}(b\hat{\lambda}_n,\epsilon b\hat{\lambda}_n)$, 
        \begin{align*}
            &\mathbb{P}\big(\lim_{t\rightarrow \infty}\pi_{n}(t)=0\big)=1,\quad \forall n\in [\mathbf{n}]\setminus\{n^*,n^*+1\},\\
            &\mathbb{P}\big(\lim_{t\rightarrow \infty}\pi_{n^*}(t)+\pi_{n^*+1}(t)=1\big)=1.
        \end{align*}
\end{proposition}
\proof
Fix $n^*\in\{1,\dots,\mathbf{n}-1\}$ and $j\in[\mathbf{j}]$, and set $$\Delta^{n^*}_{j} = \left\{ (m, i) \in [n] \times [j] \mid i \neq j \text{ or } (i = j \text{ and } m < n^*) \right\}.$$ A direct computation shows $\Psi^{\mathsf{k},\bar{\mathsf{k}}}_{m,i,n^*,j}(\kappa)<0$ for all $\kappa\in(b,\epsilon b)$ and $(m,i)\in \Delta^{n^*}_{j}$.

Condition~\ref{con:para_est_jump3} ensures $\psi^{n^*,j}_{2}(\epsilon b)<0$, and hence $\psi^{n^*,j}_{d+1}(\epsilon b)-\psi^{n^*,j}_{d}(\epsilon b)<0$. Thus $\Psi^{\mathsf{k},\bar{\mathsf{k}}}_{m,j,n^*,j}(\kappa)<0$ for all $\kappa\in(b,\epsilon b)$ and $m\geq n^*+2$.
Thus, Using analogous reasoning as in the proof of Theorem~\ref{Thm:QSR-A-F}, we obtain
\begin{align*}
    &\mathbb{Q}^j\big(\lim_{t\rightarrow \infty}\pi_{n}(t)=0\big)=1,\quad \forall n\in [\mathbf{n}]\setminus\{n^*,n^*+1\},\\
    &\mathbb{Q}^j\big(\lim_{t\rightarrow \infty}\pi_{n^*}(t)+\pi_{n^*+1}(t)=1\big)=1.
\end{align*}
By applying similar arguments as in the proof of Theorem~\ref{Thm:QSR-F}, we conclude the proof.
\hfill$\square$

\medskip

Next, we investigate the asymptotic behavior of $\pi_{n^*}(t)$ and $\pi_{n^*+1}(t)$ for $\kappa=\lambda/\hat{\lambda}_{n^*}\in(b,\epsilon b)$ under the assumptions of Proposition~\ref{Prop:Convergence_J_Multi_Gap}. 
If $\kappa$ is sufficiently close to $b$ (resp. $\epsilon b$), then by continuity $\Psi^{\mathsf{k},\bar{\mathsf{k}}}_{m,i,n^*,j}(\kappa)<0$ for all $(m,i)\neq (n^*,j)$, which implies $\mathbb{P}\big(\lim_{t\rightarrow \infty}\pi_{n^*}(t)=1\big)=1$ (resp. $\mathbb{P}\big(\lim_{t\rightarrow \infty}\pi_{n^*+1}(t)=1\big)=1$). 
We now analyze the remaining cases. Define the disjoint subsets (possibly empty)
\begin{align*}
   \mathfrak{s}_{-}:= \{j\in[\mathbf{j}]\mid\textstyle\sum_{k=1}^{N_J} \bar{\Psi}^k_{n^*+1,j,n^*,j}(\Gamma) < 0\},\\
   \mathfrak{s}_{=}:= \{j\in[\textstyle\mathbf{j}]\mid\sum_{k=1}^{N_J} \bar{\Psi}^k_{n^*+1,j,n^*,j}(\Gamma) = 0\},\\
   \mathfrak{s}_{+}:= \{j\in[\mathbf{j}]\mid\textstyle\sum_{k=1}^{N_J} \bar{\Psi}^k_{n^*+1,j,n^*,j}(\Gamma) > 0\},
\end{align*}
so that $\mathfrak{s}_{-} \cup \mathfrak{s}_{=} \cup \mathfrak{s}_{+} = [\mathbf{j}]$. Using arguments similar to Proposition~\ref{Prop:Convergence_J_bound}, we conclude
\begin{itemize}
    \item For all $j\in \mathfrak{s}_{-}$, $\mathbb{Q}^j\big(\lim_{t\rightarrow \infty}\pi_{n^*}(t)=1\big)=1$;
     \item For all $j\in \mathfrak{s}_{+}$, $\mathbb{Q}^j\big( \lim_{t \rightarrow \infty} \frac{1}{t} \log \frac{\hat{\mathsf{q}}_{n^*+1,j}(t)}{\hat{\mathsf{q}}_{n^*,j}(t)}>0\big)=1$.
    Consequently, $\mathbb{Q}^j\big(\lim_{t \to \infty} \hat{\mathsf{q}}_{n^*,j}(t) = 0\big) = 1$, since $\hat{\mathsf{q}}_{n^*+1,j}(t) \leq 1$ almost surely. Thus, $\mathbb{Q}^j\big(\lim_{t \to \infty} \pi_{n^*+1}(t) = 1\big) = 1$.
    \item For all $j\in \mathfrak{s}_{=}$, we have
    \begin{align*}
        &\mathbb{Q}^j\big(\lim_{t\rightarrow \infty}\pi_{n^*}(t)=1\big)\in(0,1), \quad \mathbb{Q}^j\big(\lim_{t\rightarrow \infty}\pi_{n^*+1}(t)=1\big)\in(0,1),\\
        &\lim_{t\rightarrow \infty}\mathbb{E}_{\mathbb{Q}^j}\big(\pi_{n^*}(t)\big)=\lim_{t\rightarrow \infty}\mathbb{E}_{\mathbb{Q}^j}\big(\pi_{n^*+1}(t)\big)=1/2,\\
        &\lim_{t\rightarrow \infty}\mathrm{Var}_{\mathbb{Q}^j}\big(\pi_{n^*}(t)\big)=\lim_{t\rightarrow \infty}\mathrm{Var}_{\mathbb{Q}^j}\big(\pi_{n^*+1}(t)\big)=1/4.
    \end{align*}
\end{itemize}
Hence, 
\begin{equation*}
    \lim_{t\rightarrow \infty} \mathbb{E}(\pi_{n^*}(t))=\lim_{t\rightarrow \infty} \sum_{j}\mathbb{P}(\mathcal{R}=j)\mathbb{E}_{\mathbb{Q}^j}(\pi_{n^*}(t))=\sum_{j\in \mathfrak{s}_{-}}\mathbb{P}(\mathcal{R}=j)+\frac{1}{2}\sum_{j\in \mathfrak{s}_{=}}\mathbb{P}(\mathcal{R}=j)
\end{equation*}
and 
\begin{equation*}
    \lim_{t\rightarrow \infty} \text{Var}(\pi_{n^*}(t))=\lim_{t\rightarrow \infty} \sum_{j}\mathbb{P}(\mathcal{R}=j)\text{Var}_{\mathbb{Q}^j}(\pi_{n^*}(t))=\frac{1}{4}\sum_{j\in \mathfrak{s}_{=}}\mathbb{P}(\mathcal{R}=j).
\end{equation*}

Combining Proposition~\ref{Prop:Convergence_J_mul}, Proposition~\ref{Prop:Convergence_J_Multi_Gap}, and the above analysis, we conclude that, as in the diffusive case, each trajectory $\{\pi_n(t,\omega)\}$ with $\omega\in\Omega$ exhibits one of two behaviors:
\begin{enumerate}
    \item \textit{Single Convergence}: Exactly one trajectory converges to one, while all others converge to zero.
    \item \textit{Oscillatory Behavior}: Two trajectories oscillate between zero and one, while all others converge to zero.
\end{enumerate}
The distinction from the diffusive case lies in the admissible parameter set.  
In the diffusion setting, single convergence corresponds to $\lambda/\hat{\lambda}_{n}\in[a,b]$.  
In the jump setting with shot noise, the admissible range enlarges to $\lambda/\hat{\lambda}_{n}\in(a/\epsilon,\epsilon b)$ owing to the introduction of $\epsilon>1$.  
The interpretation of the oscillatory regime with finite data is analogous to the diffusion case: it may either reflect insufficient observation time to reveal single convergence, or genuine persistent oscillations between two neighboring trajectories.

To address these possibilities, we propose the following procedure for estimating $\lambda\in[\underline{\lambda}, \bar{\lambda}]$. 

\begin{algorithm}[H]
\caption{Parameter Estimation for $\theta_k$, $\zeta_{k,\bar{k}}$ and $\iota_k$}\label{alg:parameter_estimation_J_Mul}
\textit{Step 1:} Collect the measurement data $\{\mathsf{N}_k(t)\}_{t \leq T}$ for $k\in[N_J]$ for sufficiently large $T>0$\;

\textit{Step 2:} Given prior information $\lambda \in [\underline{\lambda}, \bar{\lambda}]$ with $\underline{\lambda}>0$ and the requirement $\lambda/\hat{\lambda}\in[\bar{a},\bar{b}]$ with $0<\bar{a}<1<\bar{b}$, determine $a$, $b$ and $\epsilon$ satisfying Conditions~\ref{con:para_est_jump2} and~\ref{con:para_est_jump3}. Define estimator sequence $\hat{\lambda}_n=(\epsilon b/a)^{n-1}\underline{\lambda}/a$ for $n \in [\mathbf{n}]$, where $\mathbf{n}=\lceil \log(\bar{\lambda}/\underline{\lambda})/\log(\epsilon b/a) \rceil$\;

\textit{Step 3:} Using $\{\hat{\lambda}_n\}_{n \in [\mathbf{n}]}$ and the measurement data, compute $\hat{\mathsf{q}}_{n,j}(T)$ and then calculate $\pi_n(T) = \sum^{\mathbf{j}}_{j=1}\hat{\mathsf{q}}_{n,j}(T)$ for $n \in [\mathbf{n}]$\;

\textit{Step 4:}\vspace{-1em}
\begin{itemize}
    \item \textit{Single Convergence:} If a unique $n^*$ exists such that $\pi_{n^*}(T)\approx 1$ and $\pi_n(T)\approx 0$ for $n \neq n^*$, then conclude that $\lambda/\hat{\lambda}_{n^*} \in [a,b]$
    \item \textit{Two Oscillatory Trajectories:} If both $\pi_{n^*}(T)$ and $\pi_{n^*+1}(T)$ remain non-negligible while others vanish, two explanations are possible:
      \begin{enumerate}
        \item[1.] $\lambda/\hat{\lambda}_{n^*}$ belongs to or close to $(b,\epsilon b)$, 
        \item[2.] Insufficient data. 
      \end{enumerate}
      In this case, refine the admissible interval from $[\underline{\lambda}, \bar{\lambda}]$ to $(a\hat{\lambda}_{n^*}/\epsilon, \epsilon b\hat{\lambda}_{n^*+1})$, update $(a,b,\epsilon)$ to $(\mathsf{a},\mathsf{b},\varepsilon)$ satisfying Conditions~\ref{con:para_est_jump2} and~\ref{con:para_est_jump3}, and recompute with the new estimator sequence then return to \textit{Step~3}. Persistent oscillation indicates \textit{insufficient data}. 
      \item \textit{Insufficient Data:} If more than two $\pi_n(T)$ remain significant, the data are insufficient. Collect new measurements and return to \textit{Step~1}.
\end{itemize}
\end{algorithm}

\subsection{Non-identifiability of $\mu$ and $\nu$}

Under the QND measurement scheme (Assumption~\ref{asm:qnd}), the Hamiltonian $H$, the perturbation-induced noise operators $\{A_k\}_{k\in[N_P]}$, and the measurement operators are simultaneously block-diagonalizable. By Theorem~\ref{Thm:QSR-F}, the asymptotic behavior of the quantum state $\rho(t)$ is therefore independent of $H$ and $\{A_k\}_{k\in[N_P]}$. Moreover, the measurement records $Y_k(t)$ and $\mathsf{N}_k(t)$ depend on $H$ and $\{A_k\}_{k\in[N_P]}$ only indirectly through $\rho(t)$. Hence, these parameters are not identifiable under this QND scheme.



It is nevertheless instructive to examine what occurs if our estimation procedure is formally applied to $\mu$ or $\nu$. Suppose that all parameters required by Condition~\ref{con:robust_ASME}, which ensures robust stability of the estimated filter, are known. Then for every $n\in[\mathbf{n}]$, we have $\hat{\eta}_{k,n}=\eta_k$, $\hat{\gamma}_{k,n}=\gamma_k$, and similarly for the other quantities. In this setting, we obtain
\begin{align*} 
\bar{\Phi}^k_{m,i,n,j}(\eta, \gamma) &= 2\eta_k \gamma_k \left( \Re\{l_{k,i}\} - \Re\{l_{k,j}\} \right)^2 = \Phi^k_{i,j}(\eta, \gamma) \geq 0, \\
\bar{\Psi}^k_{m,i,n,j}(\Gamma) &= -\Gamma_{k,j} \left( 1 - \frac{\Gamma_{k,i}}{\Gamma_{k,j}} + \log \frac{\Gamma_{k,i}}{\Gamma_{k,j}} \right) = \Psi^k_{i,j}(\Gamma) \geq 0. 
\end{align*}
However, for any $m\neq n$, one finds $\bar{\Phi}^k_{m,i,n,i}(\eta, \gamma)=\bar{\Psi}^k_{m,i,n,j}(\Gamma)=0$, violating the strict positivity required in Condition~\ref{con:robust_ASME}. Thus the parameters $\mu$ and $\nu$ are not identifiable. Artificially assigning surrogate values to $\hat{\eta}$, $\hat{\gamma}$, or $\hat{\Gamma}$ can restore positivity, but such values bear no relation to the true system parameters. In this case, the estimation algorithm converges to these ad hoc quantities rather than to the physical parameters of interest.


\section{Conclusion}

We have developed a robust and computationally efficient framework for parameter estimation in open quantum systems governed by jump–diffusion stochastic master equations under QND measurements. By extending the stability theory of quantum filters to incorporate both initial-state mismatch and parameter uncertainty, we constructed a family of reduced-order filters enabling consistent parameter estimation over continuous domains, with significantly reduced complexity.

Future work will extend this framework beyond the QND setting to general measurement schemes, and to infinite-dimensional systems where reduced-order representations must be carefully adapted. These directions are essential for broadening the applicability of robust quantum parameter estimation to realistic experimental platforms.

\section*{Acknowledgements}
The authors thank Valery Ugrinovskii and Matthew R. James for very enlightening discussions.

\section*{Declarations}

\begin{itemize}
\item \textbf{Funding} This research was supported by the Australian Research Council Future Fellowship Funding Scheme under Project FT220100656 and the Discovery Project Funding Scheme under Projects DP200102945, DP210101938, DP240101494.
\item \textbf{Conflict of interest} The authors have no conflict of interest to declare that are relevant to the content of this article.
\item \textbf{Data availability} Data sharing not applicable to this article as no datasets were generated or analysed during the current study.
\end{itemize}






\begin{appendices}

\section{Infinitesimal generator}\label{secA1}




\begin{lemma}\label{Lemma:GeneratorSME}
For any twice continuously differentiable function $V: \mathcal{S}(\mathbb{H})\times \mathbb{R}_{+}  \rightarrow \mathbb{R}$, the corresponding infinitesimal generator with respect to SME~\eqref{Eq:J-D SDE} is given by
\begin{align}
\mathscr{L}V(\rho,t)
&=\partial_t V(\rho,t)
+\big\langle \mathfrak{D}_\rho V(\rho,t),\,\mathcal{L}(\rho)\big\rangle_{\mathrm{HS}}
+\frac12\sum_{k=1}^{N_D} \mathfrak{D}_\rho^2 V(\rho,t)\big[\mathcal{G}_k(\rho),\mathcal{G}_k(\rho)\big]\nonumber\\
&\quad+\sum_{k=1}^{N_J}\Big(V\!\big(\mathcal{J}_k(\rho)/\mathcal{T}_k(\rho)\big)-V(\rho)
-\big\langle \mathfrak{D}_\rho V(\rho,t),\,\mathcal{Q}_k(\rho)\big\rangle_{\mathrm{HS}}\Big)\,\mathcal{T}_k(\rho),
\label{Eq:GeneratorSME}
\end{align}
where $\mathfrak{D}_\rho V$ is the Fréchet derivative of $V$ at $\rho$ and $\mathfrak{D}_\rho^2V$ the continuous symmetric bilinear form.
\end{lemma}

\proof
Let $K$ be a finite constant such that $c> \|C_kC_k^*\|$ for all $k$. Due to Cauchy Schwarz inequality and the fact $\Tr(\rho^2)<1$ for all $\rho\in\mathcal{S}(\mathbb{H})$, we have $\Tr(C_k\rho C^*_k)\leq \|C_kC_k^*\|< c$. Then, we can rewrite SME~\eqref{Eq:J-D SDE} as follows,
\begin{align*}
    d\rho(t) = \mathcal{L}(\rho(t-))dt&+\sum^{N_D}_{k=1}\mathcal{G}_{k}(\rho(t-))dW_k(t) + \sum^{N_J}_{k=1}\int_{|z|<c}\tilde{\mathcal{Q}}_{k}(\rho(t-),z)\big(N_k(dz,dt)-dz dt \big),  
\end{align*}
where $\tilde{\mathcal{Q}}_{k}(\rho,z)=\mathcal{Q}_{k}(\rho)\mathds{1}_{\{0<z<\mathcal{T}_k(\rho)\}}$. 
Apply the It\^o formula~\cite{applebaum2009levy} to $V(t,\rho_t)$, one gets
\begin{align*}
    \mathscr{L}V(\rho,t)=&\partial_t V+\langle \mathfrak{D}_\rho V,\mathcal{L}\rangle_{\mathrm{HS}}+\frac12\sum_k \mathfrak{D}_\rho^2V[\mathcal{G}_k,\mathcal{G}_k]\\
    &+\sum_{k=1}^{N_J}\int^c_{0}V(\rho+\tilde{\mathcal{Q}}_{k}(\rho,z))-V(\rho)-\langle D_\rho V,\tilde{\mathcal{Q}}_{k}(\rho,z)\rangle_{\mathrm{HS}}dz
\end{align*}
Since \begin{align*}
   & V\big(\rho+\tilde{\mathcal{Q}}_{k}(\rho,z)\big)-V(\rho)-\Tr\big(\mathfrak{D}_{\rho}V \tilde{\mathcal{Q}}_{k}(\rho,z)\big)  \\
    &= \Big(V\big(\mathcal{J}_{k}(\rho)/\mathcal{T}_{k}(\rho)\big)-V(\rho)-\Tr\big(\mathfrak{D}_{\rho}V \mathcal{Q}_{k}(\rho) \big)\Big) \mathds{1}_{\{0<z<\mathcal{T}_k(\rho)\}},
\end{align*}
and 
$\int_0^c \mathds{1}_{\{0<z<\mathcal{T}_k(\rho)\}} dz = \mathcal{T}_k(\rho)$, we obtain \eqref{Eq:GeneratorSME}.
\hfill$\square$
\end{appendices}

\bibliographystyle{plain}
\bibliography{ref_Arxiv}

\end{document}